\documentclass{article}

\usepackage{arxiv}

\usepackage[utf8]{inputenc} 
\usepackage[T1]{fontenc}    
\usepackage{hyperref}       
\usepackage{url}            
\usepackage{booktabs}       
\usepackage{amsmath,amssymb,array}
\usepackage{nicefrac}       
\usepackage{microtype}      
\usepackage{lipsum}		
\usepackage{graphicx}
\usepackage{doi}
\usepackage{multirow}
\usepackage[authoryear]{natbib}
\usepackage{verbatim}
\usepackage{enumitem}
\usepackage{rotating}
\usepackage{xcolor}
\usepackage{caption}
\usepackage[ruled]{algorithm2e}
\usepackage{booktabs}
\usepackage{soul}

\newcommand{\gr}[1]{\colorbox{gray!30}{\ensuremath{{#1}^{*}}}}

\newcommand{\bs}{\boldsymbol}

\title{On the identifiability of Bayesian factor analytic models}

\author{ Panagiotis Papastamoulis\thanks{This research was funded by the  program ``DRASI 1/Research project: 11338201'', coordinated by the Research Center of the Athens University of Economics and Business (RC/AUEB). Technical support was provided from the Computational and Bayesian Statistics Lab, Department of Statistics, Athens University of Economics and Business.  The comments and suggestions of two anonymous reviewers substantially improved the presentation and findings of the paper.} \\
	Department of Statistics\\
	Athens University of Economics and Business\\
	Athens, Greece \\
	\texttt{papastamoulis@aueb.gr} \\
	\And
	Ioannis Ntzoufras \\
	Department of Statistics\\
	Athens University of Economics and Business\\
	Athens, Greece \\
	\texttt{ntzoufras@aueb.gr} \\
}

\renewcommand{\shorttitle}{Identifiability of Bayesian FA models}

\hypersetup{
pdftitle={Identifiability},
pdfsubject={stat.CO},
pdfauthor={Panagiotis Papastamoulis, Ioannis Ntzoufras},
pdfkeywords={Markov chain Monte Carlo, Post-processing, Objective Function, Simulated Annealing, Assignment Problem},
}

\begin{document}
\maketitle

\begin{abstract}
A well known identifiability issue in factor analytic models is the invariance with respect to orthogonal transformations. This problem burdens the inference under a Bayesian setup, where Markov chain Monte Carlo (MCMC) methods are used to generate samples from the posterior distribution. We introduce a post-processing scheme in order to deal with rotation, sign and permutation invariance of the MCMC sample. The exact version of the contributed algorithm requires to solve $2^q$ assignment problems per (retained) MCMC iteration, where $q$ denotes the number of factors of the fitted model. For large numbers of factors two approximate schemes based on simulated annealing are also discussed. We demonstrate that the proposed method leads to interpretable posterior distributions using synthetic and publicly available data from typical factor analytic models as well as mixtures of factor analyzers. An R package is available online at CRAN web-page.
\end{abstract}


\section{Introduction}
\label{sec:intro}

Factor Analysis (FA) is used to explain relationships among a set of observable responses using latent variables. This is typically achieved by expressing the observed multivariate data as a linear combination of a set of unobserved and uncorrelated variables of considerably lower dimension, which are known as factors.
Let $Y_i = (Y_{i1},\ldots,Y_{ip})^\top$ denote the $i$-th observation of a random sample of $p$ dimensional observations with $Y_i\in\mathbb R^{p}$; $i= 1,\ldots,n$. Let $\mathcal N_p(\mu,\bs\Sigma)$ denote the $p$-dimensional normal distribution with mean and covariance matrix $\mu=(\mu_1,\ldots,\mu_p)\in\mathbb R^p$ and  $\bs\Sigma$, respectively, and also denote by $\bs{\mathrm{I}}_p$ the $p\times p$ identity matrix.


In the typical FA model (see \textit{The fundamental factor theorem} in Chapter II of \cite{thurstone1934vectors} or the \textit{strict factor structure} of \cite{10.2307/1912275}), $Y_i$ is expressed as a linear combination of a latent vector of factors $F_i\in\mathbb R^{q}$
\begin{equation}
\label{eq:fa}
Y_i = \mu + \bs\Lambda F_i + \varepsilon_i,\quad i = 1,\ldots,n
\end{equation}
where $q>0$ denotes the number of factors, which is assumed fixed. The $p\times q$ dimensional matrix $\bs\Lambda = (\lambda_{rj})$ contains the factor loadings, while $\mu = (\mu_1,\ldots,\mu_p)$ contains the marginal mean of $Y_i$. The unobserved vector of factors $F_i=(F_{i1}, \ldots,F_{iq})^\top$ lies on a lower dimensional space, that is, $q < p$ and we will consider the special case where it consists of uncorrelated and homoscedastic features
\begin{equation}
\label{eq:y}
F_i \sim \mathcal N_q(0_q,\bs{\mathrm{I}}_q),
\end{equation}
independent for $i = 1,\ldots,n$, where $0_q:=(0,\ldots,0)^\top$ and $\bs{\mathrm{I}}_q$ denotes the $q\times q$ identity matrix. 

The terms $\varepsilon_i$ (commonly referred to as \textit{uniquenesses} or \textit{idiosyncratic errors})   are independent from $F_i$, that is, $
\mathrm{Cov}(F_{ij},\varepsilon_{ik})=0$, $\forall j=1,\ldots,q$; $k=1,\ldots,p$ and normally distributed
 \begin{equation}\label{eq:errors}
\varepsilon_i\sim \mathcal N_p( 0_p,\bs\Sigma)
 \end{equation}
 independent for $=1,\ldots,n$. Furthermore, $\varepsilon_i$ consists of independent random variables $\varepsilon_{i1},\ldots,\varepsilon_{ip}$, that is,
 \begin{equation}
 \label{eq:diag}
\bs\Sigma = \mbox{diag}(\sigma_1^2, \ldots,\sigma_p^2).
 \end{equation}
 
The distribution of $Y_i$ conditional on $F_i$ is
 \begin{equation}
 \label{eq:x_given_y}
Y_i|F_i \sim \mathcal N_p(\mu+\bs\Lambda F_i,\bs\Sigma),\ \mbox{independent for}\  i=1,\ldots,n
 \end{equation}
and corresponding marginal distribution 
\begin{equation}
\label{eq:x_marginal}
Y_i \sim \mathcal N_p(\mu,\bs\Lambda\bs\Lambda^\top + \bs\Sigma), \quad \mbox{iid for }i = 1,\ldots,n.
\end{equation}

Without loss of generality we can assume that $\mu=0_p$, although this is not a necessary assumption.  According to Equation \eqref{eq:x_marginal}, the covariance matrix of the marginal distribution of $Y_i$ is equal to $\bs\Lambda\bs\Lambda^\top + \bs\Sigma$. Thus, the latent factors are the only source of correlation among the measurements. This is the crucial characteristic of factor analytic models, where they aim to explain high-dimensional dependencies using a set of lower-dimensional uncorrelated factors \citep{kim1978factor, bartholomew2011latent}. 

There are two sources of identifiability problems  regarding the typical FA model in Equations 
\ref{eq:fa}--\ref{eq:x_marginal}. 
The first one concerns identifiability of $\bs\Sigma$ and the second one concerns identifiability of $\bs\Lambda$. Assuming that $\bs\Sigma$ is identifiable (see Section \ref{sec:identify}), we are concerned with identifiability of $\bs\Lambda$. It is well known that the factor loadings ($\bs\Lambda$) in Equation \eqref{eq:fa} are only identifiable up to orthogonal transformations. This identifiability issue is not of great practical importance within a frequentist context: the likelihood equations are satisfied by an infinity of solutions, all equally good from a statistical perspective  \citep{lawley1962factor}. 

On the other hand, under a Bayesian setup it complicates the inference procedure, where MCMC methods are applied to generate samples from the posterior distribution $f(\bs\Lambda,\bs\Sigma,\bs F|\bs y)$. 
Clearly, the invariance property makes the posterior distribution multimodal. Provided that the MCMC algorithm has converged to the target distribution, the MCMC sample will be consecutively switching among the multiple modes of the posterior surface. Despite the fact that this identifiability problem has no bearings on predictive inference or estimation of the covariance matrix in Equation \eqref{eq:x_marginal}, factor interpretation remains challenging because both $\bs\Lambda$ and $F_i$; $i = 1, \ldots,n$ are not marginally identifiable. Therefore, the standard practice of providing posterior summaries via ergodic means, or reporting Bayesian credible intervals for factor loadings becomes meaningless due to rotation invariance of the MCMC sample.

Typical implementations of the Bayesian paradigm in FA models use inverse gamma priors on the error variances and normal or truncated normal priors on the factor loadings \citep{arminger1998bayesian, song2001bayesian}. In such cases the model is conditionally conjugate and an MCMC sample can be generated by standard Gibbs sampling \citep{gelfand}. 
However, the  factor loadings  are not marginally identifiable if $\bs \Lambda$ is not constrained. Consequently, when MCMC methods are used for estimation of the FA model, inference is not straightforward. 
\color{black} 
On the other hand,  in standard factor models, certain identifiability constraints induce undesirable properties, such as a priori order dependence in the off-diagonal entries of the covariance matrix \citep{bhattacharya2011sparse}. 
Although tailored 
methods (briefly reviewed in Section \ref{sec:identify}) for achieving identifiability and for drawing inference on sparse FA models exist \citep{CONTI201431, mavridis_ntzoufras_2014, doi:10.1080/01621459.2015.1100620, kaufmann2017identifying}, they require extra modelling effort.

\cite{AMANN2016190}  introduced an alternative approach where the rotation problem is solved ex-post. In this paper, a post-processing approach is also followed. It is demonstrated that the proposed method successfully deals with the non-identifiability of the marginal posterior distribution $f(\bs\Lambda|\bs y)$ and leads to interpretable conclusions. The number of factors ($q$) is considered fixed, nevertheless a by-product of our implementation is that it can help to reveal cases of overfitting, by simply inspecting simultaneous credible regions of factor loadings. 

We propose to correct invariance of simulated factor loadings using a two-stage post-pro\-ces\-sing approach. At first we focus on generic rotation invariance, that is, to achieve a \textit{simple structure} of factor loadings per MCMC iteration. A factor model with simple structure is  one where each measurement is related to at most one latent factor \citep{thurstone1934vectors}. Varimax rotations \citep{kaiser1958varimax} are used for this task. After this step, all measurements load at most on one factor while the rest of the loadings are small (close to zero). However,  the rotated loadings are still  not identifiable across the MCMC trace due to \textit{sign} and \textit{permutation} invariance. Sign switching stems from the fact that we can simultaneously switch the signs of $F_i$ and $\bs\Lambda$ without altering $\bs\Lambda F_i$. Permutation invariance (or \textit{column switching}, according to \cite{CONTI201431}) is due to the fact that there is no natural ordering of the columns of the factor loading matrix. Thus, factor labels can change as the MCMC sampler progresses. That being said, the second step is to correct invariance due to specific orthogonal transformations which correspond to \textit{signed permutations} across the MCMC trace.

The rest of the paper is organized as follows. Section \ref{sec:identify} presents the identifiability issues of the FA model and briefly discusses related work.  Section \ref{sec:notation} gives some background on rotations and signed permutations. The contributed method is introduced in Section \ref{sec:main}. Three approaches for minimizing the underlying objective function are described in Section \ref{RSP_strategies}.  Section \ref{sec:results} applies the proposed  method using simulated (Section \ref{sec:sim}) and  real (Section \ref{sec:real})  data. Finally, an application to a model-based clustering problem is given in Section \ref{sec:mix}. An Appendix contains  geometrical illustrations of the proposed method and additional applications and computational aspects of our method.

\section{Identifiability problems and related approaches}\label{sec:identify}
At first we review some well known results that ensure identifiability of $\bs\Sigma$ (the \textit{uniqueness problem}) and will be explicitly followed in our implementation. Given that there are $q$ factors, the number of free parameters in the covariance matrix $\bs\Lambda\bs\Lambda^\top + \bs\Sigma$ is equal to $p + pq - \frac{1}{2}q(q-1)$ \citep[see Section 5 in][]{lawley1962factor}. The number of free parameters in the unconstrained covariance matrix of $Y_i$ is equal to $\frac{1}{2}p(p+1)$. Hence, if a strict factor structure is present in the data, the number of parameters in the covariance matrix is reduced by
$$
 \frac{1}{2}p(p+1) - \left[p + pq - \frac{1}{2}q(q-1)\right] = \frac{1}{2}\left[(p-q)^2-(p+q)\right].
$$
The last expression is positive if $q < \phi(p)$ where $\phi(p):=\frac{2p + 1 - \sqrt{8p+1}}{2}$, a quantity which is known as the Ledermann bound \citep{ledermann1937rank}.  When $q < \phi(p)$ it can be shown that $\bs\Sigma$ is almost surely unique \citep{BEKKER1997255}. We assume that the number of latent factors does not exceed $\phi(p)$. 

Note however that non-identifiability is not necessarily taking place when $q>\phi(p)$. A special case of the general FA model is to assume an \textit{isotropic} error model, that is, $\bs \Sigma = \sigma^2\bs{\mathrm{I}}_p$ in Equation \eqref{eq:diag}, resulting in the \textit{probabilistic principal component analysis} framework of \cite{tipping1999probabilistic}. In such a case, the number of factors can be as large as $p-1$. 

Another aspect of the 	identifiability of $\bs \Sigma$ are instances of the so-called ``Heywood cases'' \citep{heywood1931finite} phenomenon. Under maximum likelihood strategies, it frequently happens that the optimal solution lies outside the parameter space, that is, $\sigma^2_r<0$ for one or more $r=1,\ldots,p$. This problem is explicitly avoided under a Bayesian setup, where the prior distribution of each idiosyncratic variance (typically a member of the inverse Gamma family of distributions) yields a posterior distribution with support on $(0,\infty)$. However, it may be frequently the case that the posterior distributions of idiosyncratic variances are multimodal, with one mode lying in areas close to zero. According to   \cite{bartholomew2011latent} (Section 3.12.3):  \begin{quote}
There is no inconsistency of a zero residual variance, it would simply mean that the variation of the manifest variable in question was wholly explained by the latent variable.
\end{quote}
If this is unsatisfactory, then one may bound the posterior distribution of idiosyncratic variances away from zero, using the specification of the prior parameters of \cite{fruhwirth2018sparse}.

Given identifiability of $\bs\Sigma$, a second source of identifiability problems is related to  orthogonal transformations of the matrix of factor loadings, which is the main focus of this paper. A square matrix $\bs R$ is an  \textit{orthogonal}  matrix if and only if $\bs R^{\top}=\bs R^{-1}$, that is, its inverse equals its transpose. The determinant of any orthogonal matrix is equal to $1$ or $-1$. Orthogonal matrices represent rotations which may be proper (if the determinant is positive) or improper (otherwise). Consider a $q\times q$ orthogonal matrix $\bs R$   and define $\widetilde F_i = \bs R F_i$. It follows that the representation $Y_i = \mu+\bs\Lambda\bs R^{\top}  \widetilde F_i + \varepsilon_i$ leads to the same marginal distribution of $Y_i$ as the one in Equation \eqref{eq:x_marginal}. 
Since the likelihood is invariant under orthogonal transformations, the posterior distribution will typically exhibit many modes. Essentially, the rotational invariance is a consequence of assumption \eqref{eq:y}, which  ultimately yields a marginal covariance matrix of the form $\bs\Lambda\bs\Lambda^\top + \bs\Sigma$ in \eqref{eq:x_marginal}. Assuming heteroscedastic latent factors of the form $F_i\sim\mathcal N_q(0_q, \bs\Psi)$ instead, where $\bs\Psi$ is a $q\times q$ diagonal matrix with non-identical entries, the marginal covariance matrix is $\bs\Lambda\bs\Psi\bs\Lambda^\top + \bs\Sigma$. In such a case, the posterior distribution of factor loadings is free of rotational ambiguities but not permutations.

A popular technique \citep{geweke1996measuring, fokoue2003mixtures, West03bayesianfactor,  lopes2004bayesian,lucas2006sparse,carvalho2008high, mavridis_ntzoufras_2014, papastamoulis2018overfitting, papastamoulis2019clustering} in order to deal with rotational invariance in Bayesian FA models relies on a lower-triangular expansion of $\bs\Lambda$, first suggested by \cite{anderson1956statistical}, that is:
\begin{equation}\label{eq:lambda}
\bs\Lambda = \begin{pmatrix}
\lambda_{11} & 0 & \cdots & 0\\
\lambda_{21} & \lambda_{22} & \cdots & 0\\
\vdots & \vdots & \ddots & \vdots\\
\lambda_{q1} & \lambda_{q2} & \cdots & \lambda_{qq}\\
\vdots & \vdots & \ddots & \vdots\\
\lambda_{p1} & \lambda_{p2} & \cdots & \lambda_{pq}
\end{pmatrix}.
\end{equation}
However this approach still fails to correct the invariance due to sign-switching across the MCMC trace; see,  for example, Figure 2.(b) in \cite{papastamoulis2018overfitting}. Additional constraints are introduced for addressing this issue, e.g.~by assuming that the diagonal elements are strictly positive \citep{geweke1996measuring} or even fixing all diagonal elements to 1 \citep{10.2307/1392266}.  Besides  the  upper  triangle  of  the  loading  matrix  that  is  fixed  to  zero a-priori, the remaining elements in the lower part of the matrix are also allowed to take values in areas close to zero (e.g.~this is the case when the first variable does not load on any factor). In such a case, identifiability of $\bs\Lambda$ is lost; see Theorem 5.4 in \cite{anderson1956statistical}. Of course this problem can be alleviated by suitably reordering the variables, however the choice of the first $q$ response variables is  crucial \citep{carvalho2008high}. In our approach we will not consider any constraint in the elements of the factor loadings matrix.

\cite{CONTI201431} augment the FA model with a binary matrix, indicating the latent factor on which each variable loads. They also consider an extension of the model by allowing correlation among factors. Under suitable identification criteria, a prior distribution restricts the MCMC sampler to explore regions of the parameter space corresponding to models which are identified up to column and sign switching. Then, they deal with sign and column switching by using simple reordering heuristics which are driven by  the existence of zeros in the resulting loading matrices. At each MCMC iteration, the non-zero columns are reordered such that the top elements appear in increasing order. Next, sign-switching is treated by using a benchmark factor loading  (e.g., the factor loading with the highest posterior probability of being different from zero in each column) and then switching the signs at each  MCMC iterations in order to agree with the benchmark. 

\cite{mavridis_ntzoufras_2014} place a normal mixture prior on each element of $\bs\Lambda$ and introduce an additional set of latent binary indicators which is used to identify whether an item is associated with the corresponding factor. They also reorder the items such that important non-zero loadings are placed in the diagonal of $\bs \Lambda$ in Equation \eqref{eq:lambda}. 
\cite{doi:10.1080/01621459.2015.1100620} 
identify the FA model by expanding the parameter space using an auxiliary parameter matrix which drives the implied rotation. The whole procedure is fully model driven and it is implemented through an Expectation-Maximization type algorithm. Additionally, the varimax rotation is suggested every few iterations of the algorithm to stabilize and speed up the convergence of the algorithm. 

\cite{kaufmann2017identifying} proposed a $k$-medoids clustering \citep{kaufmannbook} approach in order to deal with the rotation problem in sparse Bayesian factor analytic models, considering a point mass normal mixture prior distribution of the loading matrix (see also \cite{doi:https://doi.org/10.1002/9781119995678.ch10}). A $k$‐medoids clustering to simultaneously estimate the factor‐cluster representatives and allocate each draw uniquely to the estimated clusters is then applied, using the absolute correlation distance, which accounts for factors that are subject to sign-switching. In a final step, sign identification is achieved by switching the sign of each factor and its corresponding loadings to obtain a majority of nonzero positive factor loadings. 

\cite{AMANN2016190} fix the rotation problem ex-post for static and dynamic factor models without placing any constraint on the parameter space. This is achieved by transforming the MCMC output using a sequence of orthogonal matrices, that is, Orthogonal Procrustes rotations which  minimize the posterior expected loss. More details for this method are given in Section \ref{sec:main}, where we discuss the similarities and differences with respect to the proposed approach.

\section{Method}\label{sec:method}

\subsection{Notation and definitions}\label{sec:notation}

Let $\mathcal T_q$ denote the set of all $q!$ permutations of $\{1,\ldots,q\}$. For a given $\nu = (\nu_1,\ldots,\nu_q)\in\mathcal T_q$ consider the matrix $\bs{\dot\Lambda}$ where the $j$-th column of $\bs\Lambda$ is mapped to the $\nu_j$-th column  of $\bs{\dot\Lambda}$, for $j=1,\ldots,q$. This transformation can be expressed using a $q\times q$ \textit{permutation matrix} $\bs P$ whose $i,j$ element is equal to $$(\bs P)_{ij}=\begin{cases}
1,&\quad \mbox{if}\quad  \nu_i = j\\
0,&\quad \mbox{otherwise}
\end{cases}, i=1,\ldots,q;j=1,\ldots,q.$$
The transformed matrix is simply written as $\bs{\dot\Lambda} = \bs\Lambda \bs P$. A permutation matrix is a special case of a rotation matrix. 

For example, assume that $\bs\Lambda$ has three columns and that  $\nu=(3,1,2)\in\mathcal T_3$. This means that column 1, 2, 3 becomes column 3, 1, 2, respectively. This transformation corresponds to the $3\times 3$ permutation matrix 
\[
\bs P=\begin{pmatrix}
0 & 0 & 1\\
1 & 0 & 0\\
0 & 1 & 0
\end{pmatrix}.
\]
Obviously, $\bs\Lambda \bs P$ results to the desired reordering of columns:
\begin{align*}
\bs{\dot\Lambda}=\bs\Lambda\bs P&=
\begin{pmatrix}
\lambda_{11} & \lambda_{12} & \lambda_{13}\\
\vdots & \vdots & \vdots \\
\lambda_{p1} & \lambda_{p2} & \lambda_{p3}
\end{pmatrix}
\begin{pmatrix}
0 & 0 & 1\\
1 & 0 & 0\\
0 & 1 & 0
\end{pmatrix}
=\begin{pmatrix}
\lambda_{12} & \lambda_{13} & \lambda_{11}\\
\vdots & \vdots & \vdots \\
\lambda_{p2} & \lambda_{p3} & \lambda_{p1}
\end{pmatrix}.
\end{align*}

A \textit{signed permutation matrix} is a square matrix which has precisely one nonzero entry in every row and column and whose only nonzero entries are $1$ and/or $-1$ \cite[see e.g.~][]{snapper1979characteristic}. A $q\times  q$ signed permutation matrix $\bs Q$ can be expressed as \begin{equation}\label{eq:qsp}\bs Q=\bs S\bs P,\end{equation} where $\bs S = \mathrm{diag}(s_1,\ldots,s_q)$ is a $q\times q$ diagonal matrix with diagonal entries equal to $s_j\in\{-1,1\}$, $j=1,\ldots,q$ and $\bs P$ is a $q\times q$ permutation matrix. For example, consider that
\begin{align*}
\bs Q&=\begin{pmatrix}
0 & 0 & -1\\
-1 & 0 & 0\\
0 & 1 & 0
\end{pmatrix}
=\begin{pmatrix}
-1&0&0\\
0&-1&0\\
0&0&1
\end{pmatrix}
\begin{pmatrix}
0 & 0 & 1\\
1 & 0 & 0\\
0 & 1 & 0
\end{pmatrix}.
\end{align*}

It is evident that applying a signed permutation $\bs Q$ to a $p\times q$ matrix $\bs \Lambda$ results to a transformed matrix $\bs{\mathring\Lambda}$ arising from the corresponding signed permutation of its columns. For example
\begin{align*}
\bs{\mathring\Lambda}=\bs\Lambda \bs Q = \bs \Lambda \bs S\bs P&= 
\begin{pmatrix}
-\lambda_{12} & \lambda_{13} & -\lambda_{11}\\
\vdots & \vdots & \vdots \\
-\lambda_{p2} & \lambda_{p3} & -\lambda_{p1}
\end{pmatrix}
\end{align*} 
is generated by first switching the signs of the first two columns and then reordering the columns according to $\nu=(3,1,2)$. 

Operations and/or special matrices of factor loadings will be denoted as follows: \\
\begin{tabular}{ll}
$\bs{\widetilde\Lambda}$ & varimax rotated\\
$\bs{\dot\Lambda}$& column-permuted\\ 
$\bs{\mathring\Lambda}$& sign-permuted\\ 
$\bs\Lambda^\star$ & reference matrix of factor loadings.\\
\end{tabular}

\subsection{Rotation-Sign-Permutation post-processing algorithm}\label{sec:main}

Given a $p \times q$ matrix $\bs \Lambda$ of factor loadings, the varimax problem \citep{kaiser1958varimax}  is to find  a $q \times q$ rotation matrix $\bs\Phi$ such that the sum of the within-factor variances of squared factor loadings of the rotated matrix of loadings $\bs{\widetilde\Lambda} = \bs\Lambda \bs\Phi$ is maximized. That is, the optimization problem is now summarized by  
\begin{align}
\label{eq:varimax}
\mbox{maximize} & \quad \frac{1}{4}\sum_{j=1}^{q}\left[\left(\sum_{r = 1}^{p}\widetilde\lambda^4_{rj}\right)-\frac{1}{p}\left(\sum_{r=1}^{p}\widetilde\lambda^2_{rj}\right)^2\right]\\
\nonumber
\mbox{subject to} &\quad \bs\Phi^\top \bs\Phi = \bs{\mathrm{I}}_q\\
\nonumber
\mbox{where}&\quad \widetilde\lambda_{rj}=\sum_{k=1}^{q}\lambda_{rk}\phi_{kj},\quad r=1,\ldots,p;\ j=1,\ldots,q.
\end{align}
The original approach for solving the varimax problem was to increase the objective function by successively rotating pairs of factors \citep{kaiser1958varimax}. Subsequent developments based on matrix formulations of the varimax problem involved the simultaneous rotation of all factors to improve the  objective function \citep{sherin1966matrix, neudecker1981matrix, ten1984joint}. More recently, it has been argued \citep{rohe2020vintage} that varimax rotation provides a unified spectral estimation strategy for a broad class of modern factor models.

We used the {\tt varimax()} base function in {\tt R} to solve the varimax problem. The optional arguments {\tt normalize = FALSE} and {\tt eps = 1e-5} (default) were used. We note that setting  {\tt normalize = TRUE} has no obvious impact in our results. Assuming that we have at hand a simulated output of factor loadings $\bs\Lambda^{(t)}$, $\bs\Lambda^{(t)}=\big(\lambda_{rj}^{(t)}\big)$, $t=1,\ldots,T$,  we denote as $\bs{\widetilde\Lambda}^{(t)}=\big(\widetilde\lambda_{rj}^{(t)}\big)$, $t=1,\ldots,T$ the rotated MCMC output, after solving the varimax problem per MCMC iteration, where $T$ is the size of MCMC iterations.

After solving the varimax problem for each MCMC iteration, the second stage of our solution is to apply signed permutations to the MCMC output until the transformed loadings are sufficiently close to a reference value denoted as $\bs\Lambda^\star$. For instance we will assume that $\bs\Lambda^\star$ corresponds to a fixed matrix, however we will relax this assumption later. Let $\mathcal Q_q$ denote the set of $q\times q$ signed permutation matrices. The optimization problem is stated as
\begin{align}
\label{eq:problem1.5}
\mbox{minimize}&
\quad\sum_{t=1}^{T}||\bs{\widetilde\Lambda}^{(t)} \bs Q^{(t)} - \bs\Lambda^{\star}||^2\\
\nonumber
\mbox{subject to} &\quad \bs Q^{(t)}\in\mathcal Q_q, \quad t = 1,\ldots,T,
\end{align}
where $||\bs A|| = \sqrt{\sum_i\sum_j\alpha_{ij}^2}$ denotes the Frobenius norm on the matrix space. Notice that Equation \eqref{eq:problem1.5} subject to $\bs Q^{\top}\bs Q=\bs{\mathrm{I}}_q$ is known as the Orthogonal Procrustes problem and an analytical solution (based on the singular value decomposition of $\bs\Lambda^{\star} \left(\bs\Lambda^{(t)}\right)^{\top}$) is available \citep{schonemann1966generalized}, which is the core in the post-processing approach of \cite{AMANN2016190}. Essentially, we are solving the same problem with \cite{AMANN2016190} but using a restricted parameter space, that is, the (discrete) set of $q\times q$ signed permutation matrices $\mathcal Q_q$, instead of the set of all orthogonal matrices. 

By Equation \eqref{eq:qsp} it follows that 
\begin{align*}
||\bs{\widetilde\Lambda}\bs Q - \bs{\Lambda}^\star||^2 = ||\bs{\widetilde\Lambda}\bs S \bs P - \bs\Lambda^{\star}||^2=\sum_{r=1}^{p}\sum_{j=1}^q \left(
s_j\widetilde\lambda_{r\nu_j}-
\lambda_{rj}^{\star}\right)^2,
\end{align*}
where $\nu=(\nu_1,\ldots,\nu_q)\in\mathcal T_q$ denotes the permutation vector corresponding to the permutation matrix $\bs P$. Thus, \eqref{eq:problem1.5} can be also written as
\begin{align}\label{eq:problem2}
\min
\Psi(\bs\Lambda^\star, s, \nu )=&\sum_{t=1}^{T}\mathcal L_{s^{(t)},\nu^{(t)}}^{(t)}\\
\nonumber
=&\sum_{t=1}^{T}\sum_{r=1}^{p}\sum_{j=1}^{q} \left(s_j^{(t)} \widetilde\lambda_{r\nu_j^{(t)}}^{(t)}- \lambda^{\star}_{rj}\right)^2
\\
\nonumber
\mbox{subject to} \quad s_j^{(t)}\in\{-1,1\},& \quad t = 1,\ldots,T; j = 1,\ldots,q  \\
\nonumber
\mbox{and}\quad\nu^{(t)}\in\mathcal T_q,& \quad t = 1,\ldots,T,
\end{align}
where we have also defined 
\begin{align}
\label{eq:mathcalELL}
\mathcal L_{s,\nu}^{(t)}:= \sum_{r=1}^{p}\sum_{j=1}^{q} \left(s_j \widetilde\lambda_{r\nu_j}^{(t)}- \lambda^{\star}_{rj}\right)^2.
\end{align}
Obviously, $\Psi$ in \eqref{eq:problem2} also depends on $\{\bs{\widetilde\Lambda}\}_{t=1}^{T}$ but this is already fixed  at this second stage, hence we omit it from the notation on the left hand side of Equation \eqref{eq:problem2}. 
Clearly, the reference loading matrix $\bs\Lambda^\star=(\lambda_{rj}^\star)$ is not known, thus it is approximated by a recursive algorithm. This approach is inspired by  ideas used for solving identifiability problems in the context of Bayesian analysis of mixture models \citep{stephens, Papastamoulis:10, Rodriguez}, known as \textit{label switching}  \cite[see][for a review of these methods]{papastamoulis2016label}. 

The proposed Varimax Rotation-Sign-Permutation (RSP) post-processing algorithm aims to select $\bs\Lambda^\star, s, \nu$ such that we solve the problem \eqref{eq:problem2}. The algorithm is composed by two main steps implemented iteratively: 
the first minimizes $\Psi( \bs\Lambda^\star, s, \nu )$ with respect to $\bs\Lambda^\star$ for given values of $(s, \nu)$ (Reference-Loading Matrix Estimation, RLME, step),  while the second minimizes $\Psi( \bs\Lambda^\star, s, \nu )$ with respect to $(s, \nu)$  given $\bs\Lambda^\star$ (SP step); see Algorithm \ref{algo1} for a concise summary. 
 
 \begin{algorithm*}
        \colorbox{gray!25}{\parbox{0.95\textwidth}{
                        \SetKwInOut{Input}{Input}
                        \SetKwInOut{Output}{Output}
                        \SetKwBlock{blocknotext}{~}{end}
                        \SetKwBlock{step}{Step}{~}
                        \SetKwBlock{stepi}{Step 1: Varimax Rotation Step}{~}
                        \SetKwBlock{stepii}{Step 2: Signed Permutation Step}{~}
                        \SetKwFor{For}{for}{~}{endfor}
                        \SetKwFor{ForAll}{for~all}{~}{endfor}
                        \SetKwFor{ForEach}{for~each}{~}{endfor}
                        
                        \Input{Simulated MCMC sample of factor loadings $\big\{\bs\Lambda^{(t)}, t=1,\ldots,T\big\}$.}
                        \Output{Reordered MCMC sample of factor loadings $\big\{\bs{\mathring\Lambda}^{(t)}, t=1,\ldots,T\big\}$. }
                        \stepi{ 
                        \For{$t = 1$ \KwTo $T$}{ 
                                Implement the varimax rotation on $\bs\Lambda^{(t)}$ by solving \eqref{eq:varimax} and obtain the rotated varimax loadings $\bs{\widetilde\Lambda}^{(t)}$. }
                        }
                        \stepii{ 
                                \step(\textbf{2.1} Initialization:){
                                        \lForAll{$t\in\{1,\dots,T\}$:}{ Initialize $s^{(t)}$ and $\nu^{(t)}$}  
                                        \blocknotext(\textbf{Proposed initialization}){
                                        Set $s_j^{(t)}=1$ and $\nu^{(t)}_j=j$ (identity permutation), $t=1,\ldots,T$; $j=1,\ldots,q$
                                        }               
                                }
                                \step(\textbf{2.2}){
                                        \Repeat{
                                                no improvement in $\Psi(\bs\Lambda^{\star},s,\nu)=\sum_{t=1}^{T}\mathcal L_{s^{(t)},\nu^{(t)}}^{(t)}$ is observed}{
                                        \step(\textbf{2.2.1 RLME step}){
                                                Set
                                                $\lambda^{\star}_{rj}=\dfrac{1}{T}\sum\limits_{t=1}^{T}s_j^{(t)}
                                                \widetilde\lambda_{r\nu^{(t)}_j}^{(t)},\quad r=1,\ldots,p; j=1,\ldots,q$\\
                                                where $\widetilde\lambda_{rj}^{(t)}$ are the varimax rotated loadings (see also Equation (17) in \cite{stephens}). 
                                        }
                                       \step(\textbf{2.2.2 SP step}){
                                                \ForEach{iteration $t=1,\ldots,T$} {
                                                        Solve the problem
                                                        \begin{equation}
                                                                \label{eq:minimizationStep}
                                                                \big(s^{(t)},\nu^{(t)}\big)=\mbox{argmin}_{s,\nu}\left\{\mathcal L_{s,\nu}^{(t)}: s\in\{-1,1\}^q,\nu\in\mathcal T_q\right\}, 
                                                        \end{equation}
                                                        where 
                                                        $\mathcal L_{s,\nu}^{(t)}:= \sum_{r=1}^{p}\sum_{j=1}^{q} \left(s_j \widetilde\lambda_{r\nu_j}^{(t)}- \lambda^{\star}_{rj}\right)^2.$
                                                }
                                        }       
                                        }       
                                }
                        }
                END of algorithm.
                \caption{Varimax-RSP algorithm  (Correcting rotation, sign and permutation invariance to a MCMC sample of factor loadings)}
                \label{algo1}
        }
 }
\end{algorithm*}

For Step 2.2.1, it is straightforward to show that the minimization of $\Psi(\bs\Lambda^\star, s, \nu)$ with respect to $\bs\Lambda^*$ for given values of $(s, \nu)$ is obtained by 
\[\lambda^{\star}_{rj}=\frac{1}{T}\sum_{t=1}^{T}s_j^{(t)}
\widetilde\lambda_{r\nu^{(t)}_j}^{(t)},\]
for $r=1,\ldots,p$, $j=1,\ldots,q$. Step 2.2.2 is composed by the Sign-Permutation (SP) step which  minimizes $\Psi(\bs\Lambda^\star, s, \nu)$ with respect to  $(s, \nu)$ for a given reference loading matrix $\bs\Lambda^\star$. 

Let $\bs R^{(t)}$ denote the  varimax rotation matrix at Step 1, and $\bs S^{(t)}, \bs P^{(t)}$ denote the  sign and permutation matrices obtained at the last iteration of Step 2.2 of the RSP algorithm, for MCMC draw $t=1,\ldots,T$. The overall transformation of the factor loadings $\bs\Lambda^{(t)}$ and latent factors $F_i^{(t)}$ is 
\begin{align}
\bs{\mathring\Lambda}^{(t)} &= \bs\Lambda^{(t)}\bs R^{(t)}\bs S^{(t)}\bs P^{(t)}\\
{\mathring F}_i^{(t)} &= F_i^{(t)}\bs R^{(t)}\bs S^{(t)}\bs P^{(t)},\quad i=1,\ldots,n
\end{align}
for MCMC draw $t=1,\ldots,T$. Notice that $\mathring F_i^{(t)}$  is obtained immediately, since $\bs R^{(t)}$, $\bs S^{(t)}$ and $\bs P^{(t)}$ are already derived only from the loadings.

A geometrical illustration of the proposed method is given in Appendix \ref{sec:toy}, using a toy-example with $q=2$ factors.
Step 2 can also serve as a stand alone algorithm, that is, without the varimax rotation step (an application is given in Appendix \ref{sec:multipleChains}, for the problem of comparing multiple MCMC chains). Alternative strategies for minimizing expression \eqref{eq:minimizationStep}, are provided in Section \ref{RSP_strategies} which follows. 

\subsection{Computational Strategies for the Sign-Permutation (SP) step}
\label{RSP_strategies}

The optimal solution of \eqref{eq:minimizationStep} can be found exactly  by solving the assignment problem 
\cite[see][]{burkard} using a full enumeration of the total $2^q$ combinations $s^{(t)}\in\{-1,1\}^q$. 
Although this computational strategy avoids the evaluation of the objective function for all $2^qq!$ feasible solutions,  it still requires to solve $2^q$ assignment problems. Thus, for models with many factors (say $q>10$) we propose an approximate algorithm based on simulated annealing \citep{Kirkpatrick83optimizationby} with two alternative proposal distributions. 


For typical cases of factor models (e.g.~$q\leqslant 10$), the minimization in \eqref{eq:minimizationStep} can be performed exactly within reasonable computing time, by performing the following two-step procedure: 
\begin{itemize} 
        \item  compute $\min_{\nu\in\mathcal T_q}\left\{\mathcal L^{(t)}_{s,\nu}\right\}$, for each of the $2^q$ possible sign configurations $s$, and then 
        \item find the $(s,\nu)$ that correspond to the overall minimum. The first minimization requires to solve a special version of the transportation problem, known as the assignment problem \cite[see][]{burkard}.
\end{itemize}   

For a \textit{given} sign matrix $S=\mathrm{diag}(s_1,\ldots,s_q)\in\mathcal S$, the minimization problem is stated as the  assignment problem
\begin{align}\label{eq:transportation}
\min_{\delta_{ij}\in\{0,1\},\ i,j=1,\ldots,q} &
\quad \sum_{i=1}^{q}\sum_{j=1}^{q}\delta_{ij}c_{ij}\\
\nonumber
\mbox{subject to} &\quad \sum_{i=1}^{q}\delta_{ij}=1,\quad\forall j=1,\ldots,q\\
\nonumber
&\quad \sum_{j=1}^{q}\delta_{ij}=1,\quad\forall i=1,\ldots,q\end{align}
where the $q\times q$ cost matrix $\bs C=(c_{ij})$ of the assignment problem is defined as 
\[
c_{ij}=\sum_{r=1}^{p}\left(s_j\widetilde\lambda_{rj}
-\lambda_{ri}^{*}\right)^2,\quad  i,j=1,\ldots,q
\]
and the corresponding binary decision variables $\delta_{ij}$ are defined as
$$\delta_{ij}:=\begin{cases}
1,&\quad \mbox{if index $i$ is assigned to index $j$}\\
0,&\quad \mbox{otherwise}
\end{cases}.$$
Two popular methods for solving discrete combinatorial optimization problems is the ``Hungarian algorithm'' \citep{kuhn1955hungarian} and the ``branch and bound'' method \citep{little1963algorithm}.
We used the library {\tt lpSolve}  \citep{lpSolve} in {\tt R} in order to solve the assignment problem \eqref{eq:transportation}, which uses the latter technique. In brief, ``branch and bound'' first solves the problem without the integer restrictions. In case of non-integer solutions, the model is split in sub-models and iteratively optimized, until an integer solution is found.  	This solution is then remembered as the best-until-now solution. Finally this procedure is repeated until no better solution can be found.

\begin{algorithm*}
        \colorbox{gray!25}{\parbox{0.95\textwidth}{
                        \SetKwInOut{Input}{Input}
                        \SetKwInOut{Output}{Output}
                        \SetKwBlock{blocknotext}{~}{end}
                        \SetKwBlock{step}{Step}{~}
                        \SetKwBlock{stepi}{Step 1: Varimax Rotation Step}{~}
                        \SetKwBlock{stepii}{Step 2: Signed Permutation Step}{~}
                        \SetKwFor{For}{for}{~}{endfor}
                        \SetKwFor{ForAll}{for~all}{~}{endfor}
                        \SetKwFor{ForEach}{for~each}{~}{endfor}
                        
                        \step(A.1){
                        \ForEach{$s=\left(s_1,\ldots,s_q\right)\in\{-1,1\}^{q}$}{
                                Find the permutation $\nu(s)$ that minimizes $\left\{\mathcal L_{s,\nu}^{(t)}:\nu\in\mathcal T_q\right\}$ by solving the assignment problem \eqref{eq:transportation}.}
                        }
                        \step(A.2){
                        Set $\left(s^{(t)},\nu^{(t)}\right)=\mbox{argmin}_{s,\nu}\left\{\mathcal L_{s,\nu(s)}^{(t)}: s\in\{-1,1\}^q\right\}$.} 
                        END of SP Step.
                        \caption{Sign-Permutation (SP) Step -- Exact Scheme (Scheme A)}
                        \label{algo2}
                }
        }
\end{algorithm*}

The exact SP approach is summarized in Algorithm \ref{algo2} and denoted as Scheme A. This scheme is elaborate and requires the evaluation of the problem over all possible sign combinations. 
More specifically, it requires to solve $2^q$ assignment problems in order to find the overall minimum. As we have already mentioned,  this approach is  more efficient than a brute force algorithm that requires $2^qq!$ evaluations of the objective function. It can be applied in a reasonable way for models with $q\leq 10$ factors but its implementation becomes forbidden in terms of computational time for models with factors of higher dimension. 
For this reason, for models with large number of factors, we also propose strategies based on simulated annealing \citep{Kirkpatrick83optimizationby}. 

Two approximate approaches for identifying the sign permutation are summarized in Algorithm \ref{algo3} denoted as Scheme B and Scheme C, respectively.
Both schemes are based on the Simulated Annealing (SA) framework, where \textit{candidate states} $(s^\star,\nu^\star)$ are proposed. The proposal is either accepted as the next state or rejected and the previous state is repeated. This procedure is repeated $B$ times, by gradually cooling down the temperature $T_b$ which controls the acceptance probability at iteration $b=1,\ldots,B$. These annealing steps are implemented for each MCMC iteration $t$, within step 2.2.2 of Algorithm \ref{algo1}. 
In the following, two different versions of the SA approach (Schemes B and C) are introduced and described in detail.

\begin{algorithm*}
        \colorbox{gray!25}{\parbox{0.95\textwidth}{
                        \SetKwInOut{Input}{Input}
                        \SetKwInOut{Output}{Output}
                        \SetKwBlock{blocknotext}{~}{~}
                        \SetKwBlock{step}{Step}{~}
                        \SetKwBlock{schemeb}{SA version 1: Full SA (Scheme B)}{End-Full-SA}
                        \SetKwBlock{schemec}{SA version 2: Partial SA (Scheme C)}{End-Partial-SA}
                        \SetKwFor{For}{for}{~}{endfor}
                        \SetKwFor{ForAll}{for~all}{~}{endfor}
                        \SetKwFor{ForEach}{for~each}{~}{endfor}
                        
                        \step(\textbf{1}: Initialize){
                                Set initial values $\left(s^{(t,0)},\nu^{(t,0)}\right)=\left(s^{(t)},\nu^{(t)}\right)$ and let $\mathcal L^{(t,0)}=\mathcal L^{(t)}_{s^{(t)},\nu^{(t)}}$.
                        }
                        \step(\textbf{2}){
                                \For{$b=1$ \KwTo $B$}{
                                        \blocknotext(\textbf{(a)} Propose a candidate state \mbox{$(s^{\star},\nu^{\star})$} using one of the following schemes:){
                                                \schemeb{
                                                        Randomly switch the sign of one index in $s^{(t,b-1)}$, and \\ 
                                                        Permute the values of a randomly selected pair of indices in $\nu^{(t,b-1)}$.} 
                                                
                                                \vspace{1em} 
                                                \schemec{
                                                        Obtain $s^\star$ by randomly switching the sign of one index in $s^{(t,b-1)}$. \\ 
                                                        Obtain the permutation $\nu^\star:=\nu(s^\star)$ that minimizes $\left\{\mathcal L_{s^\star,\nu}^{(t)}:\nu\in\mathcal T_q\right\}$ by solving the assignment problem \eqref{eq:transportation}.}
                                                }
                                        \blocknotext(\textbf{(b)} Compute \mbox{$\mathcal L^{\star}=\mathcal L_{s^{\star},\nu^\star}^{(t)}$}.){~}
                                        \blocknotext(\textbf{(c)} Set \mbox{$\left(s^{(t,b)},\nu^{(t,b)}\right)=(s^{\star},\nu^{\star})$} and 
                                        \mbox{$\mathcal L^{(t,b)}=\mathcal L^{\star}$} with probability )
                                        {
                                                \small 
                                                \begin{equation*}
                                                \mathrm{P}\left((s^{(t,b-1)},\nu^{(t,b-1)}) \rightarrow (s^{\star},\nu^{\star})\right)=
                                                \begin{cases}
                                                \exp\left\{-\dfrac{\mathcal L^{\star} - \mathcal L^{(t,b-1)}}{T_b}\right\}& \quad\mbox{if } \mathcal L^{\star}-\mathcal L>0\\
                                                1&\quad \mbox{if } \mathcal L^{\star}-\mathcal L\leqslant 0.
                                                \end{cases}                             
                                                \end{equation*}
                                                \normalsize 
                                                otherwise set $\left(s^{(t,b)},\nu^{(t,b)}\right)=(s^{(t,b-1)},\nu^{(t,b-1)})$ and $\mathcal L^{(t,b)}=\mathcal L^{(t,b-1)}$.
                                        }
                                }  
                        } 
                       \step(\textbf{3}: Set \mbox{$\left(s^{(t)},\nu^{(t)}\right)=\left(s^{(t,B)},\nu^{(t,B)}\right)$}.){~}
                        END of SP-SA Step.
                        \caption{Sign-Permutation (SP) Step -- Simulated Annealing (SA) schemes (Scheme B and C)}
                        \label{algo3}
                }
        }
\end{algorithm*}

       In the first version of the SA scheme (Scheme B: {\it full SA}), the pair $(s^\star,\nu^\star)$ is generated independently by randomly switching one element of the current sign vector $s$ and permuting two randomly selected indexes in $\nu$. This scheme is referred to as \textit{full} simulated annealing. It is the simplest approach that can be used to generate candidate states. 
        For this reason,  it is expected to be trapped at inferior solutions in some cases. 
        
        The second version of SA (Scheme C: {\it partial} SA) attempts to overcome this problem. 
        It is a hybrid between the exact SA approach (Scheme A) and the full SA (Scheme B). 
        Therefore, it is more sophisticated than full SA but computationally more demanding. 
        Firstly, we propose a candidate sign configuration $s^\star$ using a random perturbation of the current value as in Scheme B. Next,  we proceed as in the exact approach (Scheme A), by deterministically identifying the permutation $\nu^\star$ which minimizes $\big\{\mathcal L_{s^{\star},\nu}^{(t)}:\nu\in\mathcal T_q\big\}$, given $s^\star$. We refer to this scheme as \textit{partial} simulated annealing 
        in order to emphasize that while $s^\star$ is randomly generated from the current state, $\nu^\star$ is deterministically defined given $s^{\star}$. 

The overall procedure for the two simulated annealing schemes is described in Algorithm \ref{algo3}. Clearly, one iteration of the Partial SA scheme  is more expensive computationally than one iteration of the Full SA scheme. But since the proposal mechanism in Partial SA minimizes $\mathcal L(s,\nu)$ given $s$, it is expected that it will converge faster to the solution compared to Full SA where both parameters are randomly proposed. This is empirically illustrated in Section \ref{sec:computational} of the Appendix.

For both SA schemes, the cooling schedule $\{T_b,b=1,2,\ldots\}$ is such that $T_b>0$ for all $b$ and $\lim_{b\rightarrow\infty} T_b=0$.   \cite{romeo1991theoretical} showed that a logarithmic cooling schedule of the form $T_b=\gamma/\log(b+\gamma_0)$, $b=1,2,\ldots,$
is a sufficient condition for convergence with probability one to the optimal solution. In our applications, a reasonable trade-off between accuracy and computing time was obtained with $\gamma=\gamma_0=1$ and a  total number of annealing loops $20\leqslant B\leqslant 2000$.

 \begin{table*}[t]
        \begin{tabular}{ccccccc}
        \hline 
                  &     &    & \multicolumn{4}{c}{Factors} \\ 
                  \cline{4-7} 
                  & $n$ & $p$& $F_1$ & $F_2$ & $F_3$ & $F_4$ \\
        \hline 
        Example 1 &100& 8  & $Y_1-Y_4$ & $Y_5-Y_8$ &  &  \\
        Example 2 &200& 24 & $Y_1-Y_6$ & $Y_7-Y_{12}$ & $Y_{13}-Y_{18}$ &  $Y_{19}-Y_{24}$\\
        \hline 
        \end{tabular}
        \caption{Simulation study plan of Section \ref{sec:sim} }
        \label{sim_plan}
\end{table*}

\begin{figure*}[h!]
\begin{center}
\begin{tabular}{c}
\includegraphics[scale=0.52]{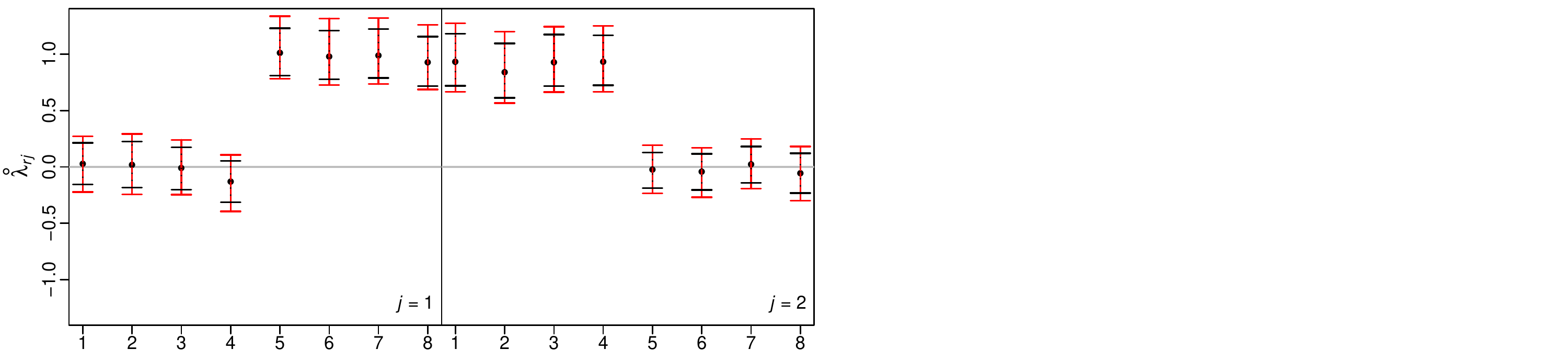}\\
\includegraphics[scale=0.52]{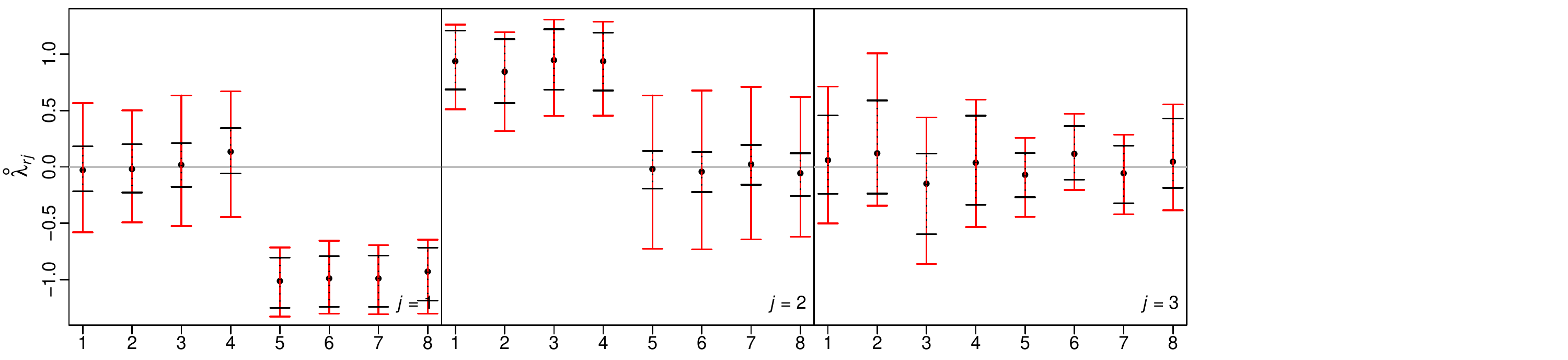}\\
\vspace{-1ex}
$r$
\vspace{-2ex}
\end{tabular}
\end{center}
\caption{\textit{Simulated data 1}: $99\%$ HPD intervals (black) and simultaneous $99\%$ credible regions (red) of reordered factor loadings, when fitting Bayesian FA models with $q=2$ (top) and $q=3$ (bottom) factors (Dataset details: $p=8$ variables and $q_{\mbox{true}}=2$ factors).}
\label{fig:sim1manyq}
\end{figure*}

\begin{figure*}[h!]
\vspace{-4ex}
\begin{center}
\includegraphics[scale=0.55]{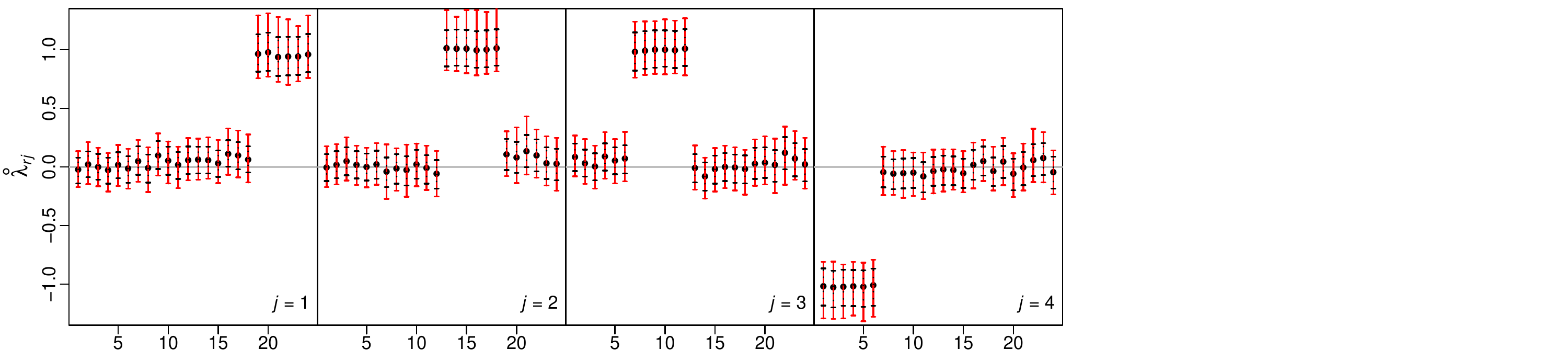}\\
\includegraphics[scale=0.55]{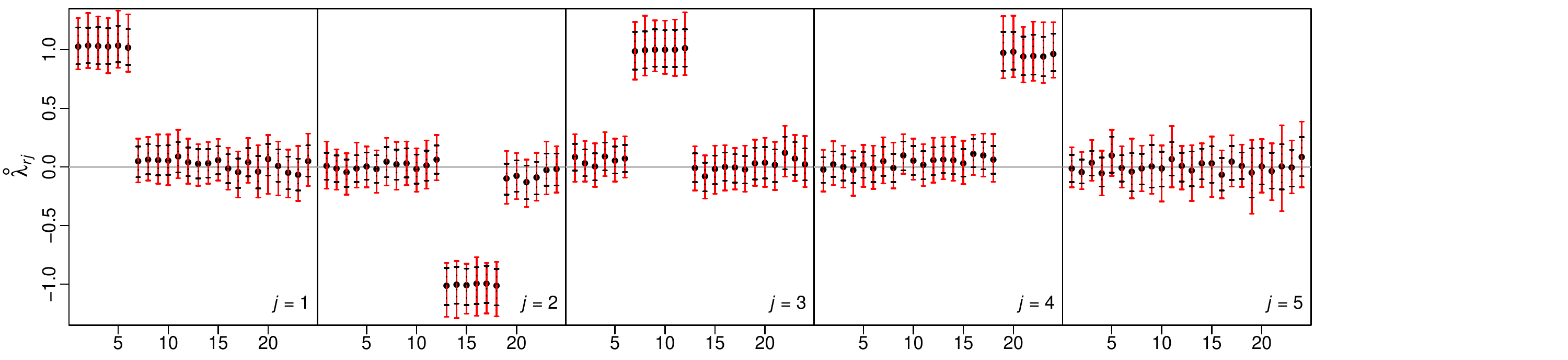}\\
\end{center}
\begin{center}
\vspace{-3ex}
$r$
\vspace{-2ex}
\end{center}

\caption{\textit{Simulated data 2}: $99\%$ HPD intervals (black) and simultaneous $99\%$ credible regions (red) of reordered factor loadings, when fitting Bayesian FA models with $q=4$ (top) and $q=5$ (bottom) factors  (Dataset details: $n=200$, $p=24$ and $q_{\mbox{true}}=4$).}
\label{fig:sim1manyq2}
\end{figure*}

\section{Applications}\label{sec:results}

\begin{figure}
\begin{tabular}{cc}
\includegraphics[width=0.5\textwidth]{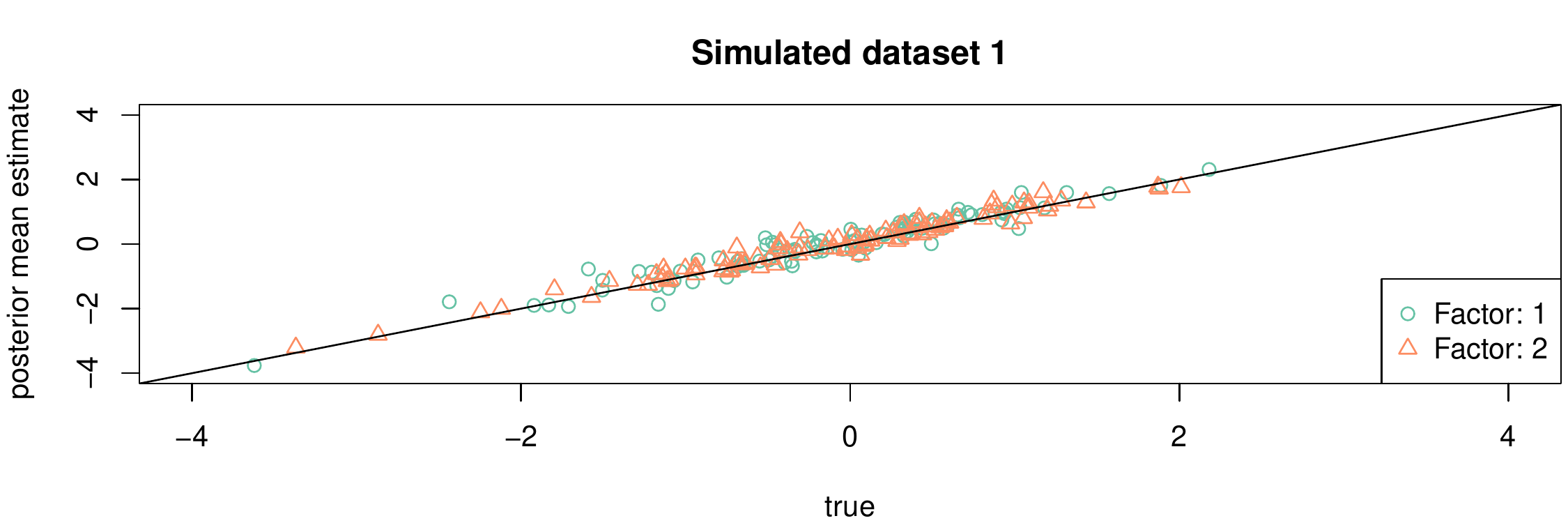}&
\includegraphics[width=0.5\textwidth]{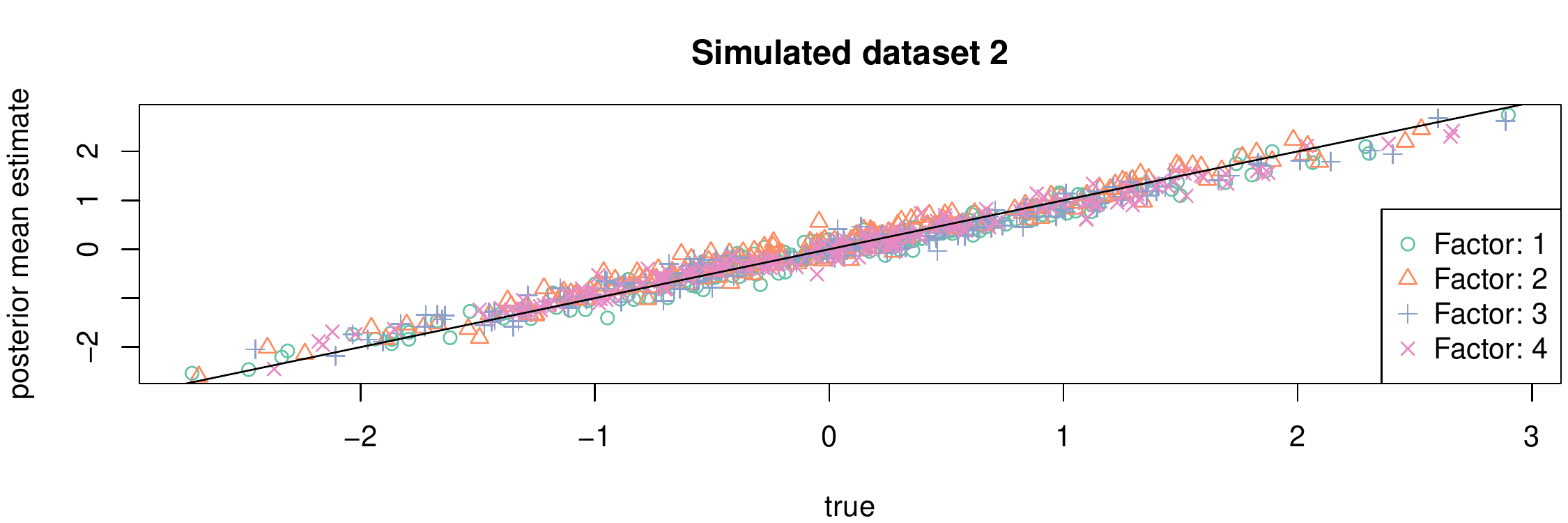}
\end{tabular}
\caption{Comparison of true factor scores $F_{ij}$, $i=1,\ldots,n$ and $j=1,\ldots,q$, with the posterior mean estimates after post-processing the MCMC sample with the RSP algorithm. The identity line is also displayed. A different symbol and color is used for each factor.}
\label{fig:factors}
\end{figure}

Section \ref{sec:sim} presents a simulation study and Section \ref{sec:real} analyzes a real dataset. Section \ref{sec:mix} deals with mixtures of factor analyzers, using the {\tt fabMix} package \citep{fabMix, papastamoulis2018overfitting, papastamoulis2019clustering}. In all cases the input data is standardized so the sample means and variances of each variable are equal to $0$ and $1$ respectively. We adopt this approach because it is easier to tune the parameters of the prior distributions, however we should note that it is not crucial to the proposed post-processing method. When the difference between two subsequent evaluations of the objective function in Equation \eqref{eq:problem2} is less than $10^{-6}Tpq$ the algorithm terminates, with $T$ denoting the size of the retained MCMC sample. The intuition behind this empirical convergence criterion is that the objective function in Equation \eqref{eq:problem2}  is a sum of $Tpq$ terms. Dividing the objective function by $Tpq$ corresponds to an average value across all $Tpq$ terms. The algorithm terminates as soon as the difference of the ``average loss'' between successive evaluations is smaller than $10^{-6}$.

In Sections \ref{sec:sim} and \ref{sec:real}, the {\tt MCMCpack} package \citep{mcmcpack, mcmcpackage} in {\tt R} was used in order to fit Bayesian FA models. We have used the {\tt MCMCfactanal()} function, which applies standard Gibbs sampling \citep{gelfand}  in order to generate MCMC samples from the posterior distribution.  The default normal priors are placed upon the factor loadings and factor scores while inverse Gamma priors are assumed for the uniquenesses, that is,
\begin{align}
\label{eq:lambda_prior}
     \lambda_{rj} &\sim \mathcal{N}(l_{0}, L_{0}^{-1}),\quad\\    
     \label{eq:s2_prior} 
         \sigma^2_{r} &\sim \mathcal{IG}(a_{0}/2, b_{0}/2)     
\end{align}
mutually independent for $r=1,\ldots,p$; $j=1,\ldots,q$. Unless otherwise stated, we use the default prior parameters of the package, that is, $l_0=0$, $L_0=0$, $a_0=b_0=0.001$. Note that the specific choices correspond to an improper prior distribution on the factor loadings and a vaguely informative prior distribution on the uniquenesses. The {\tt MCMCfactanal()} function provides the {\tt ``lambda.constraints''} option, which can be used in order to enable possible simple equality  or inequality constraints on the factor loadings, such as the typical constraints in the upper triangle of $\bs \Lambda$ described in Equation \eqref{eq:lambda}. Note that in our implementation, the {\tt ``lambda.constraints''} option was disabled, that is, all factor loadings are assumed unconstrained.

In all cases, an MCMC sample of two million iterations was simulated, following a burn-in period of $10000$ iterations. Finally, a thinned MCMC sample of $10000$ draws was retained for inference, keeping the simulated values of every $200$th MCMC iteration. Highest Posterior Density (HPD) intervals were computed by the  {\tt HDIinterval} package \citep{HDInterval}. Simultaneous credible regions were computed as described in \cite{besag1995bayesian}, using the implementation in the {\tt R} package {\tt bayesSurv} \citep{bayesSurv}. In Section \ref{sec:sim} we have also used the {\tt BayesFM} package \citep{bayesfm} in order to compare our findings  with the ones arising from the method of \cite{CONTI201431}.

\begin{figure}[p]
\begin{center}
\includegraphics[scale=0.4]{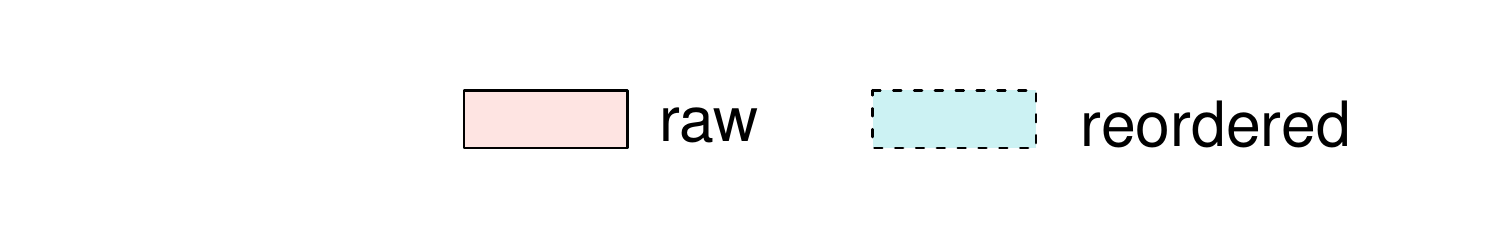}\\
\begin{tabular}{cc}
\includegraphics[scale=0.9]{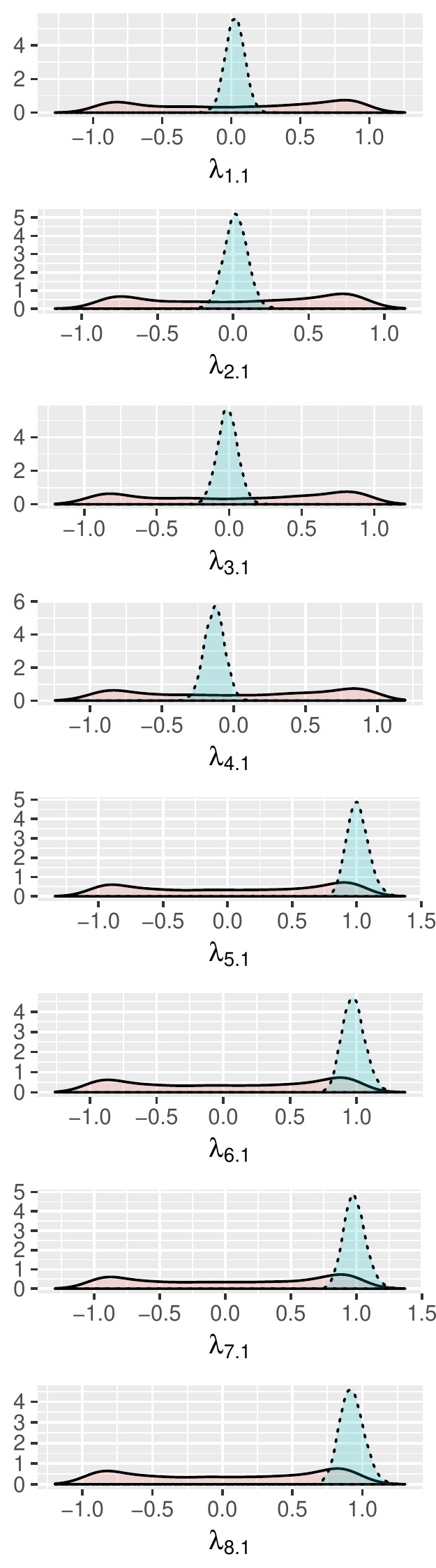}&
\includegraphics[scale=0.9]{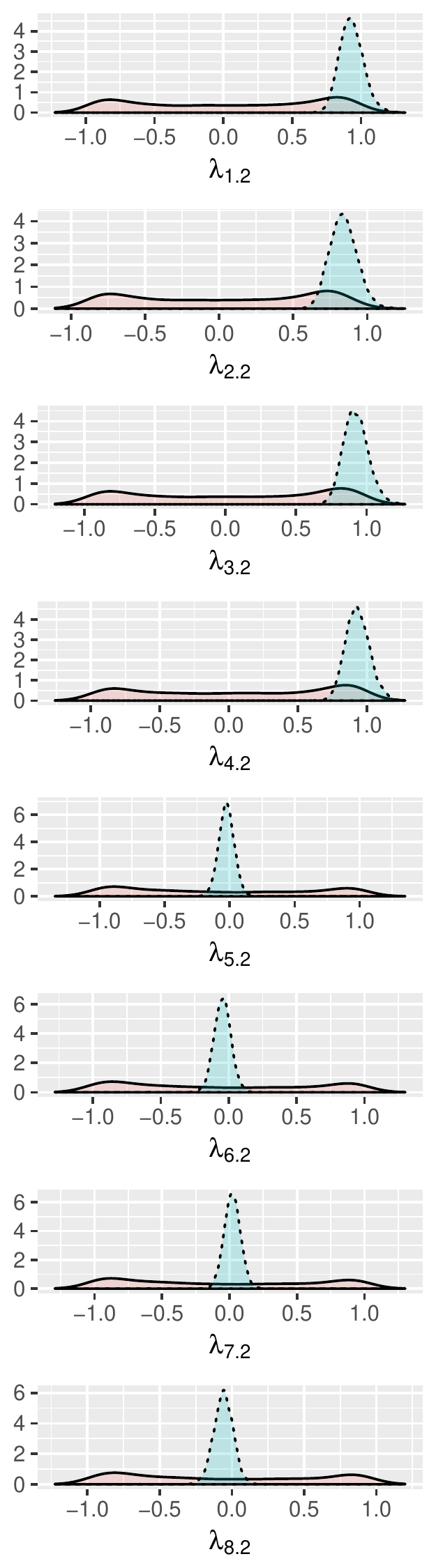}
\end{tabular}
\vspace{-3ex}
\end{center}
\caption{\textit{Simulated data 1}: Marginal posterior distribution of raw and reordered factor loadings, conditional on the true number of factors ($q=q_{\mbox{true}}=2$).}
\label{fig:sim1}
\end{figure}

\begin{figure}[p]
\begin{center}
\includegraphics[scale=0.4]{legend}\\
\begin{tabular}{c}
\includegraphics[scale=0.55]{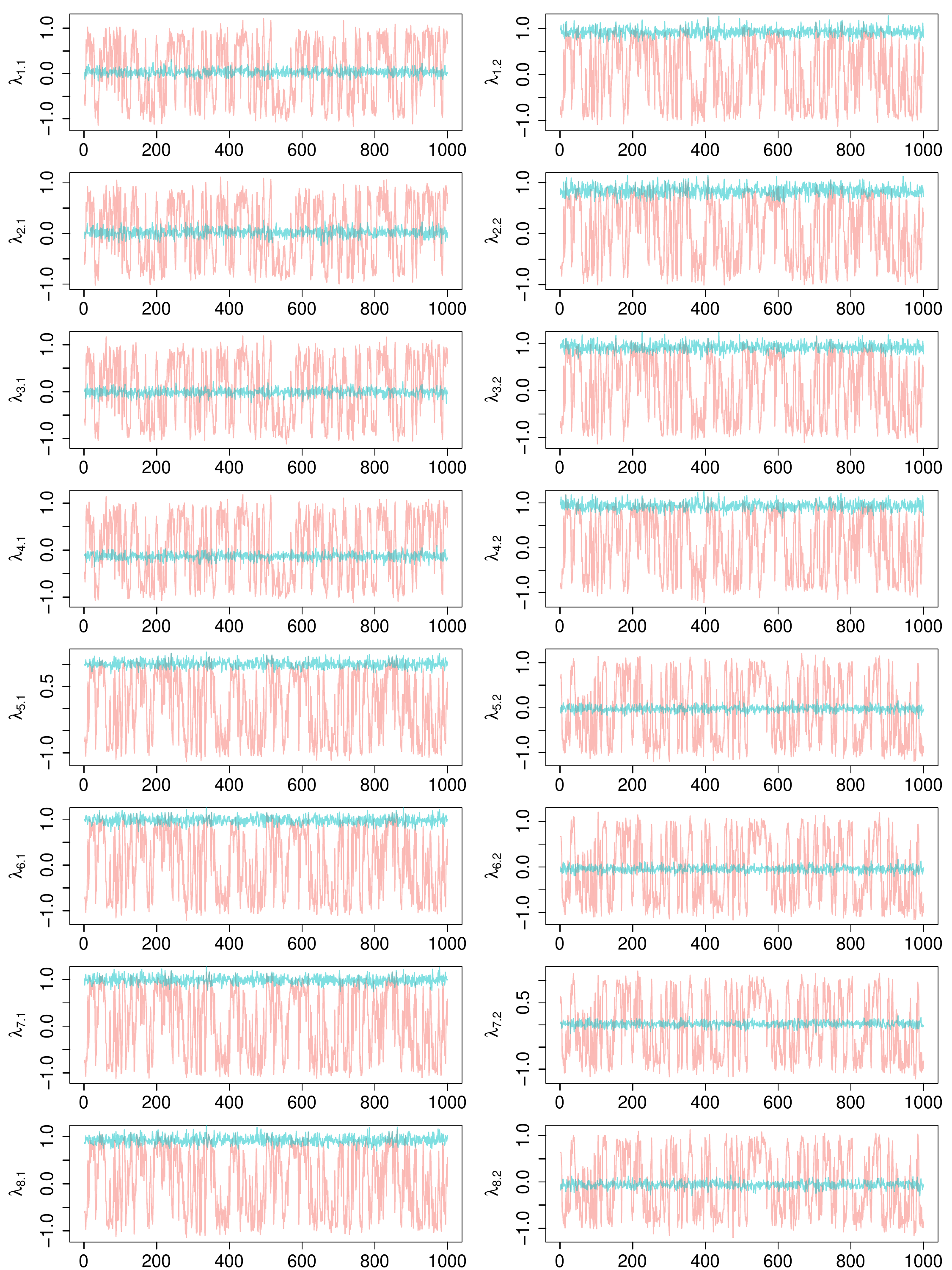}
\end{tabular}
\vspace{-2ex}
\end{center}
\caption{\textit{Simulated data 1}: MCMC trace  of raw and reordered factor loadings, conditional on the true number of factors (Thinned sample of 1000 iterations; $q=q_{\mbox{true}}=2$).}
\label{fig:sim1trace}
\end{figure}

\begin{figure}[h]
\centering
\begin{tabular}{cc}
\includegraphics[scale=0.5]{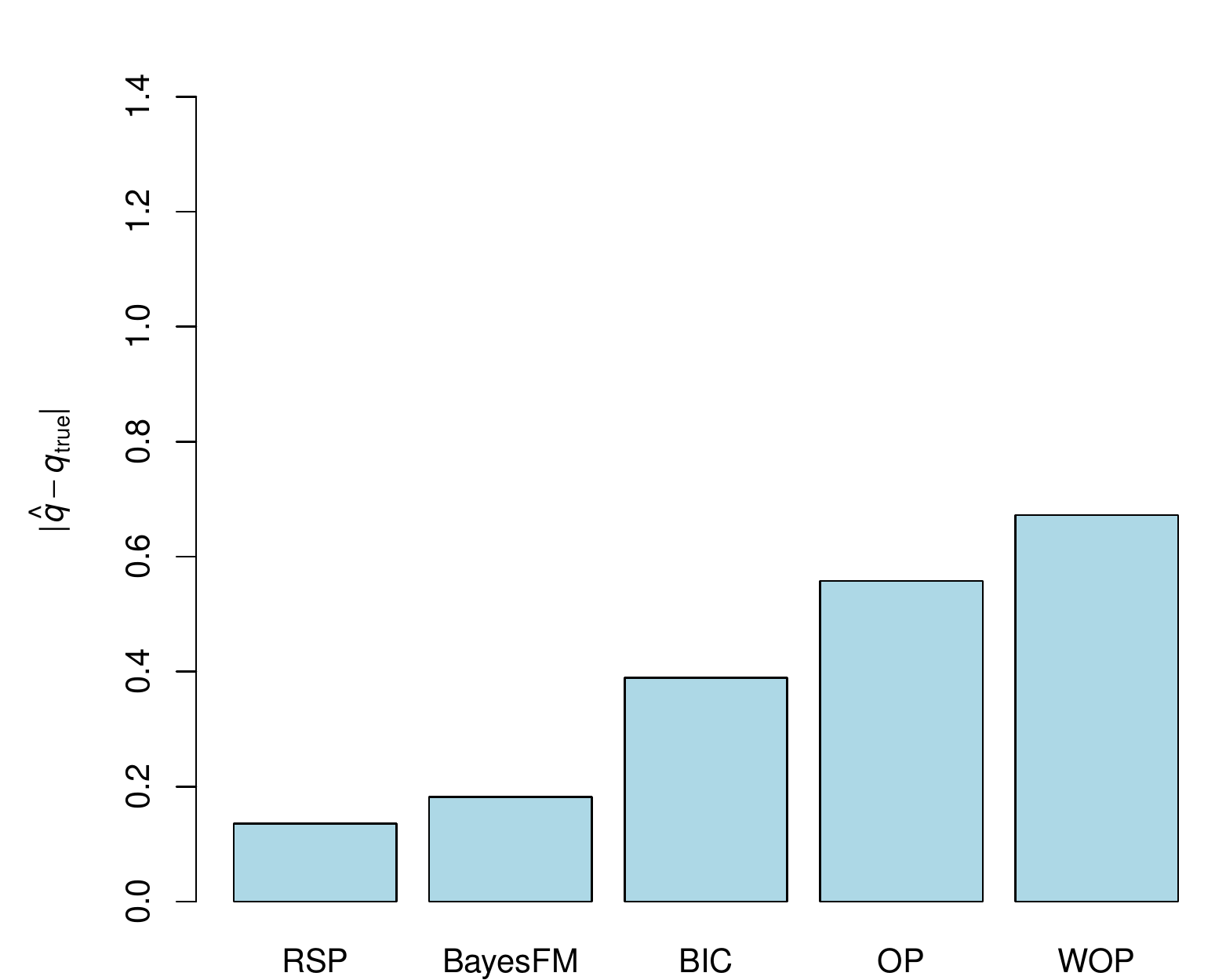}&
\includegraphics[scale=0.5]{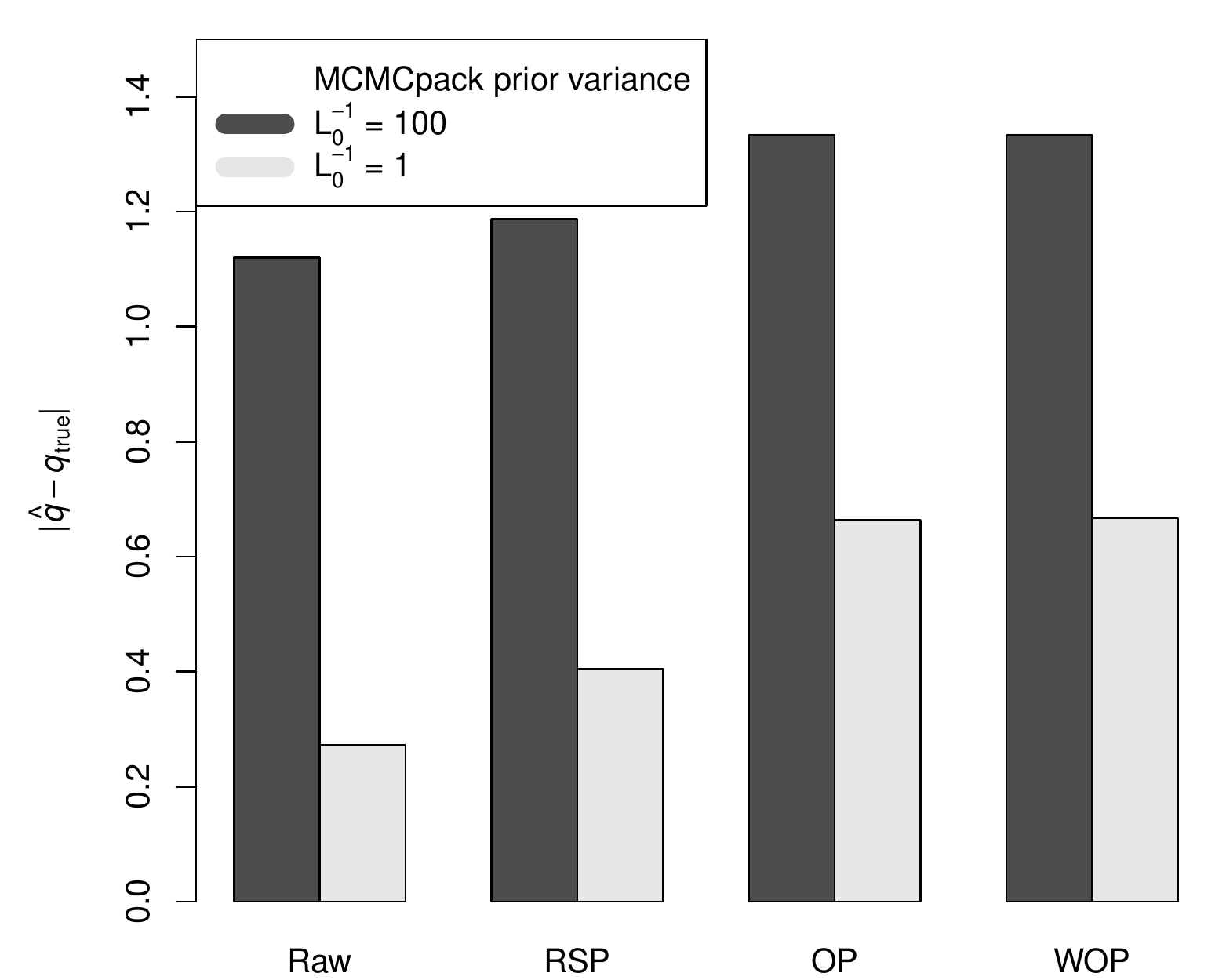}\\
(a) & (b)
\end{tabular}
\caption{The $y$ axis corresponds to the mean absolute error  for selecting the number of factors in the simulation study of Section \ref{sec:sim}. (a): For RSP (proposed method), OP (orthogonal Procrustes) and WOP (weighted orthogonal Procrustes) $\hat q$ denotes the ``effective'' number of columns of $\bs{\mathring\Lambda}$ when fitting a model with  $q_{max} > q_{\mbox{true}}$ factors.  {\tt BayesFM} denotes the selection of the number of factors according to the method of \cite{CONTI201431} and BIC denotes the selection of $q$ according to the Bayesian Information Criterion from the raw output of {\tt MCMCpack}, when considering models with $q=1,\ldots,q_{max}$ factors. (b): Selection of the number of factors resulting from marginal likelihood estimation according to bridge sampling when applied to the raw output of {\tt MCMCpack} (Raw) and the reordered output of {\tt MCMCpack} according to  RSP, orthogonal Procrustes (OP) and weighted orthogonal Procrustes (WOP). }
\label{fig:qSelection}
\end{figure}

\begin{table}[h]
\begin{center}
\begin{tabular}{ccccccccc}
\toprule
&&\multicolumn{3}{c}{$q=3$}&\multicolumn{4}{c}{$q=4$}\\
\midrule
&& 1 & 2& 3 & 1 & 2& 3 & 4\\
\midrule
\multirow{3}{*}{visual}&$Y_1$&$-0.28$    & $0.19$   &\gr{0.64}& $0.47$ & $0.41$  & $0.12$  & $0.26$  \\ 
&$Y_2$&$-0.16$    & $0.08$   &\gr{0.49}& $0.14$ &\gr{0.52}& $0.07$  & $0.15$ \\ 
&$Y_3$&$-0.28$    & $0.11$   &\gr{0.63}& $0.21$ &\gr{0.67}& $0.08$  & $0.26$ \\ 
\midrule
\multirow{3}{*}{verbal}&$Y_4$&\gr{-0.89} & $0.07$   & $0.16$  & $0.11$ & $0.15$  & $0.07$  & \gr{0.89} \\ 
&$Y_5$&\gr{-0.84} & $0.18$   & $0.11$  & $0.15$ & $0.07$  & $0.16$  & \gr{0.84} \\ 
&$Y_6$&\gr{-0.84} & $0.07$   & $0.16$  & $0.08$ & $0.16$  & $0.07$  & \gr{0.83} \\ 
\midrule
\multirow{3}{*}{speed}&$Y_7$&$-0.18$    & \gr{0.78}& $-0.07$ & $0.05$ & $-0.04$ &\gr{0.83}& $0.17$ \\ 
&$Y_8$&$-0.03$    & \gr{0.83}& $0.24$  & $0.24$ & $0.16$  &\gr{0.77}& $0.03$ \\ 
&$Y_9$& $-0.26$   & \gr{0.54}&\gr{0.45}& $0.48$ & $0.26$  &\gr{0.45}& $0.24$ \\
\bottomrule
\end{tabular}
\end{center}
\caption{
{\it Grant-White school dataset}: RSP Estimated posterior means of factor loadings 
for the 3 and 4 factor models. \\
{\small \it Notes: 
        Gray boxes and asterisks: loadings with simultaneous $99\%$ credible region that does not contain zero. \\
        First column of 4-Factor model ($q=4$): is redundant since for all loadings the zero value is a reasonable posterior value.}
}
\label{tab:grantWhite}
\end{table}

\subsection{Simulation study}\label{sec:sim}

At first we illustrate the proposed approach using two simulated datasets: in dataset 1 there are $n=100$ observations, $p=8$ variables and $q_{\mbox{true}}=2$ factors, while in dataset 2 we set $n=200$, $p=24$ and $q_{\mbox{true}}=4$. The association patterns between the generated variables and the assumed underlying factors are summarized in Table \ref{sim_plan}. 
For more details on the simulation procedure, the reader is referred to \cite{papastamoulis2018overfitting,papastamoulis2019clustering}.

Figures \ref{fig:sim1manyq} and \ref{fig:sim1manyq2} display the $99\%$ highest posterior density intervals and a $99\%$ simultaneous credible region of reordered factor loadings, when fitting Bayesian FA models with $q_{\mbox{true}}\leqslant q\leqslant q_{\mbox{true}}+1$. In particular, each panel contains the intervals for each column of the $p\times q$ matrix $\bs{\mathring\Lambda}$. Observe that for $q>q_{\mbox{true}}$ there are $q-q_{\mbox{true}}$ columns of $\bs{\mathring\Lambda}$ with all  intervals including zero. For dataset 1, a detailed view of the marginal posterior distributions of raw and reordered factor loadings when the number of factors is equal to its true value ($q=2$) is shown in Figure \ref{fig:sim1}. The corresponding (thinned) MCMC trace is shown in  Figure \ref{fig:sim1trace}. Note the broad range of the posterior distributions of the raw factor loadings (shown in red), which is a consequence of non-identifiability. On the other hand, the bulk of the posterior distributions of the reordered factor loadings is concentrated close to $0$ or $\pm 1$. Figure \ref{fig:factors} diplays the true factors $F_{ij}$, $j=1,2$ used to generate the data versus the posterior mean estimates from the post-processed sample for simulated data 1. Note that a final signed permutation has also been applied to the estimates in order to maximize their similarity to the ground truth.

The inspection of the simultaneous credible regions reveals possible overfitting of the model being used in each case. This is indeed the case with the 3-factor model at Example 1 and the 5-factor model at Example 2. Clearly, the simultaneous credible regions in Figures \ref{fig:sim1manyq} and \ref{fig:sim1manyq2} are able to detect that there is one redundant column in the matrix of factor loadings. Although this is not a proper Bayesian model selection scheme, we observed that this procedure can successfully detect cases of overfitting, provided that the number of factors is indeed larger than the ``true'' one. 

In order to further assess the ability of this approach in detecting over-fitted factor models, we have simulated synthetic datasets from \eqref{eq:fa} with $18\leqslant p\leqslant 24$. The true number of factors was set equal to $2\leqslant q_{\mbox{true}} \leqslant 6$. For each simulated dataset, the sample size ($n$) is chosen at random from the set $\{100, 200,300\}$, the idiosyncratic variances are randomly drawn from the set $\sigma^2_r = \sigma^2$, $r = 1,\ldots,p$ with $\sigma^2$ randomly drawn from the set $\{400, 800, 1200\}$. We generated 30 synthetic datasets for each value of $q_{\mbox{true}}$, that is, 150 datasets in total.

Let $q_0$ denote the number of redundant columns of $\bs{\mathring\Lambda}$ for a FA model with $q$ factors, that is, the number of columns of $\bs{\mathring\Lambda}$ where at least one interval in the $(p\times q)$-dimensional $99\%$ Simultaneous Credible Region (SCR) does not contain 0. Now define the number of ``effective'' columns of $\bs{\mathring\Lambda}$ as $\hat q = q - q_0$. We have also considered the same technique for estimating the number of factors when reordering the {\tt MCMCpack} output with the Orthogonal Procrustes rotations approach (weighted or not) as described in \cite{AMANN2016190} with the addition of a final varimax rotation to the reordered output, a possibility explicitly suggested at the end of Section 3 of \cite{AMANN2016190}. In order to give an empirical comparison of $\hat q$ with model selection approaches for estimating the number of factors, we compare our findings with the Bayesian Information Criterion using the same output from the {\tt MCMCpack} package as well as with the stochastic search method of \cite{CONTI201431} as implemented in the {\tt R} package {\tt BayesFM} \citep{bayesfm}. We have also considered estimation of the number of factors according to the bridge sampling estimator of the marginal likelihood \citep{JSSv092i10, doi:10.1080/01621459.2020.1773833}. Note that all methods, excluding {\tt BayesFM}, are applied to the MCMC output of {\tt MCMCpack}, considering two setups for the prior variance variance of the factor loadings: $L_0^{-1}=100$ (this choice corresponds to a vague prior distribution) and $L_0^{-1} = 1$ (this choice is weakly informative when comparing with the posterior range of factor loadings). The default prior settings were considered into {\tt BayesFM}.

We generated MCMC samples for FA models with  $1\leqslant q\leqslant q_{max}=8$ factors. Our results are summarized in terms of the Mean Absolute Error $|\hat q- q_{\mbox{true}}|$ across the 150 simulated datasets in Figures \ref{fig:qSelection}.(a) and  \ref{fig:qSelection}.(b). The results in Figure \ref{fig:qSelection}.(a) are averaged across the two different choices of the prior parameter $L_0$. We conclude that the number  of ``effective'' columns of $\bs{\mathring\Lambda}$ arising from the proposed method (RSP) outperforms OP and WOP and that it is fairly consistent with the active (that is, the number of non-zero columns in the matrix of factor loadings) number of factors inferred by {\tt BayesFM}. BIC tends to underestimate the number of factors resulting in a larger Mean Absolute Error compared to RSP or {\tt BayesFM}, a behaviour which was observed for the larger values of $q_{\mbox{true}}$. 

The selection of the number of factors according to the marginal likelihood estimate as implemented via the bridge sampling method is largely impacted by the prior distribution, as illustrated in Figure \ref{fig:qSelection}.(b). In the case of the vague prior distribution ($L_0^{-1}=100$) it tends to underestimate the number of factors. Improved results are obtained when the prior variance of factor loadings is small ($L_0^{-1}=1$). Note also that the marginal likelihood estimator performs better when the raw output is considered. This is expected since the distribution of interest is invariant over orthogonal rotations and the data brings no information about an ordering of the factor loadings. A similar behaviour is also obtained when considering marginal likelihood estimation of finite mixture models \citep{marin2008approximating}.

Overall, the proposed method outperforms the method of \cite{AMANN2016190} which is based on (weighted) orthogonal Procrustes rotations. This is particularly the case for the estimates arising from the effective number of columns of the reordered factor loadings corresponding to factor models which overestimate $q$. The OP/WOP method is highly affected by the number of factors of the overfitted model (here $q_{max}=8$). We observed that it tends to overestimate the number of factors when $q_{max}$ is much larger than $q_{\mbox{true}}$ (a specific example of such a behaviour is presented in Appendix \ref{sec:p_lt_n}). Finally,  note that the RSP algorithm leads to better estimates than OP/WOP rotations of \cite{AMANN2016190} when using bridge sampling to estimate the marginal likelihood. However, the method of \cite{AMANN2016190} is notably faster (time comparisons are reported in Table \ref{tab:p_lt_n} of Appendix \ref{sec:p_lt_n}).

Further simulation studies  are presented in the Appendices. Appendix \ref{sec:p_lt_n} presents an example with $p>n$,
Appendix \ref{sec:multipleChains} discusses the issue of comparing multiple chains, Appendix  \ref{sec:computational} compares the computational cost of the proposed schemes in high-dimensional cases with up to $q=50$ factors, while Appendix \ref{sec:k_med} applies the post-processing clustering approach of \cite{kaufmann2017identifying} in one of our simulated datasets. 


\subsection{The Grant-White school dataset}\label{sec:real}

In this example we use measurements on nine mental ability test scores of seventh  and eighth grade children from two different schools (Pasteur and Grant-White)  in Chicago. The data were first published in \cite{holzinger1939study} and they are publicly available through the {\tt lavaan} package \citep{lavaan} in {\tt R}. This is a well-known dataset used in the LISREL \citep{joreskog1999lisrel}, AMOS \citep{arbuckle2010ibm} and Mplus \citep{muthen2019mplus} tutorials to illustrate a three-factor model for normal data. Variables 1-3 (visual perception, cubes and lozenges) denote ``visual perception'', variables 4-6 (paragraph comprehension, sentence completion and word meaning) are related to  ``verbal ability'', and variables 7-9 (speeded addition, speeded counting of dots, speeded discrimination straight and curved capitals) are connected to ``speed''.
Following \cite{joreskog1969general, mavridis_ntzoufras_2014}, we used the subset of 145 students from the Grant-White school.

\begin{figure}[p]
\begin{center}
\includegraphics[scale=0.4]{legend}\\
\begin{tabular}{ccc}
\includegraphics[scale=0.9]{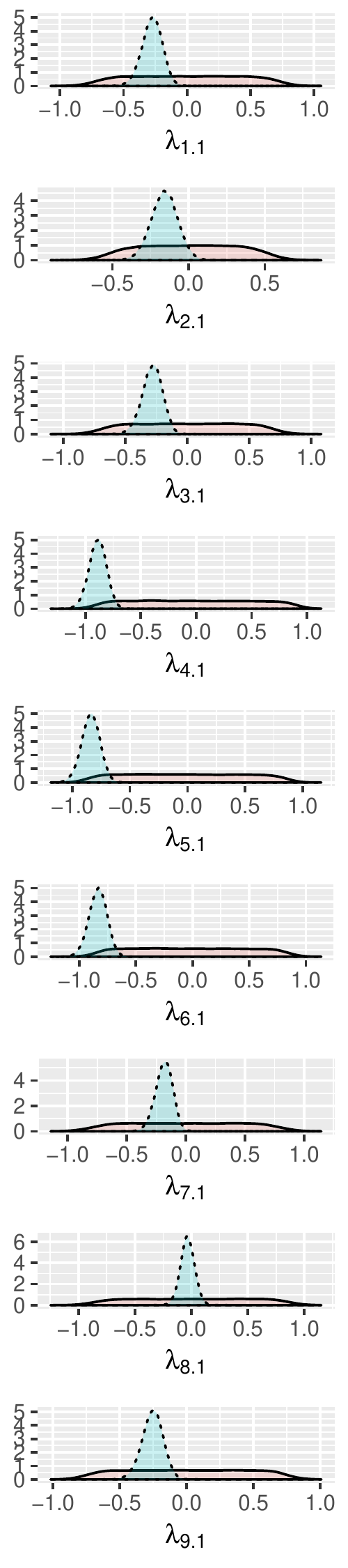}&
\includegraphics[scale=0.9]{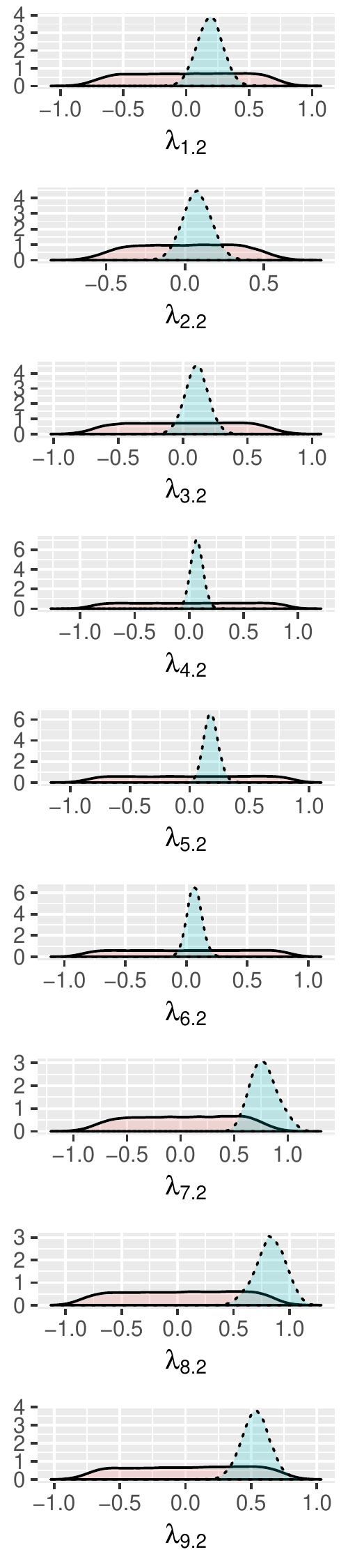}&
\includegraphics[scale=0.9]{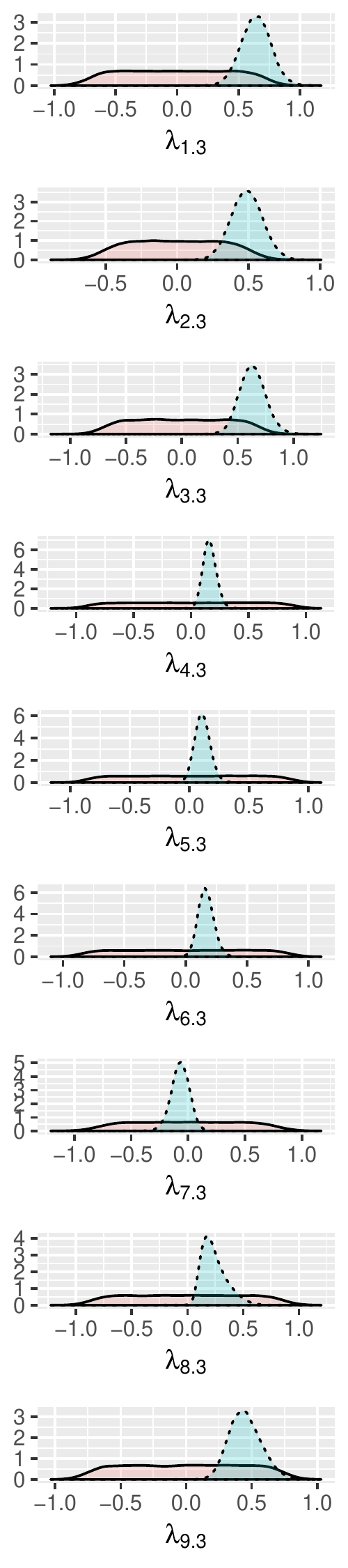}
\end{tabular}
\vspace{-2ex}
\end{center}
\caption{\textit{Grant-White school dataset}: Marginal posterior distribution of raw and reordered factor loadings, using a $q=3$ factor model.}
\label{fig:grantwhitePosterior}
\end{figure}

We fitted factor models consisting of $q=3$ and $q=4$ factors. The corresponding posterior mean estimates of reordered factor loadings are displayed in Table \ref{tab:grantWhite}. When using a model with $q=3$ factors we conclude that Factor 1 is mostly associated with variables 4-6, that is, the ``verbal ability" group. Factor 2 is associated with variables 7-9, that is, the ``speed'' group. Factor 3 is mostly associated with variables 1-3, that is, the ``visual perception" group. Notice however that variable 9 is also loading on the 3rd factor. These points are coherent with the analysis of \cite{mavridis_ntzoufras_2014}. When using a model with $q=4$ factors, the simultaneous $99\%$ credible region contains zero for all loadings of the first column of $\bs{\mathring\Lambda}$, so there is evidence that there is one redundant factor.  
 The raw and reordered outputs for the model with $q=3$ factors is shown in Figure \ref{fig:grantwhitePosterior}.

\begin{table}[p]
\centering
\begin{tabular}{rrrrrrrrrrrr}
  \toprule
& & \multicolumn{9}{c}{Variables} \\
\cline{3-12} 
&\\[-0.5em]
& & $Y_1$ & $Y_2$ & $Y_3$ & $Y_4$ & $Y_5$ & $Y_6$ & $Y_7$ & $Y_8$ & $Y_9$ & $Y_{10}$ \\ 
  \midrule
\multirow{2}{*}{Cluster 1}&Factor 1 &1.0 & 1.0 & 1.0 & 1.0 & 1.0 & 0.0 & 0.0 & -0.0 & -0.0 & -0.0 \\ 
& Factor 2 & 0.0 & -0.0 & 0.0 & 0.0 & -0.0 & -1.0 & -1.0 & -1.0 & -1.0 & -1.0\\ 
\midrule
\multirow{2}{*}{Cluster 2}& Factor 1 &0.0 & -0.0 & -0.0 & -0.0 & -0.0 & 0.5 & 0.5 & 0.5 & 0.5 & 0.5\\ 
& Factor 2 & -0.0 & 0.0 & -0.0 & 0.0 & -0.0 & 0.0 & -0.0 & -0.0 & 0.0 & -0.0 \\ 
   \bottomrule
\end{tabular}
\caption{True factor loadings values for the simulated dataset with 2 clusters (up to a multiplicative constant).}
\label{tab:mfa}
\end{table}

\begin{figure}[p]
\centering
\begin{tabular}{cc}
\includegraphics[scale=0.6]{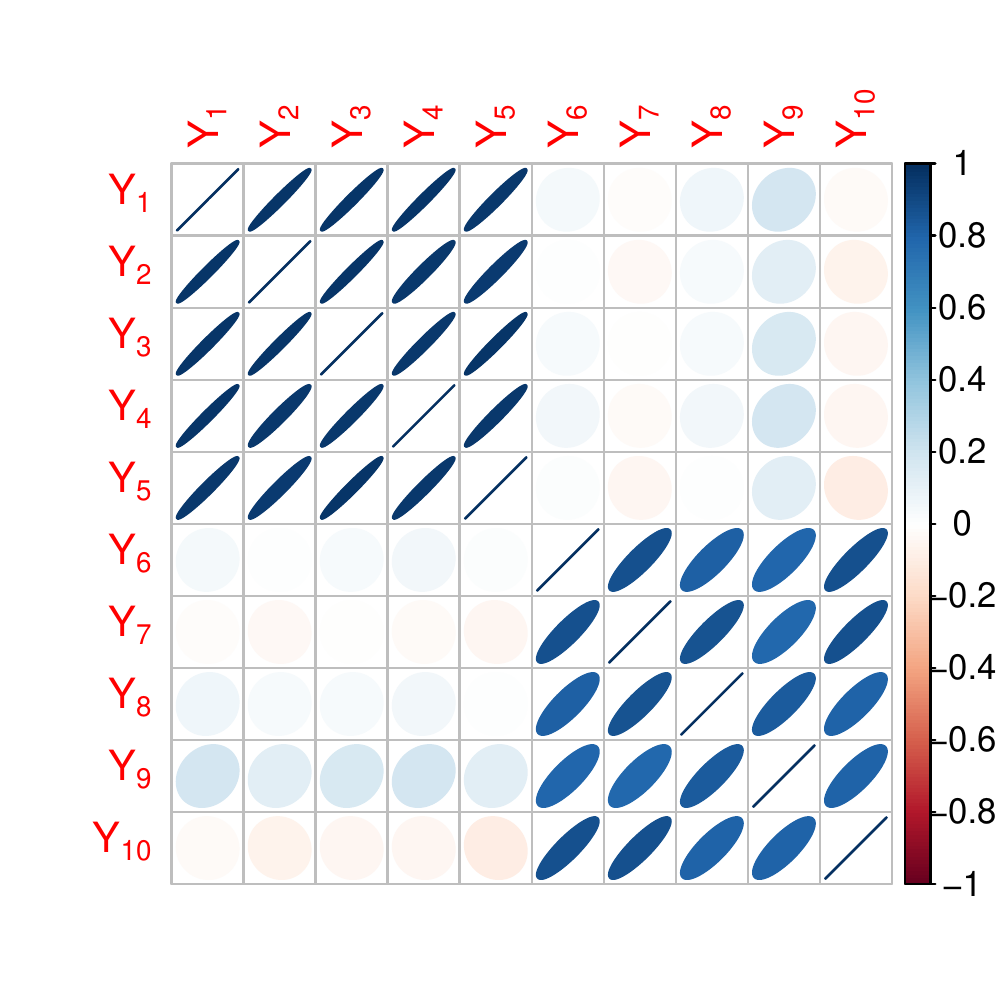}&
\includegraphics[scale=0.6]{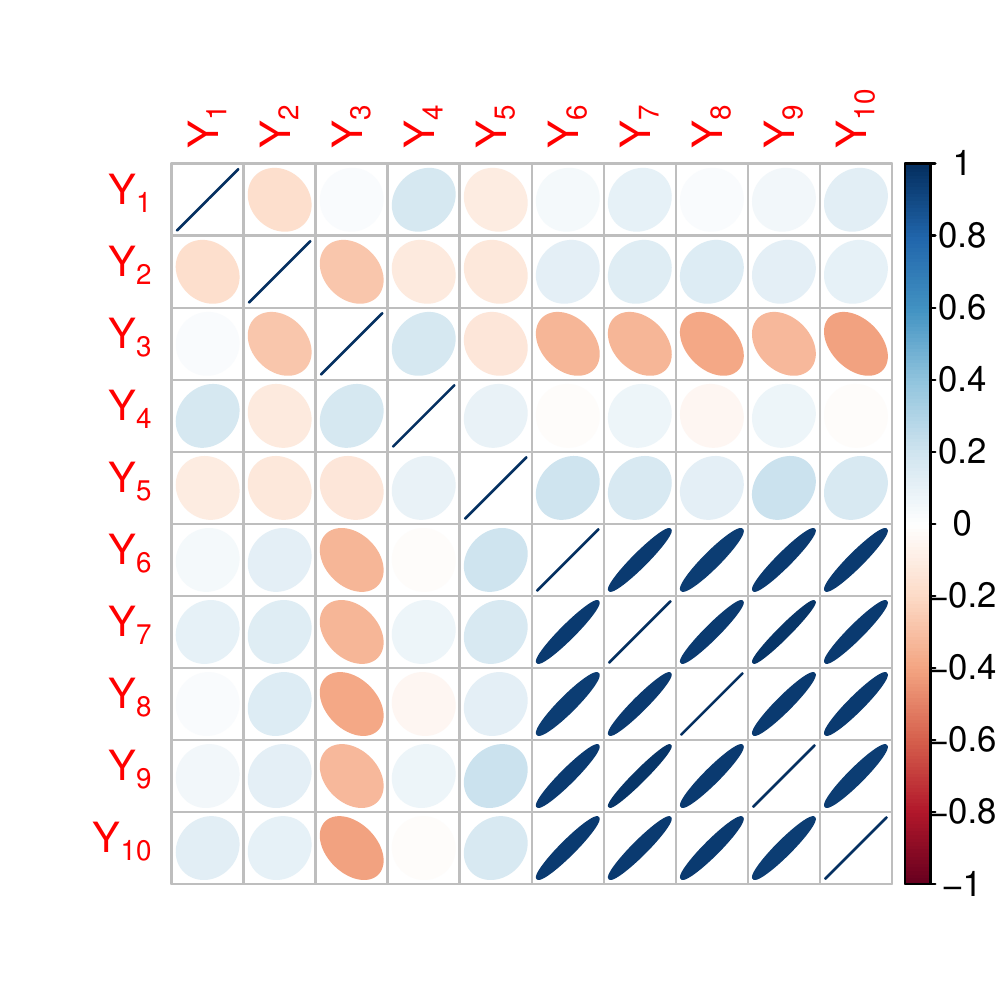}\\
Cluster 1 & Cluster 2
\end{tabular}
\caption{Correlation matrix per cluster for the simulated dataset.}
\label{fig:mfacor}
\end{figure}

\begin{figure}[p]
\begin{tabular}{cc}
\includegraphics[scale=0.61]{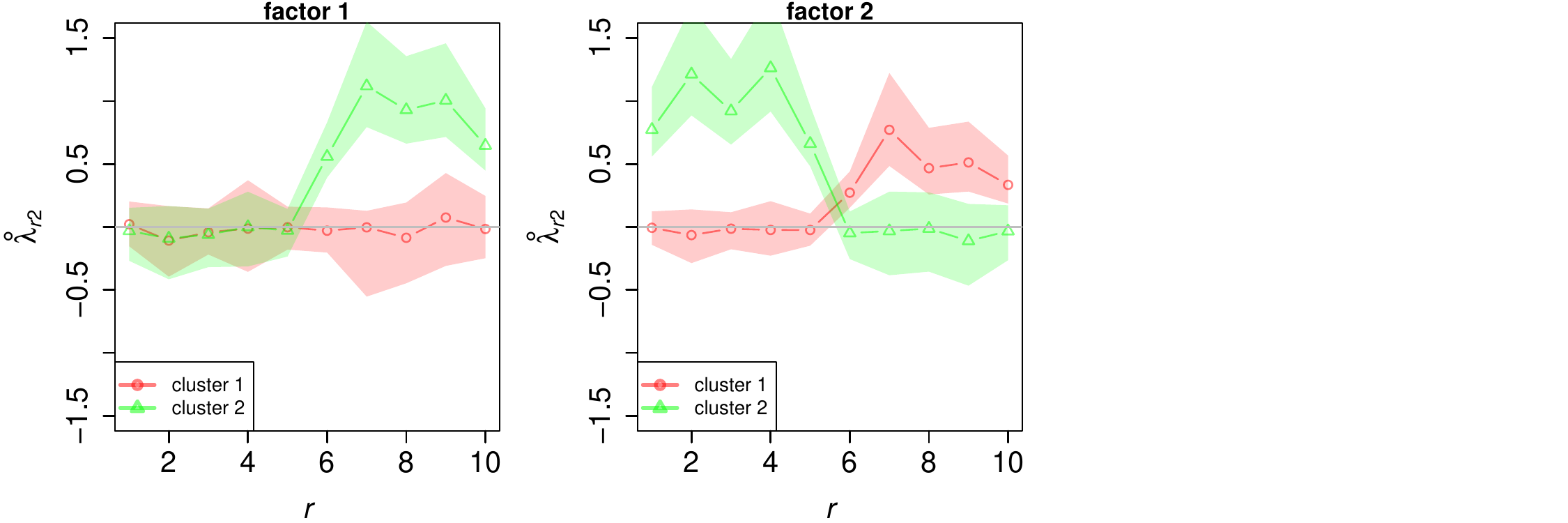}\\
$q=2$\\
\includegraphics[scale=0.61]{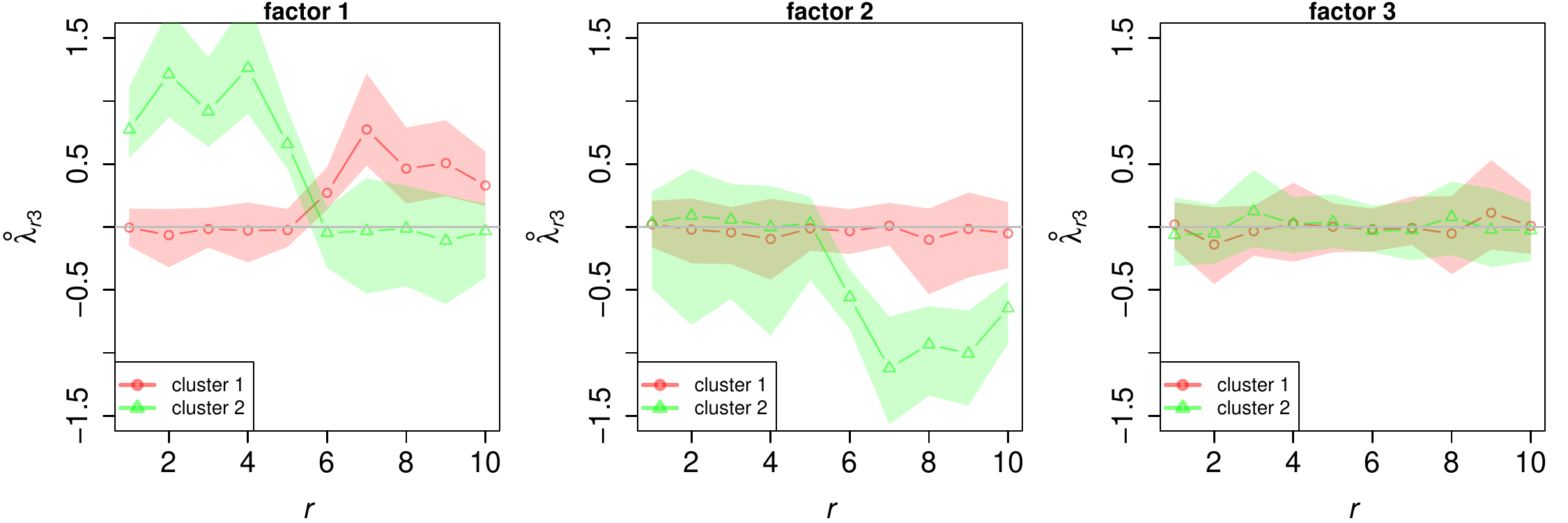}\\
$q=3$
\end{tabular}
\caption{Post-processed factor loadings per cluster for the simulated dataset with $K=2$ clusters, when fitting Bayesian Mixtures of Factor Analyzers with $q$ factors.}
\label{fig:mfacr}
\end{figure}

\subsection{Mixtures of Factor Analyzers}\label{sec:mix}

Mixtures of Factor Analyzers \citep{ghahramani1996algorithm, mclachlan2003modelling,fokoue2003mixtures, mclachlan2011mixtures, McNicholas2008, mcnicholas2016mixture, mal-etal:mod, mal-etal:ide, fru-mal:fro, murphy2018, papastamoulis2018overfitting, papastamoulis2019clustering} are generalizations of the typical FA model, by assuming that Equation \eqref{eq:x_marginal} becomes
\begin{equation}
\bs x_i \sim \sum_{k = 1}^{K}w_k\mathcal N_p\left(\bs\mu_{k},\bs\Lambda_{k}\bs\Lambda_{k}^\top + \bs\Sigma_{k}\right), \mbox{ iid } i = 1,\ldots,n\label{eq:mixture}
\end{equation}
where $K$ denotes the number of mixture components. The vector of mixing proportions $\bs w := (w_1,\ldots,w_K)$ contains the weight of each component, with $0\leqslant w_k\leqslant 1$; $k = 1,\ldots,K$ and $\sum_{k=1}^{K}w_k = 1$. Note that the mixture components are characterized by different parameters $\bs\mu_k,\bs\Lambda_k,\bs\Sigma_k$, $k = 1,\ldots,K$.  Thus, MFAs are particularly useful when the observed data exhibits unusual characteristics such as heterogeneity. We will assume that the number of factors is common across components, although this may not be the case for the true generative model (note that this need not be the case for the fitted model either, see e.g.~\cite{murphy2018}). The reader is referred to \cite{papastamoulis2019clustering} for details of the prior distributions. A difference is that now the factor loadings are assumed unconstrained, in contrast to the original modelling approach of \cite{papastamoulis2019clustering} where the lower triangular expansion in Equation \eqref{eq:lambda} was enabled.

We considered a simulated dataset of $n=100$ and $p=10$-dimensional observations with $K=2$ clusters. The real values of factor loadings per cluster are shown in Table \ref{tab:mfa}. The correlation matrix per cluster is shown in Figure \ref{fig:mfacor}. Notice that the 1st cluster consists of 2 factors, while cluster 2 consists of 1 active factor since the second column is redundant. This is a rather challenging scenario because the posterior distribution suffers from many sources of identifiability problems: At first, all component-specific parameters (including the factor loadings per cluster) are not identifiable due to the label switching problem of Bayesian mixture models. Next, the factor loadings within each cluster are not identifiable due to rotation, sign and permutation invariance. 

The {\tt fabMix}  package \citep{fabMix, papastamoulis2018overfitting, papastamoulis2019clustering} was used in order to produce an MCMC sample from the posterior distribution of the MFA model, using a prior parallel tempering scheme with 4 chains and a number of MCMC iterations equal to 100000, following a burn-in period of 10000 iterations. A thinned MCMC sample of 10000 iterations was retained for inference. 

We considered overfitted mixture models with $K_{\max}=5$ components under all constrained {\tt pgmm} parameterisations \citep{McNicholas2008} of the marginal covariance in \eqref{eq:mixture}. Two different factor levels were fitted, that is, models with $q= 2$ and $q=3$ factors. In both cases, the most-probable  number of clusters is 2, that is, the true value, and the constraint $\bs \Sigma_1=\ldots,\bs\Sigma_K$ (the UCU model in the {\tt pgmm} nomenclature) is selected according to the Bayesian Information Criterion \citep{schwarz1978}. Furthermore, the model with $q=2$ factors is identified as optimal according to the BIC, though we present results for both values of factor levels. 
The raw output of the MCMC sampler is first post-processed according to the Equivalence Classes Representatives (ECR) algorithm \citep{Papastamoulis:10} in order to deal with the label switching between mixture components. Next, the proposed method was applied within each cluster  in order to correct the rotation-sign-permutation  invariance of factors. The resulting simultaneous $99\%$ credible region of the reordered output of factor loadings is displayed in Figure \ref{fig:mfacr}. 

More specifically, when the number of factors is set equal to 2, there is one cluster (coloured red) where the simultaneous credible region of all loadings of the factor labelled as ``factor 1'' contains zero, while the loadings of the factor labelled as ``factor 2'' are different than zero for all variables $r\geqslant 6$. This is the correct structure of loadings for cluster 2 in Table \ref{tab:mfa}. In addition, there is another cluster (coloured green) where the simultaneous credible region of all loadings of the factor labelled as ``factor 1'' does not contain zero for all $r\geqslant 6$, while the loadings of the factor labelled as ``factor 2'' are different than zero for all variables $r\leqslant 5$.  When the number of factors in the MCMC sampler is set to $q=3$ (larger than its true value by one), observe that the simultaneous credible region of the factor labelled as ``factor 3'' in  Figure \ref{fig:mfacr} contains zero. We conclude that, up to a switching of cluster labels and a signed permutation of  factors within each cluster, the proposed approach successfully identifies the structure of true factor loadings.

Further analysis based on simulated and publicly available data is presented in the Appendices. Appendix \ref{sec:mfa_ccc} deals with the case where all clusters share a common idiosyncratic variance and a common matrix of factor loadings, while Appendix \ref{sec:wave} presents results based on the publicly available Wave dataset \citep{breiman}.

\section{Discussion}\label{sec:discussion}

The problem of posterior identification of Bayesian Factor Analytic models has been addressed using a post-processing approach. 
According to our simulation studies and the implementation to real  datasets, the proposed method  leads to meaningful posterior summaries.  We  demonstrated that the reordered MCMC sample can successfully identify over-fitted models, where in such cases the credible region of factor loadings contains zeros for the corresponding redundant columns of $\bs\Lambda$.  Our method is also relevant to the model-based clustering community as shown in the applications on mixtures of factor analyzers (Section \ref{sec:mix} and Appendix \ref{sec:wave}). Comparison of multiple chains is also possible after coupling the pipeline with one extra reordering step as discussed in Appendix \ref{sec:multipleChains}.

The proposed method first proceeds by applying usual varimax rotations on the generated MCMC sample. We have also used oblique rotations \citep{hendrickson1964promax} and we obtained essentially the same answers. Then, we minimize the loss function in Equation \eqref{eq:problem2}, which is carried out in an iterative fashion as shown in Algorithm \ref{algo1}: given the sign ($s$) and permutation ($\nu$) variables, the matrix $\bs \Lambda^{*}$ is set equal to the mean of the reordered factor loadings. Given $\bs \Lambda^{*}$, $s$ and $\nu$ are chosen in order to minimize the expression in Equation \eqref{eq:minimizationStep}. In order to minimize  \eqref{eq:minimizationStep}, we solve one assignment problem (see Equation \eqref{eq:transportation}) for each value of $s$ (per MCMC iteration), as detailed in Section \ref{RSP_strategies}. This approach works within reasonable computing time for typical values of the number of  factors (e.g.~$q\leqslant 10$). 

For larger values of $q$, we propose two approximate solutions based on simulated annealing. 
Simulation  details concerning the computing time for each proposed scheme 
for models with different numbers of factors (up to $q=50$) are provided in Appendix \ref{sec:computational} . 
In these cases we have generated MCMC samples using Hamiltonian Monte Carlo techniques implemented in the {\tt Stan} \citep{JSSv076i01, rstan} programming language. 
According to these empirical findings, the two simulated annealing based algorithms are very effective, rapidly decreasing the objective function within reasonable computing time. 
Nevertheless, the partial simulated annealing algorithm should be preferred since it reaches solutions close to the true minimum faster than the full annealing scheme. 
This finding was expected since the proposal mechanism in Partial SA is more elaborate compared to the completely random proposal in Full SA. 

As discussed after Equation (10), the proposed method solves a discretized version of the Orthogonal Procrustes problem which is the basis for the OP/WOP method of \cite{AMANN2016190}. The refinement of the search space under our method leads to improved results as concluded by our simulation studies, but at the cost of an increased computing time (see Table \ref{tab:p_lt_n} for an example). The final varimax rotation in the method of \cite{AMANN2016190} is essential when looking at the simultaneous credible intervals of factor loadings in order to obtain columns of $\bs \Lambda$ that do not include zero. Hence, the results are significantly worse without the varimax rotation step because in such a case, the resulting factors will not in general correspond to a ``simple structure''. Regarding the estimation of the number of factors according to marginal likelihood estimation, the results obtained by the OP/WOP method are essentially the same with or without the varimax rotation step (that is, overestimate the number of factors regardless of the inclusion of the additional varimax rotation step).

Finally, the authors are considering the implementation of the proposed approach in combination with Bayesian variable selection methods such as stochastic search variable selection -- SSVS \citep{george_mcculloch_93,mavridis_ntzoufras_2014}, Gibbs variable selection -- GVS \citep{dellaportas_etal_02} and/or reversible jump MCMC -- RJMCMC \citep{green95}. 
The implementation of the method might solve not only identifiability problems but also provide more robust results for Bayesian variable selection methods where the specification of the prior distribution is crucial due the Lindley--Bartlett paradox. 
Moreover, in this paper, our proposed method is used as a post-processing tool for the estimation of the posterior distribution of factor loadings within each model. 
For Bayesian variable selection, the implementation of the Varimax-RSP  algorithm within each MCMC might be influential for the selection of items and factor structure. 
For this reason, a thorough study (theoretical and empirical) and comparison between the post-processing and the within-MCMC implementation of the method is needed.

\bigskip
\begin{center}
{\large\bf SUPPLEMENTARY MATERIAL}
\end{center}

\begin{description}

\item[Appendices:]
{\ref{sec:toy}}: Toy example: Geometrical illustration for $q=2$ factors. \\ 
                  {\ref{sec:p_lt_n}}: An example with $p>n$. \\
                  { \ref{sec:multipleChains}}: Comparison of multiple chains. \\
                  { \ref{sec:computational}}: Computational benchmark: Performance comparison in high dimensional factor analytic models.\\ 
                  { \ref{sec:k_med}}: Comparison with $k$-medoids reordering.\\ 
                  {\ref{sec:mfa_ccc}}: Mixtures of factor analyzers: further simulation study.\\
                  {\ref{sec:wave}}: Illustration of mixtures of factor analyzers in publicly available data: The Wave dataset.\\
                  {\ref{sec:exchangeRates}}: Analysis of the exchange rate returns dataset \citep{10.2307/1392266}.

All benchmarks and time comparisons reported in the Appendix were implemented in a Linux workstation with the following specifications: Processor: intel Core i7-8700K CPU @ 3.7GHz $\times$ 12, Memory: 15.6 Gb, OS: Ubuntu 20.04.2 LTS, 64 bit.

\item[Contributed software:] R-package {\tt factor.switching} \citep{arxivVersion} containing code implementing the methodology described in the article:  \url{http://CRAN.R-project.org/package=factor.switching}

\item[Reproducibility:] \url{https://github.com/mqbssppe/factor_switching} This repository contains scripts that reproduce the results for both simulated and real data.

\end{description}

\appendix

\section*{Appendix}

\renewcommand{\thesubsection}{\Alph{subsection}}
\addcontentsline{toc}{section}{Appendices}
\renewcommand{\theequation}{\thesubsection.\arabic{equation}}
\renewcommand\thefigure{\thesubsection.\arabic{figure}}
\renewcommand\thetable{\thesubsection.\arabic{table}}

\setcounter{equation}{0}
\setcounter{figure}{0}
\setcounter{table}{0}

\subsection{Toy example: Geometrical illustration for $q=2$ factors}\label{sec:toy}

\begin{figure*}
\centering
\begin{tabular}{p{6.2cm}@{~~~~~}p{6.2cm}}
\includegraphics[scale=0.45]{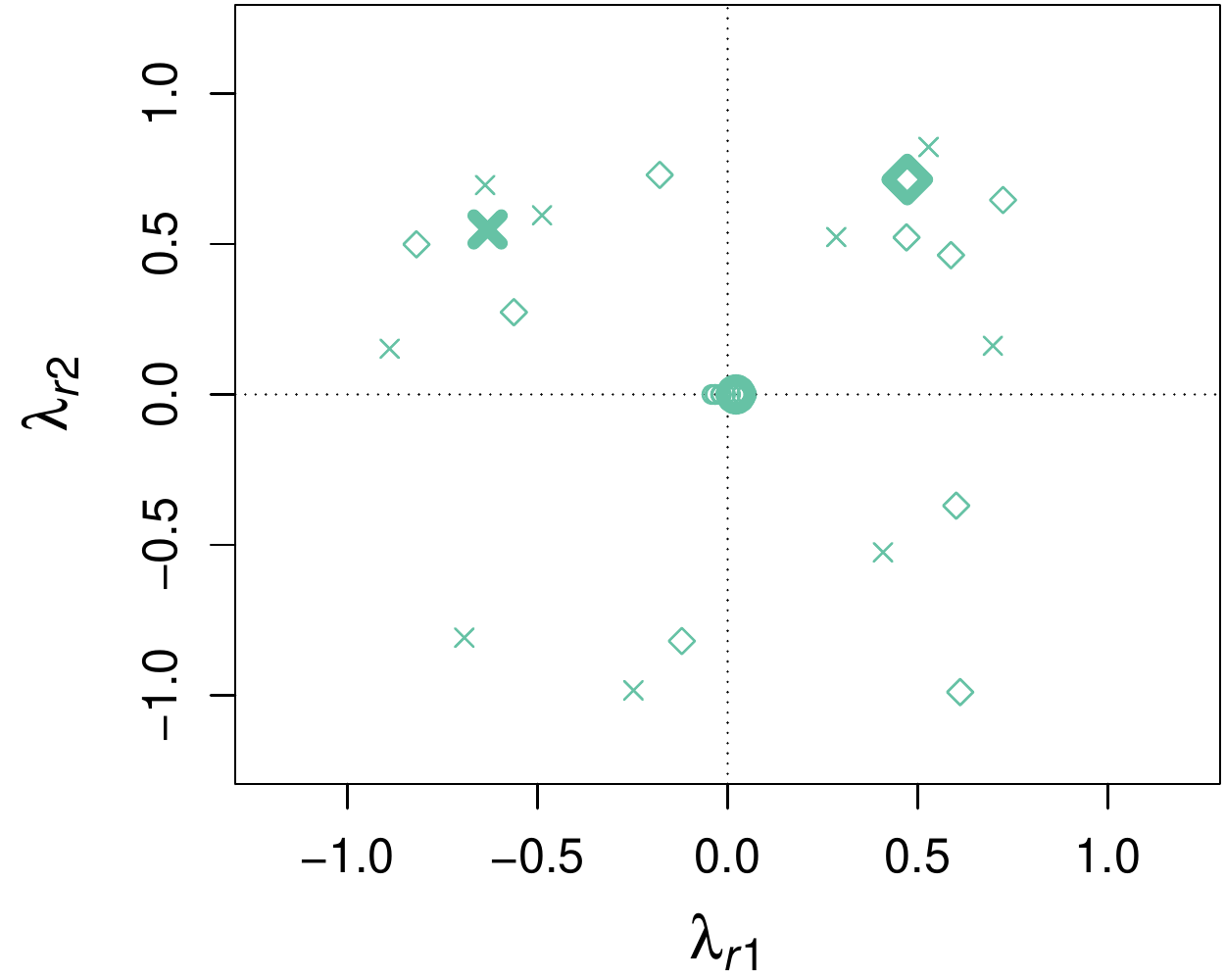} &
\includegraphics[scale=0.45]{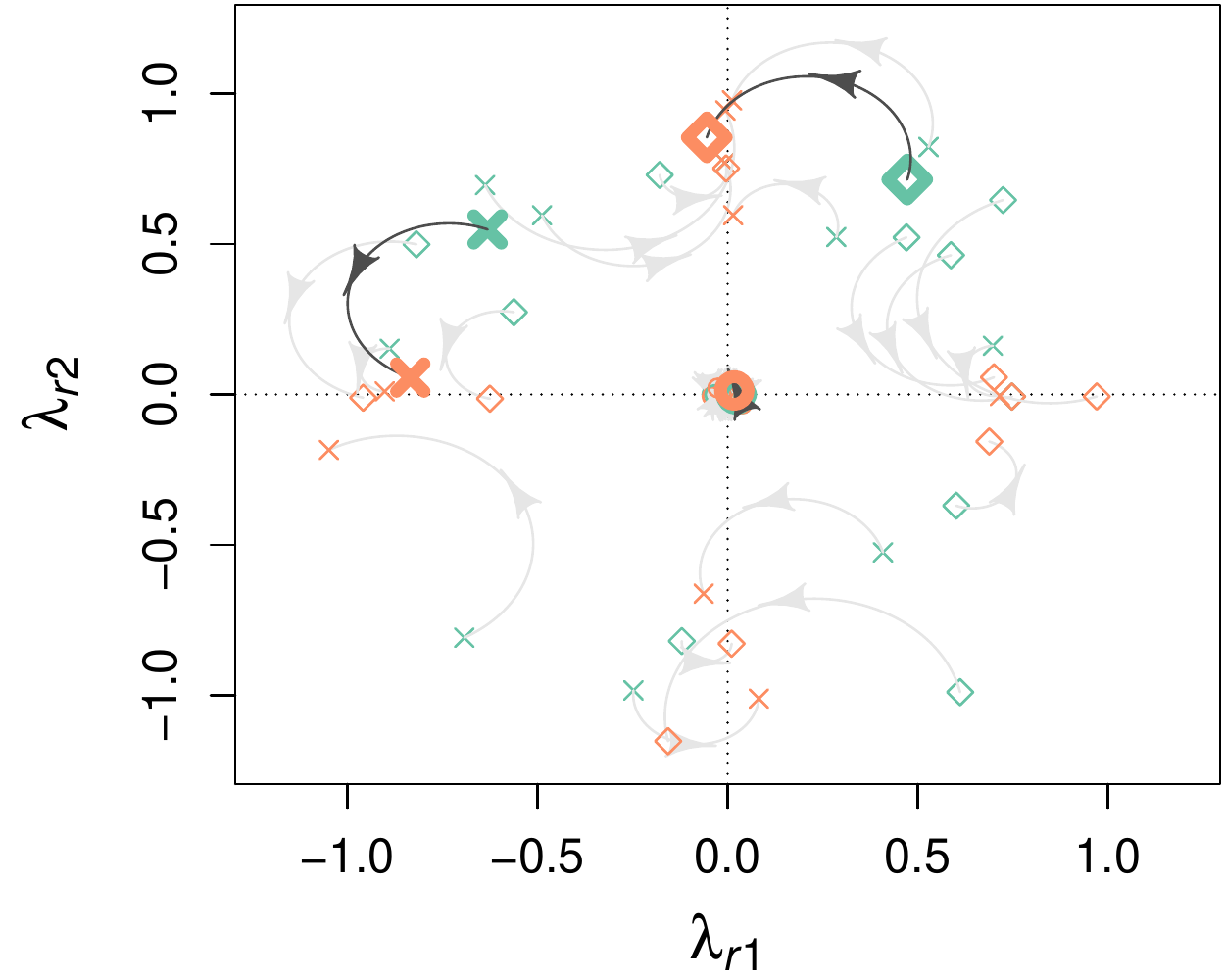} \\
\noindent
(a) 10 raw MCMC draws of factor loadings $\lambda_{rj}$ for 3 variables ($r=1,2,3$) and two factors ($j=1,2$) & 
\noindent
(b) Illustration of  varimax rotations on the raw MCMC output\\
\includegraphics[scale=0.45]{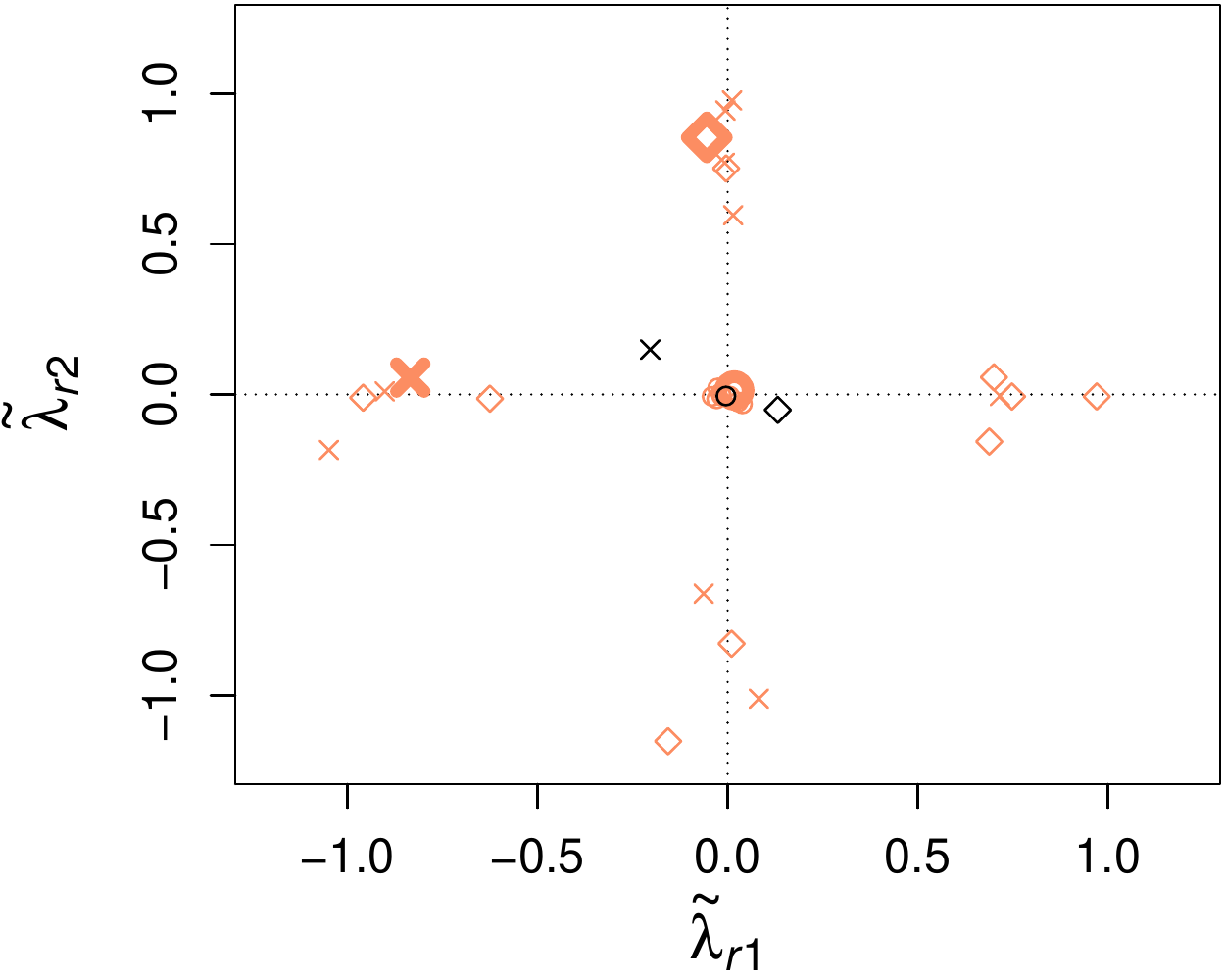} &
\includegraphics[scale=0.45]{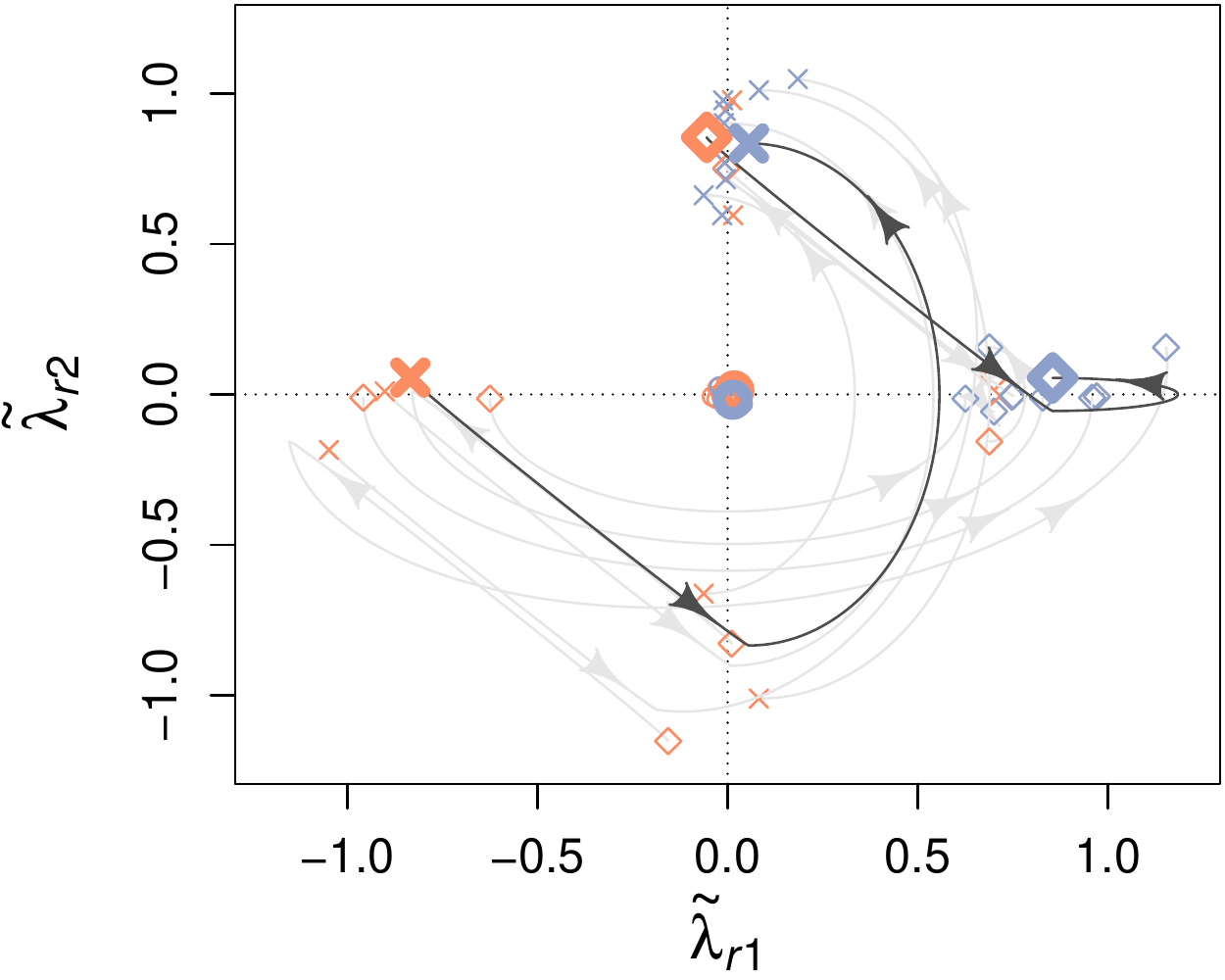}\\
\noindent
(c) Rotated factor loadings according to varimax rotations & \
\noindent
(d) Illustration of signed permutations on the varimax rotated MCMC output\\
\includegraphics[scale=0.45]{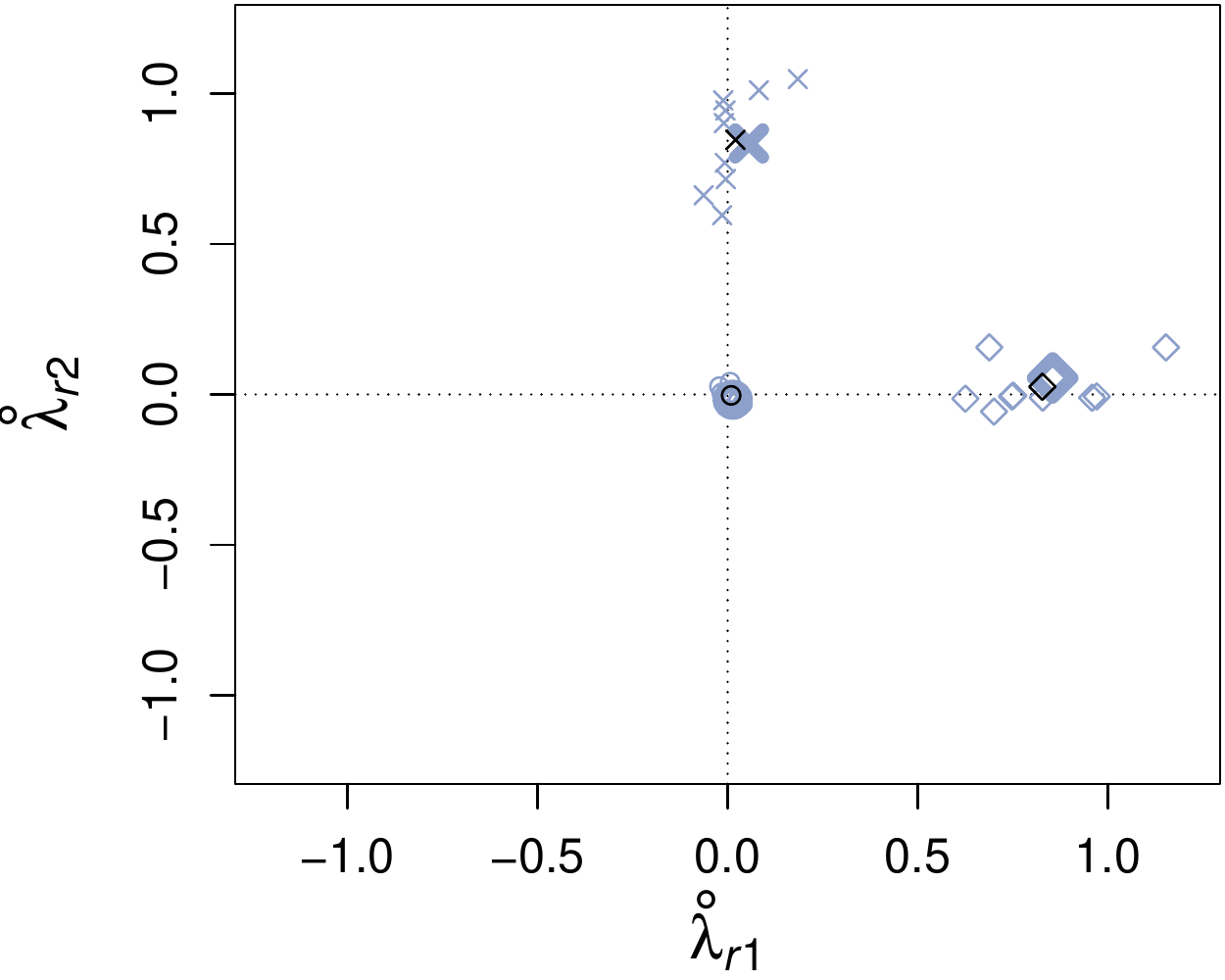} &
\includegraphics[scale=0.45]{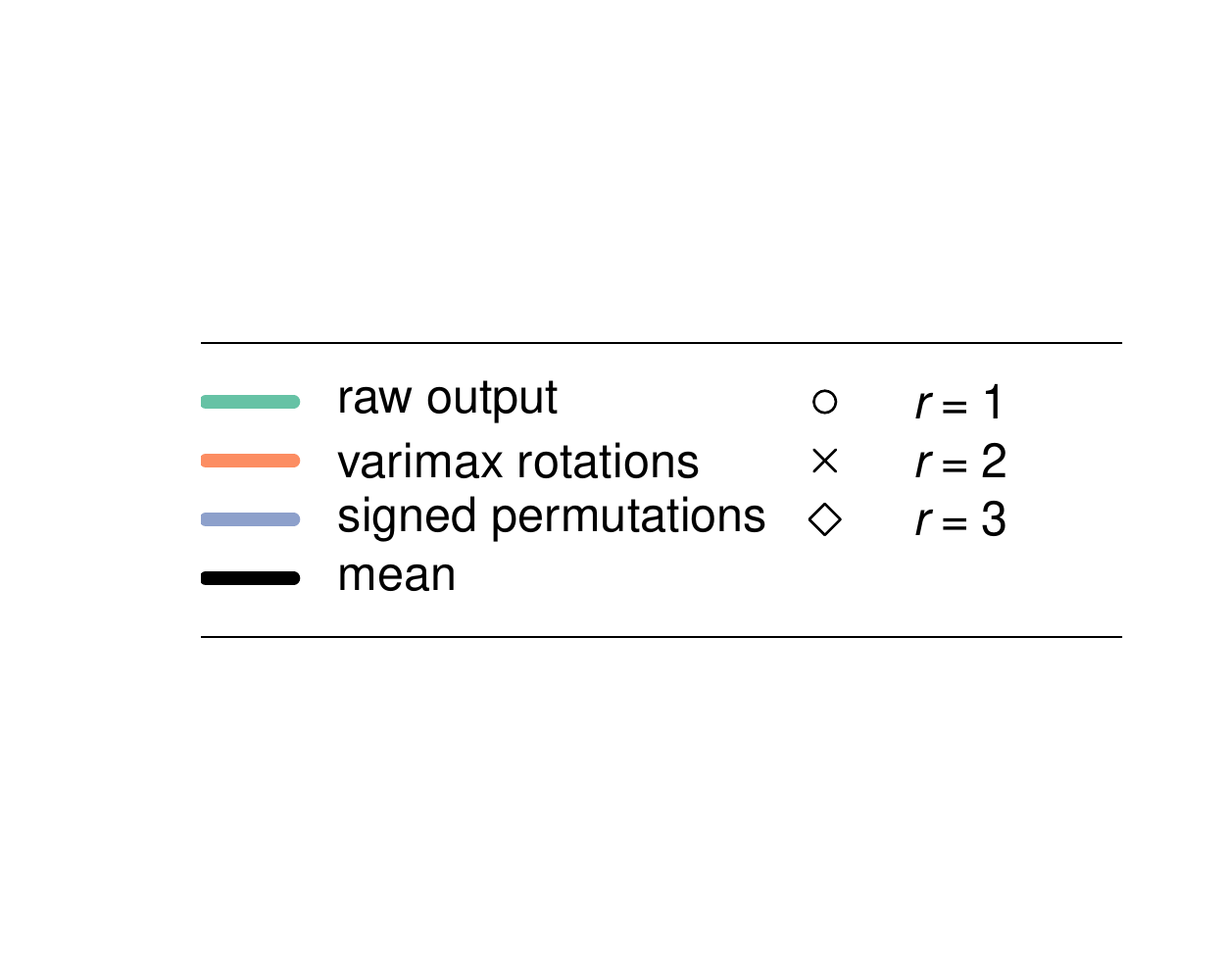} \\
\noindent
(e) Final reordered MCMC output of factor loadings. & 
\end{tabular}
\caption{Graphical demonstration of the evolution of the RSP algorithm for $T=10$ MCMC draws. The highlighted symbols in each panel correspond to a particular MCMC draw. 
The black symbols in panels (c) and (e) denote the average loadings $(\lambda^\star_{r1},\lambda^\star_{r2})$ per variable across the 10 MCMC draws. 
\label{fig:illustration}
}
\end{figure*}

A geometrical illustration of the proposed method is provided in Figure \ref{fig:illustration}, for the special case of $q=2$ factors. Figure \ref{fig:illustration}.(a) shows the scatterplot of the ordered pairs $\big(\lambda_{r1}^{(t)},\lambda_{r2}^{(t)}\big)$ for variable $r = 1, 2, 3$ (corresponding to distinct symbols), where  $t =1,\ldots,10$ denotes a given MCMC draw.  The large variability of the
 MCMC draws suggests that the output of factor loadings is not identifiable due to rotation, sign and permutation invariance. For illustration purposes, a particular MCMC draw is emphasized, with values equal to
\[
\bs\Lambda= \begin{pmatrix}
 0.02 & 0.00 \\
 -0.63 & 0.55 \\
 0.47 & 0.71 
\end{pmatrix},
\]
where each row of the matrix above corresponds to the enlarged symbols in Fig \ref{fig:illustration}.(a). 
Firstly, we apply usual varimax rotations to the whole MCMC output, as shown in Figure \ref{fig:illustration}.(b). The rotated values after this step are displayed in Figure \ref{fig:illustration}.(c). For example, $\bs\Lambda$ is transformed to
\[
\bs{\widetilde\Lambda} = \begin{pmatrix}
0.02 & 0.01\\
-0.84 & 0.06 \\
 -0.05 & 0.86 
 \end{pmatrix}.
\]
Note that a simple structure is achieved for each MCMC iteration: each variable loads to at most one factor. However, the factor loadings are still unidentified across the MCMC draws due to sign and permutation invariance.

The final step is to apply Algorithm \ref{algo1},  as shown in Figure \ref{fig:illustration}.(d). In the initialization step of Algorithm \ref{algo1}, the reference matrix of factor loadings is equal to 
\[\bs\Lambda^{\star}=\begin{pmatrix}
-0.00 & -0.01 \\
-0.20 & 0.15 \\
0.13 & -0.05 
\end{pmatrix}\]
and its rows correspond to the black points shown in Figure \ref{fig:illustration}.(c). The objective function at the initialization step is equal to
$\sum_{t=1}^{10}\mathcal L_{s^{(t)},\nu^{(t)}}^{(t)}\approx 13.76$.
After 1 iteration of Steps 1 and 2 of Algorithm \ref{algo1} the reordered factor loadings correspond to the blue points  in Figure \ref{fig:illustration}.(d). 

In this case, for each MCMC draw, the transformation consists of a permutation of the index set $\mathcal T_2 = \{1,2\}$ (first segment of the curved arrows) and a sign switching (second segment of the curved arrows). The first transformation means that, for MCMC draw $t$, $\left(\widetilde\lambda_{r1}^{(t)},\widetilde\lambda_{r2}^{(t)}\right)$ is transformed to 
\[\mbox{$\left(\dot\lambda_{r1}^{(t)}, \dot\lambda_{r2}^{(t)}\right) = \left(\widetilde\lambda_{r2}^{(t)},\widetilde\lambda_{r1}^{(t)}\right)$ if $\nu^{(t)} = (2,1)$}\] or \[\mbox{$\left(\dot\lambda_{r1}^{(t)}, \dot\lambda_{r2}^{(t)}\right) = \left(\widetilde\lambda_{r1}^{(t)},\widetilde\lambda_{r2}^{(t)}\right)$ if $\nu^{(t)} = (1,2)$},\] for $r=1,2,3$. The second part of this step is to apply a sign switch, thus $\left(\dot\lambda_{r1}^{(t)},\dot\lambda_{r2}^{(t)}\right)$ may be transformed to \[\mbox{$\left(\mathring\lambda_{r1}^{(t)},\mathring\lambda_{r2}^{(t)}\right)=
\left(-\dot\lambda_{r1}^{(t)},\dot\lambda_{r2}^{(t)}\right)$ if $\left(s_1^{(t)},s_2^{(t)}\right) = (-1,1)$},\] or to \[\mbox{$\left(\mathring\lambda_{r1}^{(t)},\mathring\lambda_{r2}^{(t)}\right)=
\left(\dot\lambda_{r1}^{(t)},-\dot\lambda_{r2}^{(t)}\right)$ if $\left(s_1^{(t)},s_2^{(t)}\right) = (1,-1)$},\]  or to \[\mbox{$\left(\mathring\lambda_{r1}^{(t)},\mathring\lambda_{r2}^{(t)}\right)=
\left(-\dot\lambda_{r1}^{(t)},-\dot\lambda_{r2}^{(t)}\right)$ if $\left(s_1^{(t)},s_2^{(t)}\right) = (-1,-1)$},\] or even retain the same sign, that is, \[\mbox{$\left(\mathring\lambda_{r1}^{(t)},\mathring\lambda_{r2}^{(t)}\right)=
\left(\dot\lambda_{r1}^{(t)},\dot\lambda_{r2}^{(t)}\right)$  if   $\left(s_1^{(t)},s_2^{(t)}\right) = (1,1)$},\] for  $r=1,2,3$. The processed values after this step are displayed in Figure \ref{fig:illustration}.(e) which is the final output returned by our method. Note that the processed output is switched to a simple structure which is coherent across all MCMC iterations. 

For example, the values of the emphasized MCMC draw are first permuted according according to $\nu^{(t)} = (2,1)$ and then switched according to $\left(s_1^{(t)},s_2^{(t)}\right) = (1,-1)$ which corresponds to a reflection with respect to the $x$ axis. The corresponding permutation and reflection matrices are $\bs P=
\begin{pmatrix}
0 & 1\\
1 & 0
\end{pmatrix}$ and $\bs S=\begin{pmatrix}
1 & 0\\
0 & -1
\end{pmatrix}$, respectively. Thus, the signed permutation matrix in this case is $\bs Q=\bs S\bs P=
\begin{pmatrix}
0 & 1\\
1 & 0
\end{pmatrix}\begin{pmatrix}
1 & 0\\
0 & -1
\end{pmatrix}=
\begin{pmatrix}
0 & -1\\
1 & 0
\end{pmatrix}$, implying that $\bs{\widetilde\Lambda}$ is finally transformed to
\begin{align*}
\bs{\mathring\Lambda}=\bs{\widetilde\Lambda}\bs Q=\bs{\widetilde\Lambda}\bs S\bs P&=\begin{pmatrix}
0.02 & 0.01\\
-0.84 & 0.06 \\
 -0.05 & 0.86 
 \end{pmatrix}\begin{pmatrix}
0 & 1\\
1 & 0
\end{pmatrix}\begin{pmatrix}
1 & 0\\
0 & -1
\end{pmatrix}\\
&=\begin{pmatrix}
 0.01 & -0.02\\
 0.06 & 0.84\\
 0.86 & 0.05
\end{pmatrix}.
\end{align*}

The reference matrix of factor loadings is now equal to 
\[\bs{\Lambda}^{\star}=
\begin{pmatrix}
 0.01 & -0.00 \\ 
 0.02 & 0.85 \\
 0.83 & 0.03
\end{pmatrix}
\]
and the objective function is 
$\sum_{t=1}^{10}\mathcal L_{s^{(t)},\nu^{(t)}}^{(t)}\approx 0.55$.
Subsequent iterations do not improve further this value and Algorithm \ref{algo1} terminates.

\setcounter{equation}{0}
\setcounter{figure}{0}
\setcounter{table}{0}

\subsection{An example with $p>n$}\label{sec:p_lt_n}

\begin{figure}[h]
\centering
\includegraphics[scale=0.25]{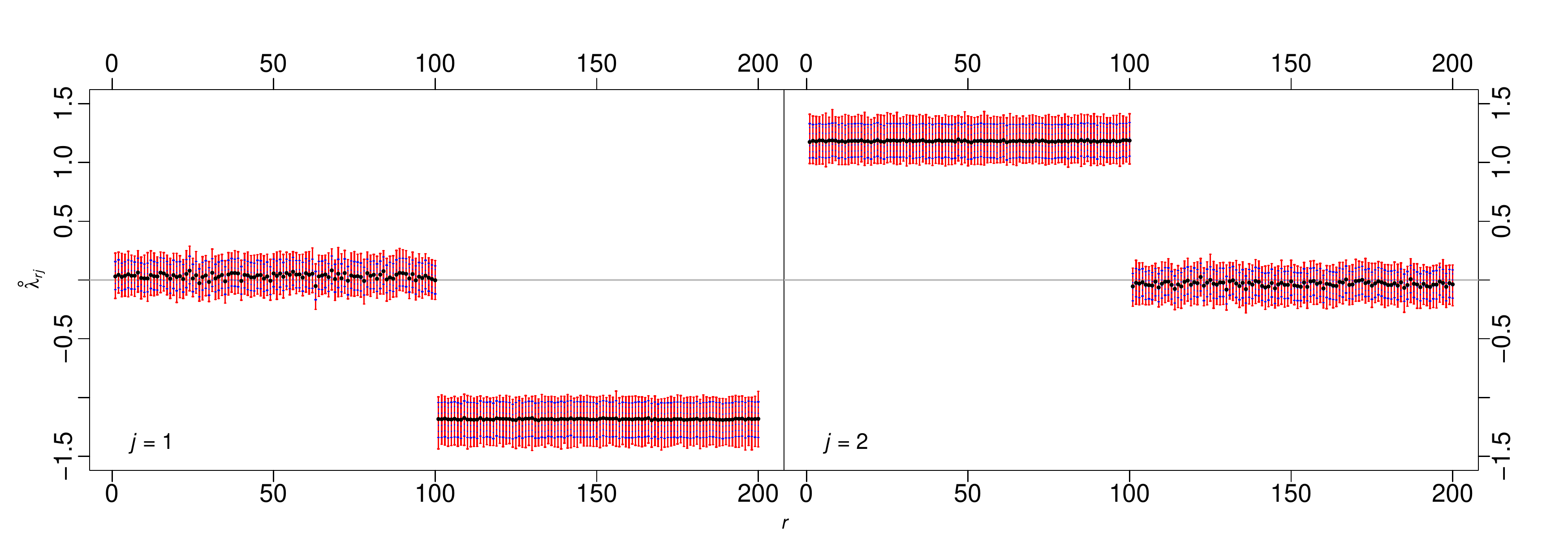}
\caption{$99\%$ credible intervals (red denotes simultaneous credible intervals and blue the individual ones) of reordered values of factor loadings according to the RSP algorithm for a synthetic dataset with $p=200$ variables and $n=180$ observations of Section \ref{sec:p_lt_n}.}
\label{fig:p_l_n}
\end{figure}

In this section we consider the estimation of a factor analytic model in the case where the number of observed variables ($p$) is larger than the sample size ($n$). Such high-dimensional settings are attracting increasing interest in the literature on factor analytic models of late (see e.g.~\cite{10.2307/23248938, 10.1093/biomet/asx030}). 

We generated a synthetic dataset with $p=200$ variables and the sample size was set to $n=180$. The ``true'' number of factors was equal to $q = 2$. The parameters of the prior distribution of the factor loadings in \eqref{eq:s2_prior} and \eqref{eq:lambda_prior} were set equal to  $a_0=b_0=0.001$, $l_0$ and $L_0 = 1$. Then, we used the {\tt MCMCpack} in order to generate  MCMC samples of $T=10000$ draws (after thinning) from factor analytic models with $q=2,3,4,5,6$ factors.  
\begin{table}[h]
\begin{tabular}{ccccccccc}
$q$ & \multicolumn{2}{c}{RSP (exact)} & \multicolumn{2}{c}{RSP (SA)} & \multicolumn{2}{c}{OP} & \multicolumn{2}{c}{WOP}\\
& $\hat q$ & time &$\hat q$ & time &$\hat q$ & time &$\hat q$ & time \\
\toprule
2 & 2 & 91.7 & 2 & 419.6 & 2 & 6.1 & 2& 28.8\\
3& 2 & 138.3 & 2 & 424.0 & 2 & 10.4 & 2&  47.7\\
4 & 2 & 380.4 & 2 & 1159.6 & 2 & 11.3 & 2& 60.3\\
5& 2 & 1821.4 & 2 &  1584.6 & 3 & 11.5 & 3& 65.1\\
6&  2 &  2460.4& 2 & 1817.3 & 3 & 11.8 & 3& 77.3\\
\end{tabular}
\caption{Results for a synthetic dataset with $p=200$ variables and $n=180$ observations of Section \ref{sec:p_lt_n}. The true number of factors is equal to $2$. Each row displays the inferred number of ``effective'' columns $\hat q$ of the factor loadings matrix when fitting factor models with $q$ factors, $q=2,3,\ldots,6$. Timings (in seconds) are averages across 10 runs.}
\label{tab:p_lt_n}
\end{table}

The results are summarized in Table \ref{tab:p_lt_n}. When the number of factors of the fitted model is $2\leqslant q \leqslant 4$ all methods indicate that the matrix of reordered factor loadings contains 2 ``effective'' columns, that is, the ``true'' number. The same continues to hold true for the proposed method (RSP - exact or SA) when in the cases where $q = 5$ or $q=6$. On the contrary, the method of Procrustes rotations (OP and WOP) of \cite{AMANN2016190} reveals an extra factor in these overfitting models. The method of Procrustes rotations is notably faster in all cases.  Figure \ref{fig:p_l_n} illustrates the resulting $99\%$ credible intervals of the post-processed output of factor loadings according to the RSP algorithm when fitting a factor model with $q=2$ factors. We conclude that variables 1-100 load at the first factor, while the remaining ones load at the second factor. This configuration agrees with the actual scenario used to generate the synthetic dataset.

\setcounter{equation}{0}
\setcounter{figure}{0}

\subsection{Comparison of multiple chains}\label{sec:multipleChains}

\begin{figure*}[p]
        \begin{center}
                \includegraphics[scale=0.7]{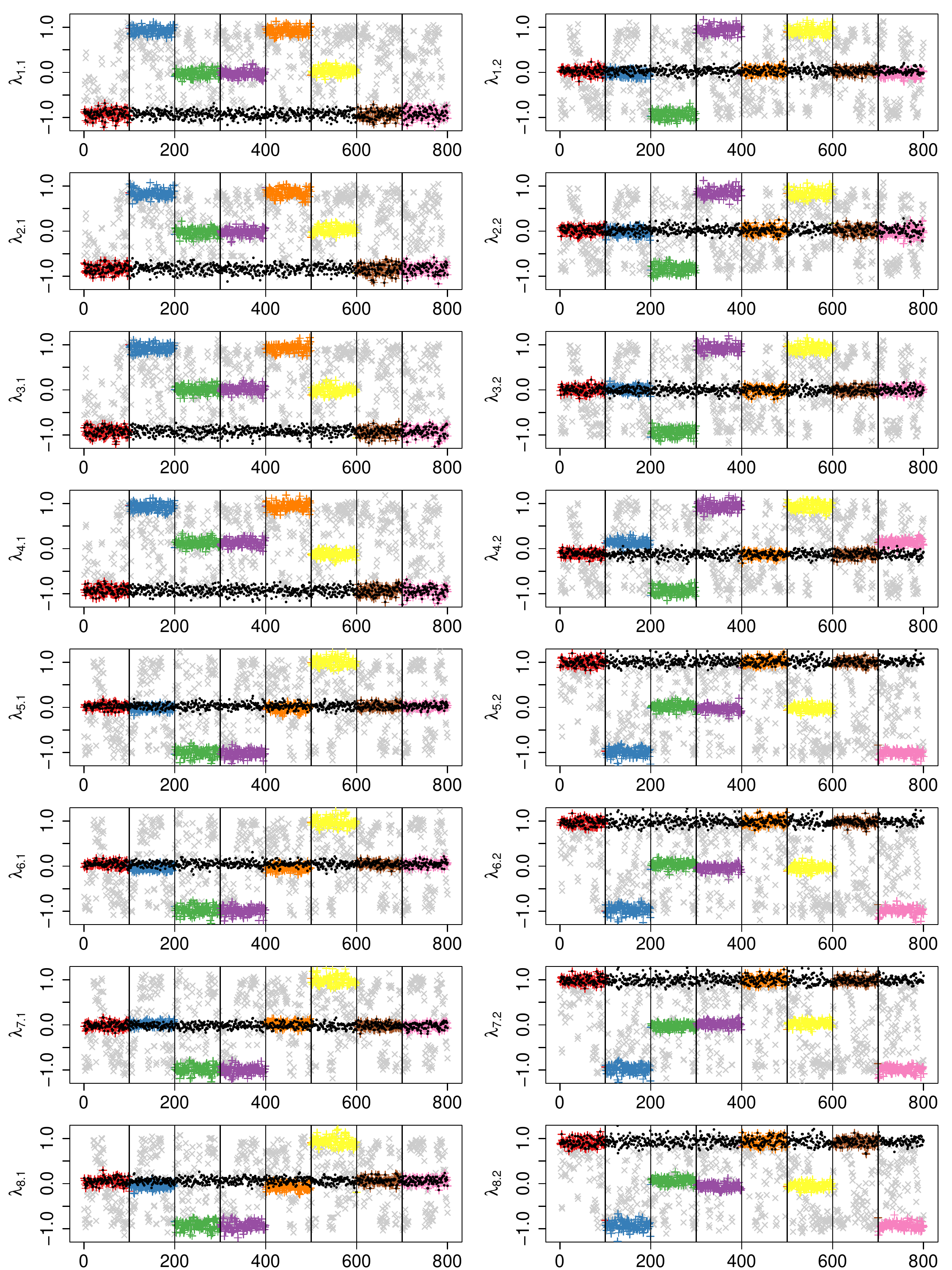}
        \end{center}
        \caption{\textit{Simulated data 1}: Trace plots of the generated loadings before and after the implementation of RSP algorithm. \\
                {\footnotesize\it Notes: Gray-coloured points: raw output of factor loadings (8 parallel chains from {\tt Stan}); Segments (vertical lines): 100 successive MCMC draws. 
                        Coloured points: RSP reordered factor loadings. 
                        Black trace: sign-permuted reordered traces making all chains comparable. }}
        \label{fig:sim1Multipletrace}
\end{figure*}

Running parallel chains  is a standard practice in MCMC applications \cite[see e.g.~][]{JSSv076i01} in order to assess convergence. Applying our method to each chain separately will make the factor loadings identifiable \textit{within} each chain. However, the post-processed outputs will not be directly comparable \textit{between} chains. Clearly, one run may be a signed permutation of another, provided that the chains have converged to their stationary distribution. Thus, it makes sense to post-process the chains in order to switch all of them  in a common region.

This reduces to find \textit{a single} sign-permutation per chain that will reorder \textit{all values} of the given chain. Let $\bs \Lambda^{\star}_{c}$ denote the $\bs \Lambda^{\star}$ matrix (that is, the matrix which contains the estimates of posterior mean of factor loadings) for chain $c=1,\ldots,C$, where  $C$ denotes the total number of parallel chains.  In order to find the final sign-permutations per chain we only have to apply Step 2 of Algorithm \ref{algo1} on $\bs \Lambda^{\star}_{c}$, $c=1,\ldots,C$ (that is, without the varimax rotations step). Let now $\mathring{\bs\Lambda}^{(t,c)}$ denote the (reordered) matrix of factor loadings for chain $c$ on iteration $t$ and $\bs Q^{(c)}$ the resulting sign-permutation for chain $c$. Then, the final step is to transform $\mathring{\bs\Lambda}^{(t,c)}$ to $\mathring{\bs\Lambda}^{(t,c)}\bs Q^{(c)}$, for all $t=1,\ldots,T$, $c=1,\ldots,C$.

We illustrate this procedure on the simulated dataset 1 (used in Section \ref{sec:sim}). We used {\tt Stan} \citep{JSSv076i01}  in order to generate 8 chains of 10000 iterations, following a burn-in period of 1000. Figure \ref{fig:sim1Multipletrace} displays the raw and post-processed output from successive segments of each chain. The first segment displays the first 100 raw and post-processed iterations of the first chain, the second segment displays the raw and reordered values of  iterations $101-200$ for the second chain, and so on. The black coloured points correspond to successive segments of the simultaneously processed chains. It is evident that all chains have been successfully switched on a common labelling. Moreover, both the point estimate and the upper limit of the $95\%$ confidence interval of the potential scale reduction factor \citep{gelman1992inference,brooks1998general} are equal to $1.00$ for all loadings, indicating that there are no convergence issues.

\setcounter{equation}{0}
\setcounter{figure}{0}

\subsection{Computational benchmark: Performance comparison in high dimensional factor analytic models}\label{sec:computational}

\begin{figure*}[p]
\centering{
\includegraphics[page=2,scale=0.5]{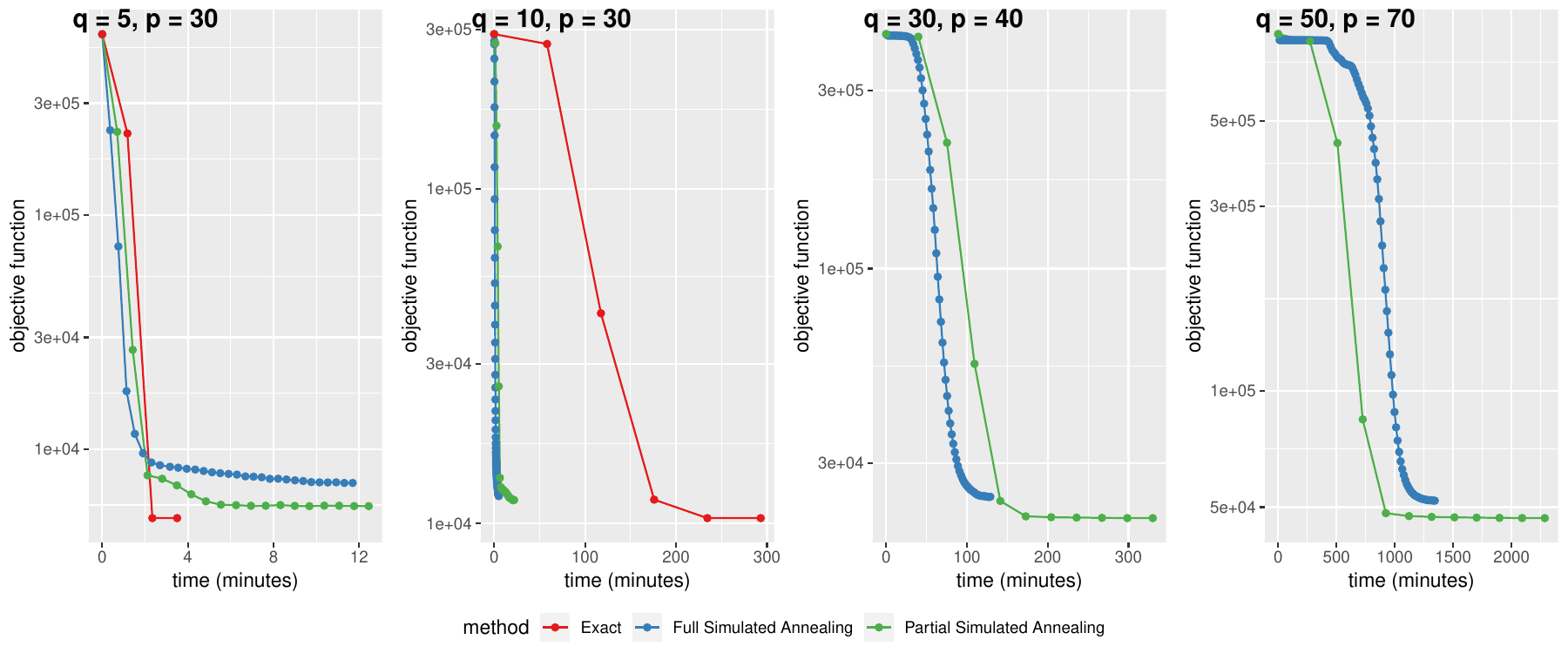}
}
\caption{Comparison of the three computational schemes into four datasets for various number of factors ($q$) and observed variables ($p$). Each point in the graphs corresponds to counts of the number of iterations of Step 2.2 of Algorithm 1. The exact version demands small number of iterations  in order to reach a specific state, compared to the simulated annealing algorithms. However, the time needed for a single iteration of the exact algorithm is elevated compared to the simulated annealing algorithms.  \\ 
        {\it \small Notes: 
        $Y$ axis: in log-scale.         
        Exact version (Scheme A) is not applied for $q>10$.  
        MCMC details: 10000 iterations using {\tt Stan}.}
\label{fig:time}
}
\end{figure*}

Figure \ref{fig:time} displays the progress of successive evaluations of the objective function \eqref{eq:minimizationStep} versus the time needed in order to reach the specific iteration of the algorithm. In all cases, the number of retained MCMC draws is equal to 10000, generated from {\tt Stan} (following a burn-in period of 10000). Note that the time required in order to generate these MCMC samples with {\tt Stan} ranges from a few minutes (for $q=5,10$) to five days (for $q=50$). 

In the partial simulated annealing scheme we used $B=20$ simulated annealing steps/repetitions for $q=5,10$ factors and $B=200$ for $q=30,50$ factors. In the full simulated annealing scheme we used $B=100, 100, 500$ and $2000$ simulated annealing steps/repetitions for $q=5,10, 30$ and $50$ factors, respectively. Observe that for typical values of the number of factors (e.g. $q=5$) the exact scheme should be preferred. As expected, the partial simulated annealing scheme performs better than the full simulated scheme and should be preferred whenever the number of factors is large (e.g.~when $q > 10$). At first, in all cases, the algorithm under the full simulated annealing scheme (blue line) converges to a worse (i.e., larger) value of the objective function compared to all other schemes. Second,  observe that when the number of factors is larger than 10, the algorithm under the full simulated annealing scheme requires a large number of iterations in order to escape from the initial values and start to descend.

\begin{figure*}[p]
\centering
\includegraphics[scale=0.45]{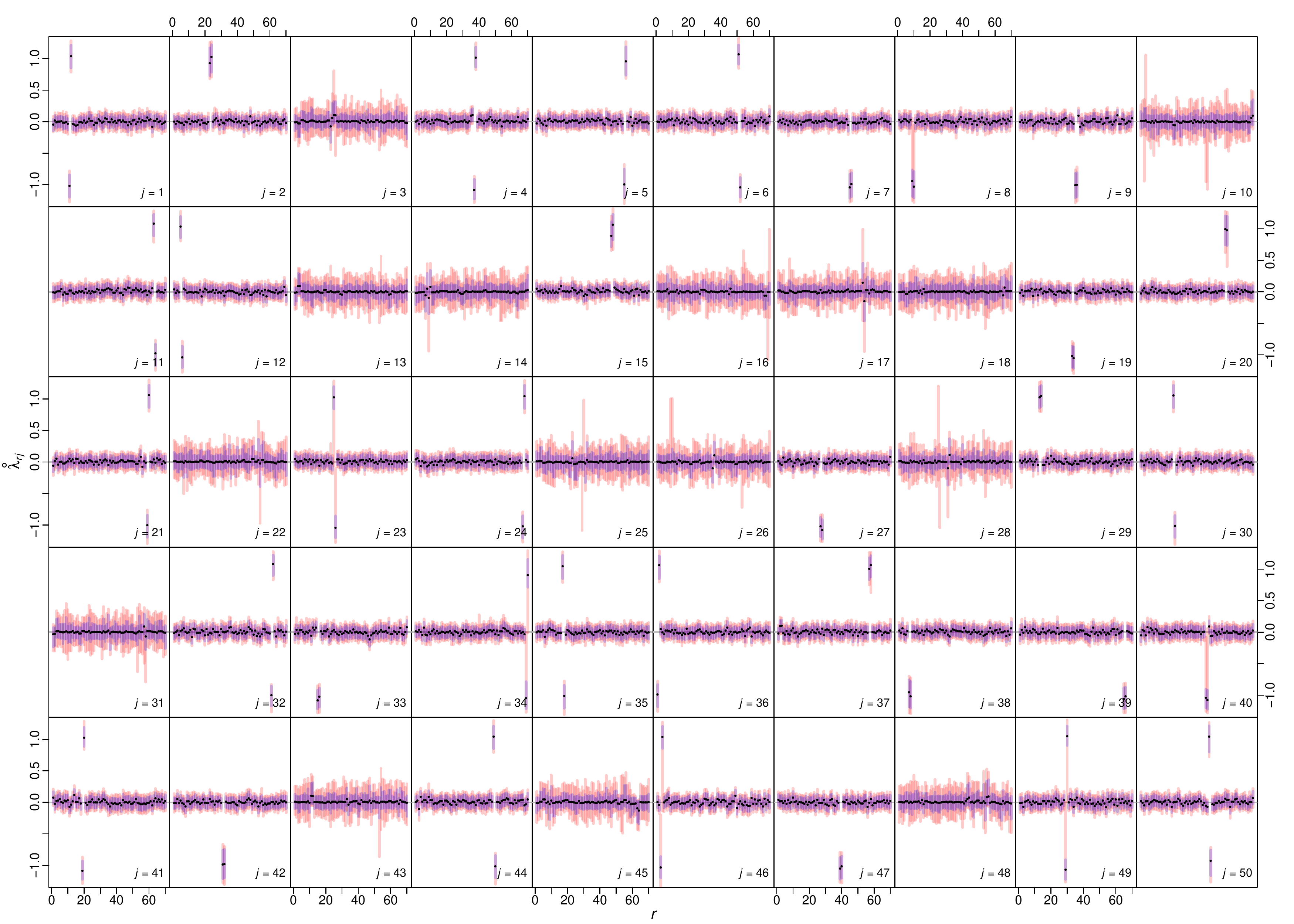}
\caption{Post-processed factor loadings according to RSP algorithm under the partial simulated annealing scheme, for a synthetic dataset of $n=500$ observations and $p=70$ variables. \\
        {\it \small Notes:  
        Fitted model: $q=50$ factors;  True factors: $q_{True}=35$. 
        Blue regions: $99\%$ Highest Density Intervals; 
        Red regions : $99\%$ Simultaneous Credible Region
        Dots: posterior means.}}
\label{fig:big}
\end{figure*}

The post-processed values for the dataset with the 50 factors are presented in Figure \ref{fig:big}, which corresponds to the output returned by the partial simulated annealing algorithm. In this case, the number of factors used to generate the specific dataset is equal to $35$, thus, when fitting a 50-factor model, there should be 15 redundant columns in the resulting matrix of factor loadings. A careful inspection of the $99\%$ Highest Density Intervals (illustrated in blue) reveals that there are 15 panels where all intervals contain zero. The same holds for the $99\%$ simultaneous Credible Region illustrated in red, that is, the panels corresponding to factor $j \in\{3, 10, 13, 14, 16, 17, 18, 22, 25, 26, 28, 31, 43, 45, 48\}$. The same holds for the $99\%$ HPD intervals when using the full simulated annealing scheme (which converged to an inferior solution), but the $99\%$ simultaneous credible region contain zero for 29 factors instead of 15 (results not shown). 

\setcounter{equation}{0}
\setcounter{figure}{0}

\subsection{Comparison with $k$-medoids reordering}\label{sec:k_med}

\begin{figure*}[p]
\centering
\includegraphics[scale=0.6]{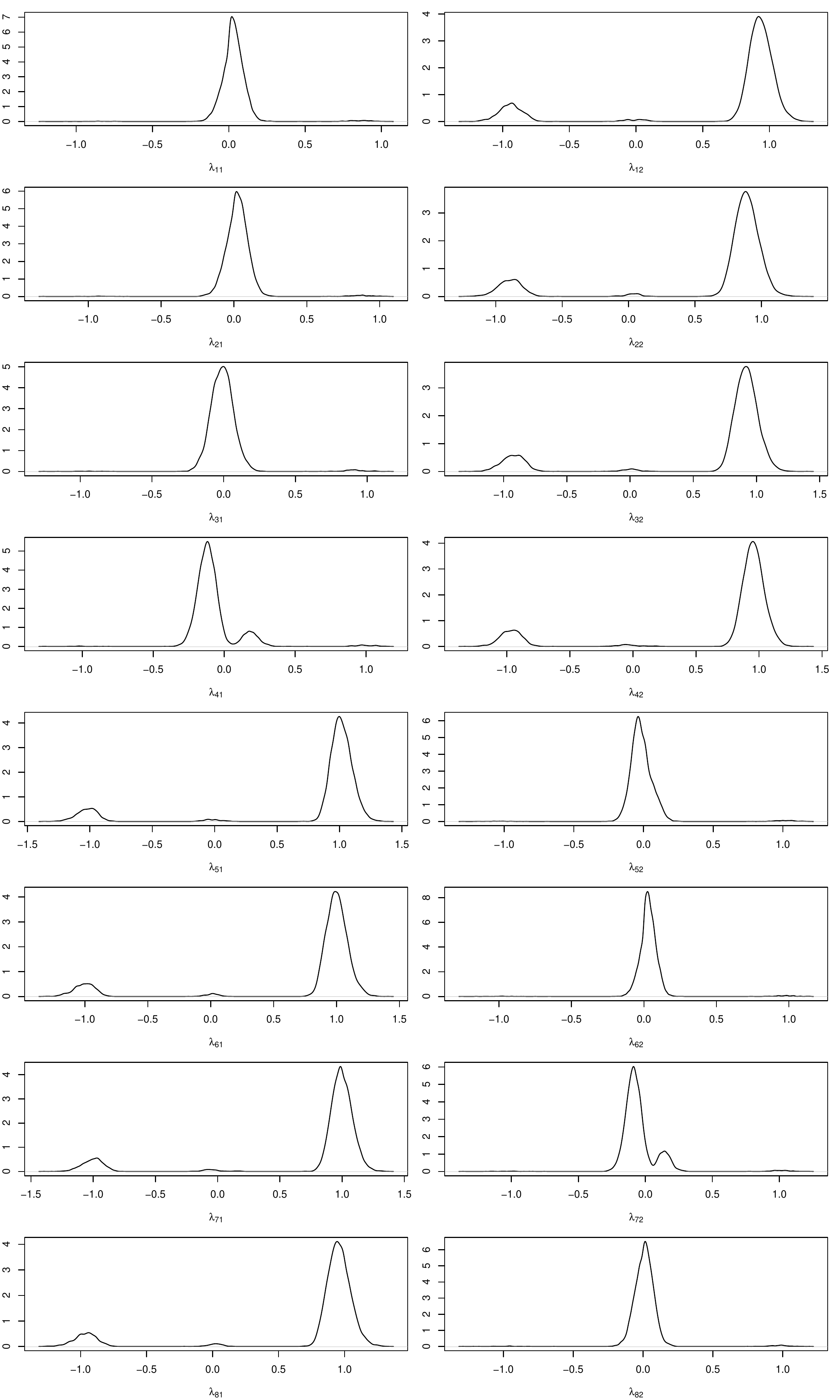}
\caption{Marginal posterior distribution of reordered factor loadings according to the $k$-medoids clustering approach of  \cite{kaufmann2017identifying}, conditional on the true number of factors $q=q_{\mbox{true}}=2$ for the simulated dataset 1 in Section \ref{sec:sim} of the main manuscript.}
\label{fig:kmedoids}
\end{figure*}

We applied the method of \cite{kaufmann2017identifying} (briefly discussed in the second to last paragraph of Section \ref{sec:identify}) in the MCMC output (after varimax rotations) for the simulated dataset 1 of Section \ref{sec:sim} and the results are displayed in Figure \ref{fig:kmedoids}. The same MCMC output was used as input as the one that generated  Figure \ref{fig:sim1} in the main manuscript. Clearly, the estimated marginal posterior distribution of reordered factor loadings is multimodal which indicates that this simpler method is not working well in our case where the model is not sparse. Recall that  (see last paragraph of Section \ref{sec:identify}) a crucial characteristic in the model of \cite{kaufmann2017identifying} is the fact that the generated  matrices of factor loadings contain zeros and this is not the case in our implementation.  Another problem with this approach is computational: a huge amount of memory is required in order to compute the distance matrix among the generated MCMC draws. In practice, the load becomes prohibitively large when considering a few tens of thousands of MCMC iterations. 

\setcounter{equation}{0}
\setcounter{figure}{0}

\subsection{Mixtures of factor analyzers: Simulation study}\label{sec:mfa_ccc}

\begin{figure*}[p]
\centering
\begin{tabular}{c}
\includegraphics[scale=0.6]{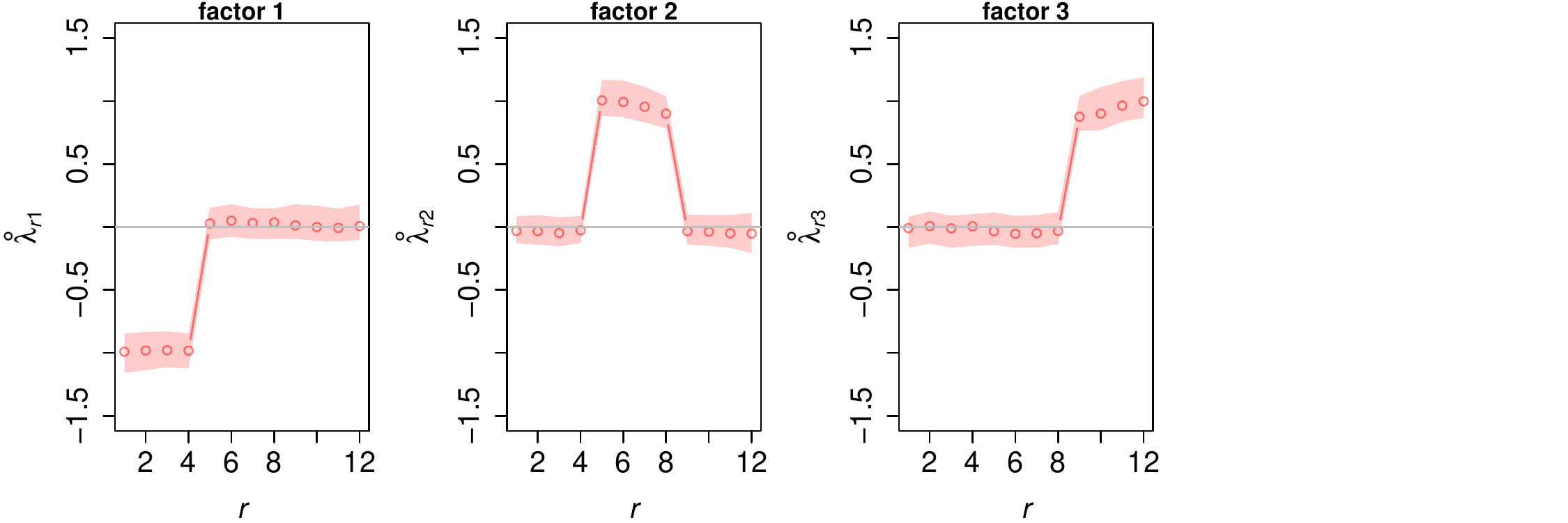}\\
\includegraphics[scale=0.6]{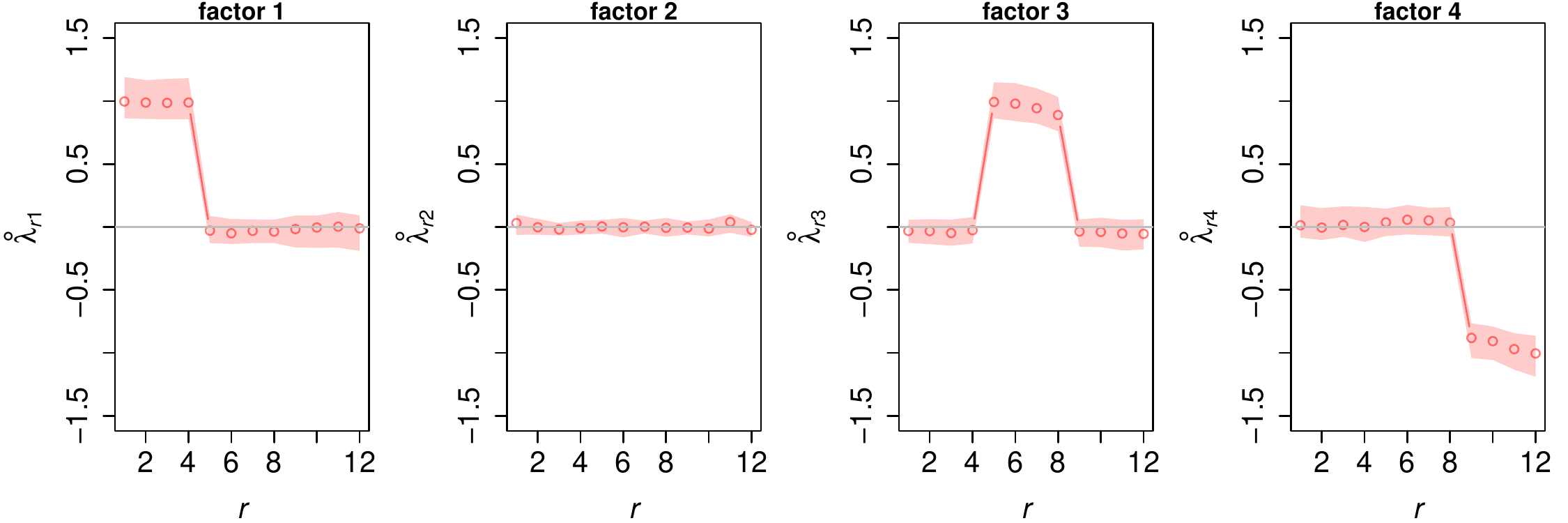}
\end{tabular}
\caption{Simulation study in Appendix \ref{sec:mfa_ccc}: Posterior means and $99\%$ HPD intervals for the reordered factor loadings of the CCC model (common loadings per cluster), when fitting MFA models with $q=3$ (first row) and $q=4$ (second row) factors.}
\label{fig:ccc_model}
\end{figure*}

A synthetic dataset of $n=200$ $p=12$-dimensional observations was generated from a two-component MFA model where the true number of factors is equal to $q=3$.  Following the suggestion of a reviewer, we have considered that  the factor loadings matrix as well as the idiosyncratic variance are common to all clusters, that is, 
\begin{align*}
\bs\Lambda_1 &= \bs\Lambda_2=\bs\Lambda\\
\bs\Sigma_1 &= \bs\Sigma_2= \sigma^2 I_p
\end{align*}
According to the real value for $\bs\Lambda$, variables 1-4 load to factor 1, variables 5-8 load to factor 2 and variables 9-12 load to factor 3. The mixing proportions are almost equal.

In order to estimate the MFA model, the {\tt fabMix} package was used. We considered that the number of factors ranges between 1 and 4 and fitted the 8 {\tt pgmm}  parameterizations available in {\tt fabMix}. Note that the number of clusters is estimated using overfitting mixture models, while the number of factors as well as the parameterization is estimated using BIC. The selected model corresponds to the ``CCC'' parameterization with $K = 2$ clusters and $q = 3$ factors, corresponding to the settings used to generate the data, that is, common factor loadings and common isotropic idiosyncratic variance per cluster.

We present results of our reordering approach which correspond to the selected parameterization for $q=3$ (that is, the true number) and $q=4$ factors.  The reordered factor loadings are displayed in Figure  \ref{fig:ccc_model}. Recall that in this model the factor loadings are shared between the two clusters.  In the first row, which corresponds to the ``true'' number of factors we see that variables 1-4 load to factor 1, variables 5-8 load to factor to 2 and variables 9-12 load to factor 3. These results are coherent with the underlying simulation scenario described previously. In the second row of Figure \ref{fig:ccc_model}, ($4$ factor model) the same variable-to-factor relationships are displayed for factors 1, 3 and 4, while the simultaneous credible region for the loadings of the second factor is centered around zero and this indicates the presence of one redundant column in the factor loadings matrix. 

\setcounter{equation}{0}
\setcounter{figure}{0}
\subsection{Mixtures of factor analyzers: The Wave dataset}\label{sec:wave}

\begin{figure*}[p]
        \begin{center}
                \begin{tabular}{c}
                        \includegraphics[scale=0.55]{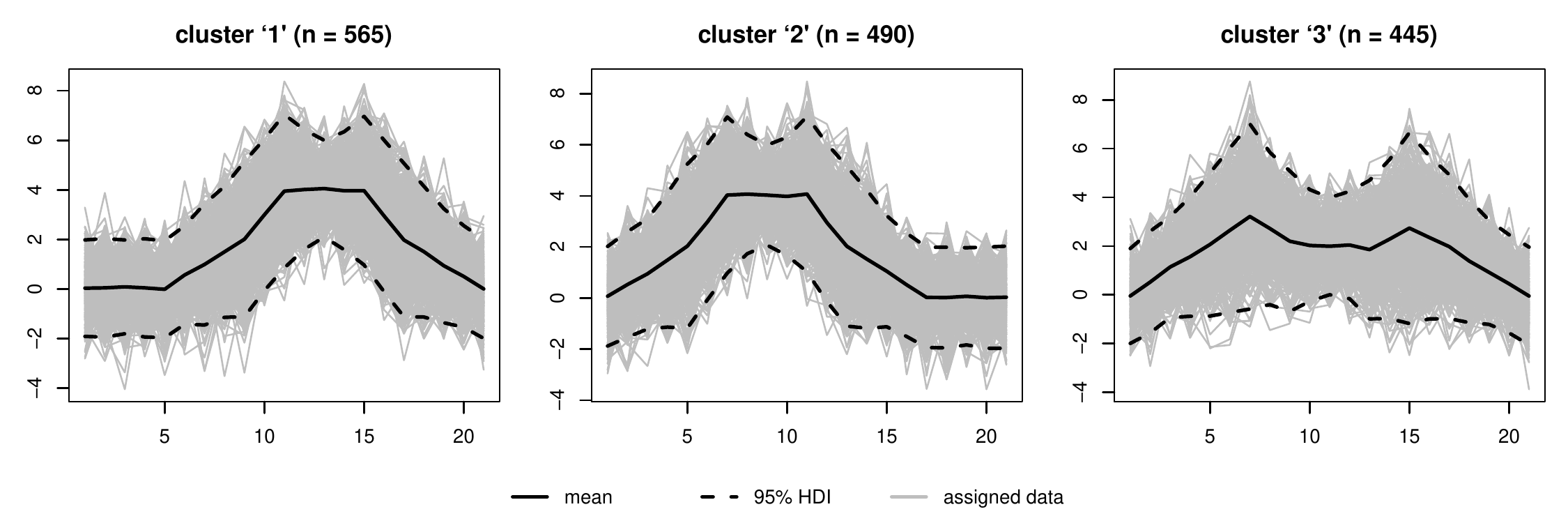}\\
                        \includegraphics[scale=0.55]{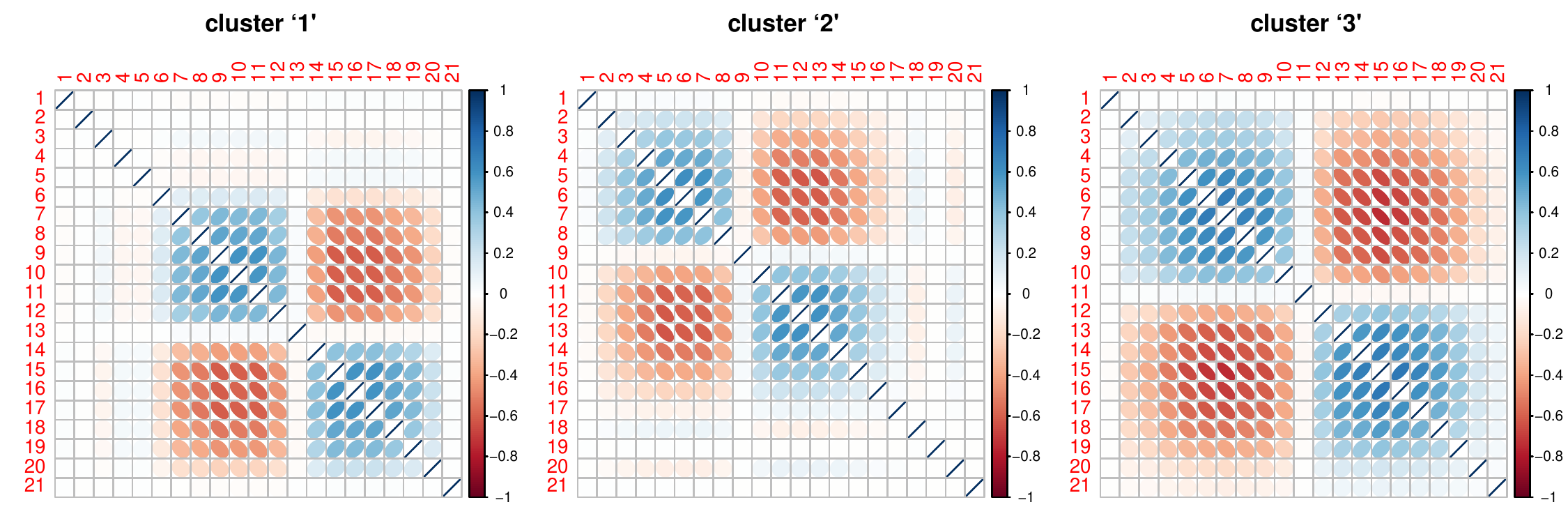}
                \end{tabular}
        \end{center}
        \caption{{\it Wave dataset}: Assigned (21-dimensional) observations  per cluster (1st row) and estimated correlation matrix per cluster (2nd row), according to a Bayesian mixture of factor analyzers with $K= 3$ clusters and $q = 1$ factors.}
        \label{fig:wave1}
\end{figure*}

\begin{figure*}[p]
        \begin{center}
                \includegraphics[scale=0.5]{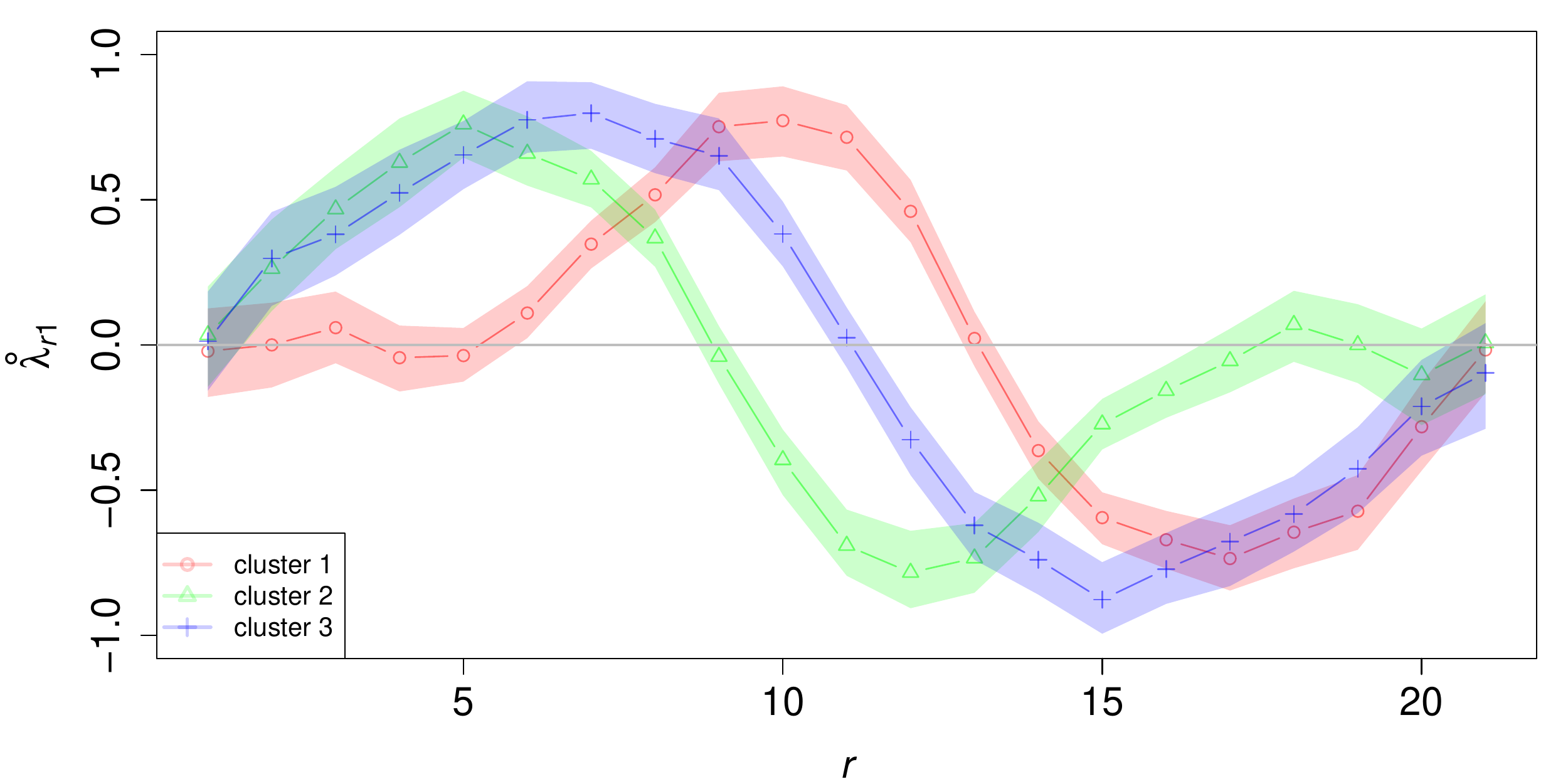}
        \end{center}
        \caption{{\it Wave dataset}: Posterior means and $99\%$ HPD intervals for the reordered factor loadings per cluster.}
        \label{fig:wave2}
\end{figure*}

We used the Wave dataset, available in the {\tt fabMix} package in {\tt R}. The dataset is generated from the Waveform Database Generator \citep{breiman} and consists of $p=21$ variables, all of which include noise, and there are 3 underlying classes of waves.  \cite{papastamoulis2018overfitting, papastamoulis2019clustering} fitted various parameterizations of the general model in Equation \eqref{eq:mixture} assuming an unknown number of clusters and using the Bayesian Information Criterion \citep{schwarz1978} for choosing $q$. The selected model corresponds to $K=3$ clusters and $q = 1$ factors, under the constraint that $\bs\Sigma_1=\bs\Sigma_2=\bs\Sigma_3$, see Table 4 in \cite{papastamoulis2019clustering}. The {\tt fabMix}  package \citep{fabMix} was used in order to produce an MCMC sample from the posterior distribution of the MFA model, using a prior parallel tempering scheme with 8 chains and a number of MCMC iterations equal to 20000. The clustered data as well as the inferred correlation matrix per cluster, conditional on $K=3$ and $q = 1$, is shown in Figure \ref{fig:wave1}. The posterior means and $99\%$ HPD intervals of the reordered factor loadings per cluster are presented in Figure \ref{fig:wave2}. 

\setcounter{equation}{0}
\setcounter{figure}{0}
\setcounter{table}{0}
\subsection{Exchange rate returns dataset}\label{sec:exchangeRates}

In this section we consider the returns on weekday closing spot rates for several currencies relative  to the U.S. dollar during the period from January 1, 1992,  to October 31, 1995 (1000 observations), used by \cite{10.2307/1392266}.  The currencies are the German mark  (DEM), British pound (GBP), Japanese yen (JPY), French franc (FRF), Canadian dollar (CAD), and Spanish peseta
 (ESP). Following \cite{10.2307/1392266}, we analyze the one-day-ahead returns, that is, $y_{ir} = s_{ir}/s_{i-1,r}-1$ where $s_{ir}$ denotes the original measurement at time-point $i=1,\ldots,1000$ and currency $r=1,\ldots,6$.

\begin{table*}[ht]
\centering
\begin{tabular}{clllll}
\toprule
Currencies & \multicolumn{2}{c}{2 factor model} & \multicolumn{3}{c}{3 factor model}\\
\midrule
DEM & \gr{-0.95(0.03)} & \gr{\hspace{0.77em}0.19(0.04)} & \gr{-0.92(0.05)} & \gr{\hspace{0.77em}0.20(0.09)}  & $0.06(0.18)$\\
GBP & \gr{-0.80(0.03)} & $\ -0.01(0.06)$ &\gr{-0.73(0.10)} &  $\ \ \ 0.07(0.10)$ & $0.30(0.26)$\\
JPY & \gr{-0.51(0.04)} &  \gr{\hspace{0.77em}0.48(0.09)}& \gr{-0.44(0.10)} & \gr{\hspace{0.77em}0.61(0.28)} &$0.00(0.14)$\\
FRF & \gr{-0.96(0.03)} &  \gr{\hspace{0.77em}0.18(0.04)} & \gr{-0.93(0.05)} &\gr{\hspace{0.77em}0.20(0.09)} &  $0.07(0.18)$\\  
CAD & $\ \ \ 0.03(0.03)$  & \gr{-0.34(0.06)} & $\ \ \ 0.04(0.03)$ & \gr{-0.40(0.25)} &$0.06(0.09)$ \\
ESP & \gr{-0.85(0.03)} & $-0.01(0.05)$ & \gr{-0.80(0.10)} &  $\ \ \ 0.08(0.08)$ &$0.26(0.22)$\\
\bottomrule
\end{tabular}
\caption{Estimates of posterior means (standard deviations) arising from the post-processed MCMC output of factor loadings of the exchange rate returns dataset of Section \ref{sec:exchangeRates} for 2 and 3 factor models. }
\label{tab:exchange}
\end{table*}

 \begin{figure*}[ht]
 \begin{tabular}{l}
 \includegraphics[scale=0.55]{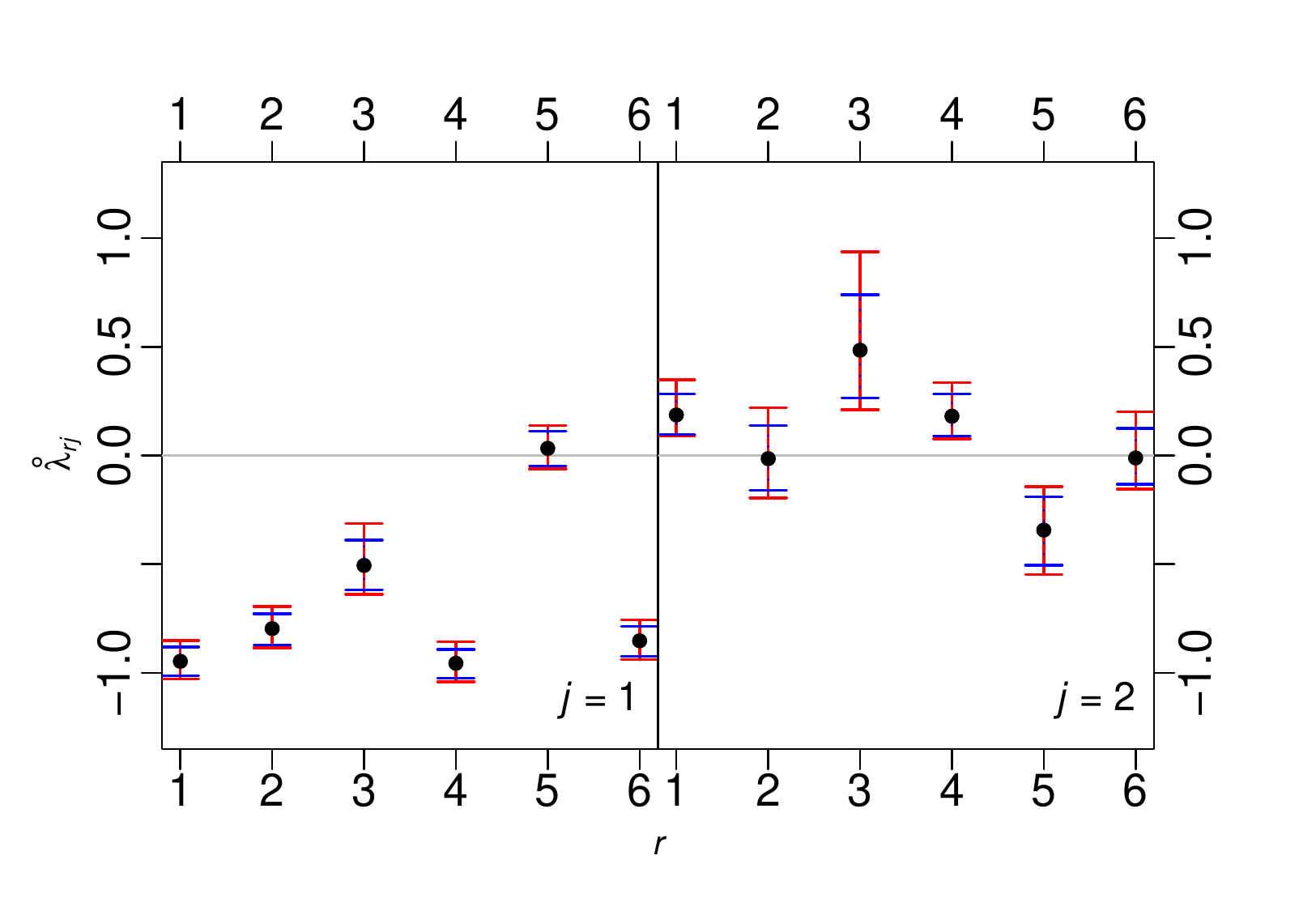}\\
  \includegraphics[scale=0.55]{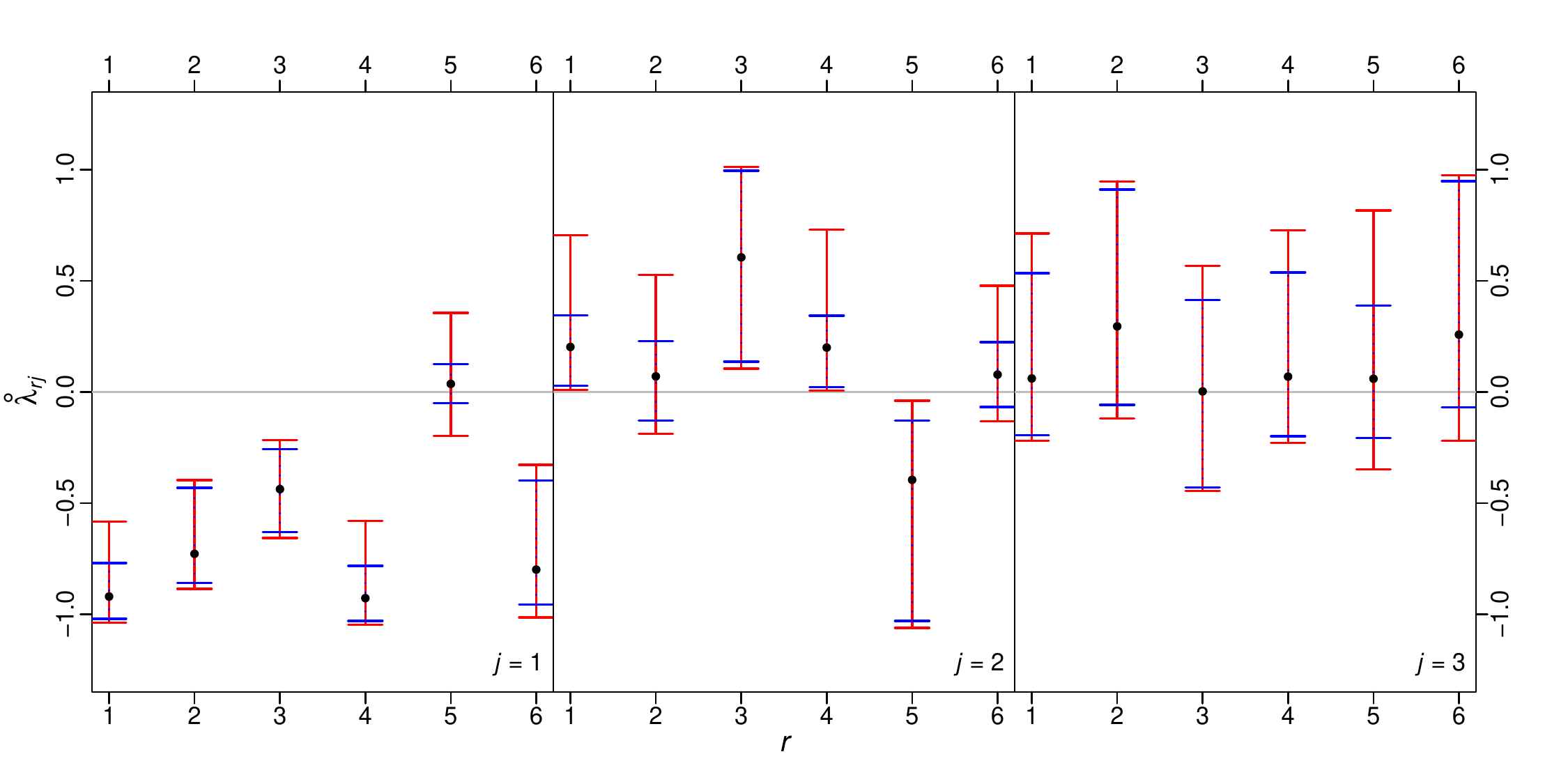}\\
 \end{tabular}
 \caption{Exchange rate returns dataset: $99\%$ HPD intervals (black) and simultaneous $99\%$ credible regions (red) of reordered factor loadings, when fitting Bayesian FA models with $q=2$ (top) and $q=3$ (bottom) factors.}
  \label{fig:exchange}
 \end{figure*}

As pointed out by an anonymous reviewer, this is an example  where the difference between small (``zero'') and large (``non-zero'') loadings is less pronounced compared to our previous analyses. Thus, the task here is to check whether such a structure is revealed in the post-processed MCMC output of factor loadings using a typical factor model. It is to be noted that  \cite{10.2307/1392266} modelled the dataset using dynamic factor models, which seems a more appropriate choice for such datasets. But we will focus on inference regarding the factor loadings,  which are also constant across time in the dynamic factor model of  \cite{10.2307/1392266}. 
 \cite{10.2307/1392266} present results for $q=3$, however in the Discussion section of the same paper is mentioned that a two factor model might be plausible for this type of data. We have used {\tt MCMCpack} to produce an MCMC sample of $20000$ iterations (after burn-in), for $q=2$ and $q=3$ factor models. The raw MCMC sample of factor loadings is post-processed according to the RSP algorithm. The estimates of the posterior mean for each factor loading (as well as the estimate of the posterior standard deviation in parentheses) is shown in Table \ref{tab:exchange}. An  asterisk indicates cases where the $99\%$ simultaneous credible region does not contain 0. Figure \ref{fig:exchange} illustrates the individual and joint $99\%$ HPD regions for both choices of number of factors. 

There are some common findings with \cite{10.2307/1392266}, despite the fact that we do not force any constraint on the matrix of factor loadings and that the model is different.   According to \cite{10.2307/1392266} (Table 1 in page 346) the first column of the matrix of factor loadings positively weights the first factor for  all currencies but CAD. Observe that, up to a sign-switching, this is the also the case for the first factor in Table \ref{tab:exchange}: zero is not contained in all currencies but CAD and moreover all of them (except CAD) have the same (negative) sign. This holds true for both models (that is, with 2 or 3 factors). The remaining columns are not directly comparable to the analysis of  \cite{10.2307/1392266} due to the fact that they impose the lower-triangular expansion in Equation \eqref{eq:lambda}, while they also force all diagonal elements to be equal to 1. Note however that the loadings of the second factor are essentially zero (GBP, ESP: notice that zero is contained in the simultaneous credible region), small (DEM, FRF) or moderate (JPY, CAD) values, and this fact is consistent with the task of identifying factors with less pronounced differences between small and large values of loadings. When considering $q=3$ factors, the simultaneous credible region of post-processed  values contains 0 for all loadings of the 3rd factor and this is an indication that a typical factor model with 3 factors may be over-parameterized (we emphasize once again that we do not use a dynamic factor model).

\bibliographystyle{unsrtnat}
\bibliography{main}  






\end{document}